%% file: main.tex
\documentclass[acmsmall,screen,nonacm]{acmart}
\AtBeginDocument{%
  }

\usepackage{amsmath}
\usepackage{algorithmic}
\usepackage{graphicx}
\usepackage{textcomp}

\usepackage{tabularx}
\usepackage{xltabular}
\usepackage{booktabs}
\usepackage{multirow}

\usepackage{tcolorbox}
\usepackage{colortbl}
\usepackage{csquotes}
\usepackage{hyperref}
\usepackage{paralist}
\usepackage{enumitem}
\usepackage{paralist}
\usepackage{listings}
\usepackage{relsize}
\usepackage{mdframed}
\usepackage{tikz}
\usepgflibrary{shapes.geometric}
\usepackage{ulem}
\usepackage{subcaption}
\usepackage{flushend}
\usepackage{xspace}
\usepackage{todonotes}
\usepackage{subcaption}

 \newcommand{\edit}[1]{#1}

\newcommand\parhead[1]{\vspace{1mm}\noindent\textbf{{#1}}}

\usepackage{xparse}
\NewDocumentCommand{\interview}{ >{\SplitList{,}} m }
{%
	{(\ProcessList{#1}{\myitem})}%
}

\newcommand\myitem[1]{I\textsubscript{#1}\let\myitem\myitema}
\newcommand\myitema[1]{,\,I\textsubscript{#1}}

\lstdefinelanguage{json}{
	numbers=left,
	numberstyle=\tiny,
	frame=leftline,
	rulecolor=\color{black},
	showspaces=false,
	showtabs=false,
	breaklines=true,
	basicstyle=\ttfamily\scriptsize,
	upquote=true,
	keywordstyle=\color{blue},
	keywords={name, description, specifications, elements, aspects, counteracts, checks, detects, analyzes},
	otherkeywords={applies-to, specified-by}
}

\usepackage[capitalize]{cleveref}
\Crefname{figure}{Figure}{Figures}
\crefname{figure}{Fig.}{Figs.}

\usepackage{framed}

\newlength{\leftbarwidth}
\newlength{\leftbarsep}
\renewenvironment{leftbar}{%
	\MakeFramed {\advance \hsize -\width \FrameRestore }%
}{%
	\endMakeFramed
}
\newcounter{ObservationIdx}
\newenvironment{observation}%
{\begin{leftbar}
		\refstepcounter{ObservationIdx}
		\noindent\textsc{\textbf{Finding\,\theObservationIdx.}}\@ }%
	{\end{leftbar}\vspace{-4pt}}

\newcommand{\validation}[2]{\vspace{.5em}\noindent\begin{tikzpicture}
    \node[align=center,draw,thin,minimum width=\columnwidth,inner sep=1.5mm] (titlebox)%
    {\parbox{\dimexpr \linewidth-2\fboxsep-2\fboxrule}{\looseness=-1\noindent\textit{#2\vspace{0.1cm}}}};
    \node[label=left:{\colorbox{white}{\small Validation Results for the #1}}] (W) at (titlebox.south east) {};%
    \end{tikzpicture}\vspace{-15pt}}

\newcommand{\recommendation}[1]{\vspace{.5em}\noindent\begin{tikzpicture}
    \node[align=center,draw,thin,minimum width=\columnwidth,inner sep=1.5mm] (titlebox)%
    {\parbox{\dimexpr \linewidth-2\fboxsep-2\fboxrule}{\looseness=-1\noindent\textit{#1\vspace{0.1cm}}}};
    \node[label=left:{\colorbox{white}{\small Recommendation}}] (W) at (titlebox.south east) {};%
    \end{tikzpicture}}

\newcommand{\model}{SecLan model\xspace}
\newcommand{\securityRelation}{\textit{If Compromised, also Compromised}\xspace}

\newcommand{\dsls}[0]{66\xspace}
\newcommand{\analyzers}[0]{36\xspace}
\newcommand{\checks}[0]{559\xspace}
\newcommand{\experts}[0]{22\xspace}
\newcommand{\validated}[0]{18\xspace}
\newcommand{\interviews}[0]{9\xspace}

\begin{document}

\title{Can I Check What I Designed? Mapping Security Design DSLs to Code Analyzers}

\author[S. Peldszus]{Sven Peldszus}
\orcid{0000-0002-2604-0487}
\affiliation{%
  \institution{Chalmers\,$|$\,University of Gothenburg}
  \city{Gothenburg}
  \country{Sweden}
}
\affiliation{%
  \institution{Ruhr University Bochum}
  \city{Bochum}
  \country{Germany}
}
\affiliation{%
  \institution{IT University of Copenhagen}
  \city{Copenhagen}
  \country{Denmark}
}
\email{sven.peldszus@gu.se}

\author[F. Reiche]{Frederik Reiche}
  \orcid{0000-0002-5993-0558}
  \affiliation{%
  \institution{Karlsruhe Institute of Technology}
  \city{Karlsruhe}
  \country{Germany}
}
\email{frederik.reiche@kit.edu}

\author[K. Hermann]{Kevin Hermann}
\orcid{0009-0004-6207-4045}
\affiliation{%
  \institution{Ruhr University Bochum}
  \city{Bochum}
  \country{Germany}
}
\email{kevin.hermann@rub.de}

\author[S. Corallo]{Sophie Corallo}
  \orcid{0000-0002-1531-2977}
  \affiliation{%
  \institution{Karlsruhe Institute of Technology}
  \city{Karlsruhe}
  \country{Germany}
}
\email{sophie.corallo@kit.edu}

\author[T. Berger]{Thorsten Berger}
\orcid{0000-0002-3870-5167}
\affiliation{%
  \institution{Ruhr University Bochum}
  \city{Bochum}
  \country{Germany}
}
\email{thorsten.berger@rub.de}
\affiliation{%
  \institution{Chalmers\,$|$\,University of Gothenburg}
  \city{Gothenburg}
  \country{Sweden}
}

\author[R. Heinrich]{Robert Heinrich}
  \orcid{0000-0003-0779-9444}
  \affiliation{%
  \institution{Ulm University}
  \city{Ulm}
  \country{Germany}
}
\email{robert.heinrich@uni-ulm.de}

\renewcommand{\shortauthors}{Peldszus et al.}

\begin{abstract}
When assessing the potential impact of code-level vulnerabilities, e.g., discovered by automated analyzers, it is essential to consider them in the context of the system’s security design. However, this is a challenging task due to the abstraction gap between security design, often specified using security DSLs, and implementation. As we will show, even security experts lack a complete understanding of this relationship. Intrigued by this gap—and the general disconnect between secure design and secure implementation—we present a study of \dsls design-level security DSLs and \checks security checks from \analyzers code-level analyzers. We identify what concepts are common to both and capture them in the SecLan model, which has been validated by \experts security experts. Based on this, we investigate the relationship between DSLs and analyzers quantitatively and explore it qualitatively together with \interviews security experts. We learn that there are few commonalities between design-level and implementation-level security; security checks are often described by overly general weaknesses, resulting in many non-obvious potential relationships between security DSLs and analyzers; and even security experts are overwhelmed by this complexity. We provide an empirical basis that helps practitioners and researchers better understand the gap and serves as a first step toward bridging it.
\end{abstract}




\maketitle
\input{sections/01_intro}
\input{sections/02_background}

\input{sections/03_methodology}
\input{sections/04_0_conceptual_model.tex}

\input{sections/05_evaluation}
\input{sections/05_1_discussion}
\input{sections/06_related}
\input{sections/07_conclusion}

\begin{acks}
We would like to thank everyone who participated in our questionnaire and survey.

This work was supported by funding from the project FeCoMASS by the Deutsche Forschungsgemeinschaft (DFG, German Research Foundation) under project number 499241390, and the topic Engineering Secure Systems of the Helmholtz Association (HGF), by KASTEL Security Research Labs (46.23.01 Methods for Engineering Secure Systems), and by the Deutsche Forschungsgemeinschaft (DFG, German Research Foundation) - SFB 1608 - 501798263, and by the Deutsche Forschungsgemeinschaft
(DFG, German Research Foundation) under Germany’s Excellence Strategy - EXC 2092 CASA - 390781972.

\end{acks}

\bibliographystyle{ACM-Reference-Format}
\bibliography{bibliography.bib}

\appendix

\end{document}

%% file: sections/01_intro.tex
\section{Introduction}
\noindent
Considering security as a core aspect during development is becoming increasingly important.
One of the most important steps is the design of a secure software architecture\,\cite{McGraw2004}. 
Since insecure design is one of the OWASP Top 10 security risks\,\cite{OWASPTop10}, architects need to consider typical threats, vulnerabilities, and weaknesses. They also need to adhere to security principles or realize secure design patterns, such as distrustful decomposition, privilege separation, or secure chain of responsibility\,\cite{Dougherty2009}. 
To this end, many domain-specific languages (DSLs)\,\cite{Wasowski2023} for secure software design have been presented in the literature\,\cite{Uzunov2012, Nguyen2015, DenBerghe2017, Mashkoor2023} (\dsls of which we analyze in this paper). These security DSLs are important means to design secure systems that meet their security requirements\,\cite{DenBerghe2017}. 
They allow specifying the security design of a system in terms of security-related decisions, such as defining what critical information is and planning in which parts of the system it will be processed\,\cite{Tuma2019a}, and reasoning about the security of the design\,\cite{Uzunov2012, Nguyen2015, DenBerghe2017, Mashkoor2023}.

\begin{figure}
    \centering
    \includegraphics[width=0.6\linewidth]{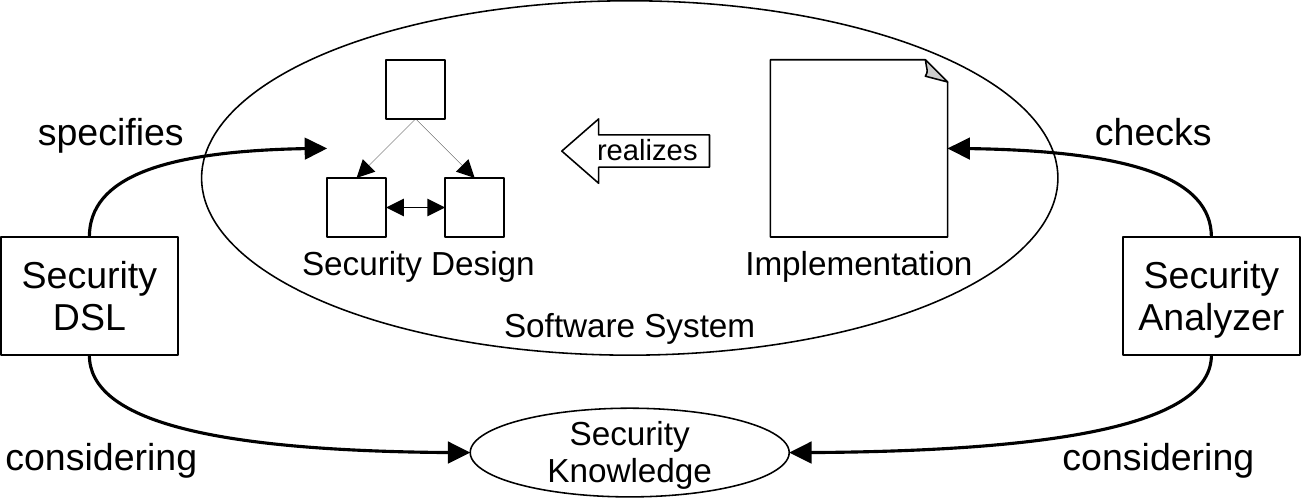}
    \caption{Overview of potential relationships between security design and implementation}
    \label{fig:concept}
\end{figure}

\edit{As shown in \cref{fig:concept}, this security design must then be realized by the system's implementation} as planned to obtain a system that is actually secure.
However, systematically considering security and ensuring proper implementation is particularly difficult for developers\,\cite{Hermann2025}. 
It is a common misconception that developers are well-informed about software security\,\cite{Ryan2023}.
As a result, implementations often contain vulnerabilities that invalidate or fail to conform to the assumptions of the security design. 
For example, improper data handling may expose information that was identified as confidential in the design to unauthorized actors---a common weakness\,\cite{cwe200}.
The security design expressed using the DSLs serves as the reference against which the system, when implemented, must be manually checked to assess its security\,\cite{Tuma2022,Peldszus2022a}.
It is necessary that security experts with deep expertise in programming and the code at hand \textit{thoroughly review the implementation to ensure that the security design is implemented correctly and securely}\,\cite{Baca2009}.
However, these reviews are laborious and error-prone\,\cite{Edmundson2013}. 

So-called security analyzers strive to automate code analysis by performing code-level security checks.
Many have been presented in the literature (\analyzers of which we analyze in this work) to identify vulnerabilities in the code\,\cite{Singh2017}.
Their checks range from detecting possible buffer overflows\,\cite{Cowan1998}, via insecure API usage\,\cite{Zhang2023}, to insecure data flows (e.g., via taint analysis, a scalable data-flow analysis technique)\,\cite{Giffhorn2008,Arzt2014a}.
Since complete security is not feasible due to resource constraints\,\cite{Hermann2025,Mazurek2022}, prioritization of the usually vast number of security check results is required to fix the vulnerabilities with the highest impact first\,\cite{Walden2014}, i.e., those that lead to violated assumptions of the planned security design.
This requires considering project-specific knowledge---that among others is captured using security DSLs---in the prioritization of vulnerabilities detected by security analyzers\,\cite{shedden2011incorporating}.
So, despite automated support, developers still need to manually map the analysis results to the security design to assess their impact.

A core problem is that the relationship between code-level security (checked by code analyzers) and the planned security design (represented by design models expressed in security DSLs) is not systematically captured\,\cite{Zhioua2014}.
\edit{While studies show that analyzer findings can be prioritized based on the general strength of their relationship to the security design\,\cite{Peldszus2025},} it is not even clear whether and to what extent the security checks provided by code analyzers relate to the \edit{concrete} security aspects modeled in security DSLs.
\edit{As shown, in \cref{fig:concept}, security DSLs and analyzers are both based on broader security knowledge. However, it is unclear which parts of this knowledge are considered by both.}
As we will show, even security experts are unaware of all possible relationships and neglect the potential impact of some vulnerabilities detected by analyzers on the security objectives of their security DSL.
Without bridging the gap between security design and code-level security, the impact of insecure coding on the overall security design is unclear.
We need to systematically map the security aspects modeled in security DSLs to the security checks provided by analyzers to identify the commonalities that can be used to relate each vulnerability to the security design---the goal of this work.



\looseness=-1
To bridge this gap, we present a study of common and popular representatives of security DSLs and static code analyzers with respect to their commonalities\edit{, e.g., what are the common parts of general security knowledge that both of them consider (see \cref{fig:concept})}.
Our results provide the basis for a holistic understanding of software security that is not limited to individual security aspects, e.g., secure data handling, and security features such as cryptography.
Based on our findings, we synthesize the SecLan model, which captures what are the common security-related aspects between security DSLs and code-level analyzers. This allows them to be described according to these common concepts and to reason about the relationships between design-time security, as addressed by security DSLs, and implementation-level security, as targeted by security analyzers. 
We contribute:

\begin{enumerate}[label=\textbf{C\arabic*:},labelsep=2pt,leftmargin=17pt]
    \item The \model of the commonalities between design-time security DSLs and code-level security checks of static analyzers, systematically extracted from the analysis of \dsls security DSLs and \edit{\analyzers} security analyzers providing \edit{\checks} security checks, validated by \edit{\experts security experts.} 

    \item A classification of all \edit{\dsls security DSLs and \analyzers security analyzers} studied according to the \model, \edit{\validated of which have been validated} by their authors.
	We provide them in an online repository\,\cite{Replication} \edit{in a machine-readable format based on a JSON-like DSL introduced in this paper, as well as a human-readable HTML export of this format, to support} replication and further studies.
    
    \item A technique for deriving explicit relationships between security DSLs and the checks of implementation-level security analyzers. This technique enables the systematic investigation of the relationship between the security design expressed in security DSLs and implementation-level security, facilitating the understanding of how the check results of security analyzers affect the security design.
    
    \item A quantitative analysis of the relationship between security DSLs and security analyzers, quantifying the relationship between design-time security aspects and implementation-level vulnerabilities.
    
    \item A qualitative exploration of the relationship between security design and implementation-level vulnerabilities that can be detected by a security analyzer with seven security experts, providing actionable results for practitioners and researchers on the relationship between design-time security and code analysis.

\end{enumerate}

\noindent 
Our work is a necessary step towards bridging the gap between design-time and code-level security.
We have identified the concepts that are common to all security checks and DSLs, derived a technique to make the relationships between the two explicit, and developed tooling to automatically extract and explore their relationships.
Using this tooling, we investigated the relationship quantitatively and qualitatively, providing insights that allow practitioners to more effectively assess the impact of analyzer findings on the security design, and help researchers bridge the gap.
Furthermore, our insights and the public release of our dataset\edit{, including the classification and description of all \dsls security DSLs and \analyzers security analyzers according to the SecLan model,} and developed tooling\,\cite{Replication} will benefit the community for future research.

%% file: sections/02_background.tex
\section{Motivation and Background}
\noindent
We now discuss the development of secure software systems. We follow the security-by-design paradigm as a best practice, where security is considered early in the development process.
We present security DSLs and static code analyzers as the, arguably, most relevant techniques for security by design. 
We discuss their limitation to motivate our study.
Our illustrative example is the electronic health records system iTrust\,\cite{Heckman2018,Meneely} for managing patient care in a hospital.
The system has been developed at North Carolina State University based on detailed requirements that comply with the US HIPAA statute for ensuring the security and privacy of patient records\,\cite{Massey2008}.
In the following, we discuss the use case of a doctor using iTrust to document office visits and to manage patient health records.

\subsection{Security Features}
To implement a secure software system, developers must plan and implement security features.
Security features are functionalities of a software system that address security issues by protecting assets, i.e., patient-related data in iTrust, preventing a security attack, or mitigating the damage caused by a security attack\,\cite{McGraw2004}.
In this context, security features increase a system's resistance, tolerance, or resilience against threats\,\cite{Allen2008} or directly aim to resist, detect, or recover from attacks\,\cite{Bass2003}.
They can be realized by applying security patterns in different stages of the software development lifecycle, which include threat assessment, risk assessment, or validation\,\cite{schumacher2013}.
Through these, they address functional security requirements to realize, e.g., authentication, access control, or other security aspects\,\cite{tsipenyuk2005,hermann2025taxonomy}.
The different levels of granularity at which security can be considered are one of the core challenges in engineering security features\,\cite{Peldszus2022a}.

\subsection{Security Standards}
\edit{
Security standards are guidelines used by organizations to plan and organize security measures for software systems.
Depending on the domain, following such standards can be legally binding to be allowed to bring a software-based product into a market.
As such, they propose security features that developers need to implement to secure a software system.
In addition, many security standards require that the organization follows certain security processes.
For example, security standards such as the Common Criteria (ISO/IEC JTC 1/SC 27 2009) \cite{CC} or ISO/SAE 21434 for road vehicles (ISO/TC 22/SC 32 2021) \cite{ISO21434}, require maintaining and tracing security features \cite{hermann2025taxonomy}.
Each control of a standard, i.e., measures to be taken or implemented to comply with the standard, commonly lists security objectives according to the CIA triad (confidentiality, integrity, availability) to state the control's goals.
In the Common Criteria, an evaluator analyzes the system for vulnerabilities as part of a vulnerability assessment (abbreviated AVA\_VAN in the Common Criteria) to test whether all security objectives of the standard are upheld.
While security standards require considering security across both design and implementation phases, they provide little guidance on how to connect these phases, leaving practitioners without a clear path to ensure end-to-end compliance. Our work addresses this gap by empirically analyzing the relationship between design-level DSLs and code-level analyzers, providing the first step toward a unified understanding that standards alone cannot deliver.
}

\subsection{Security By Design}
\begin{figure}
	\center
	\includegraphics[width=.6\linewidth]{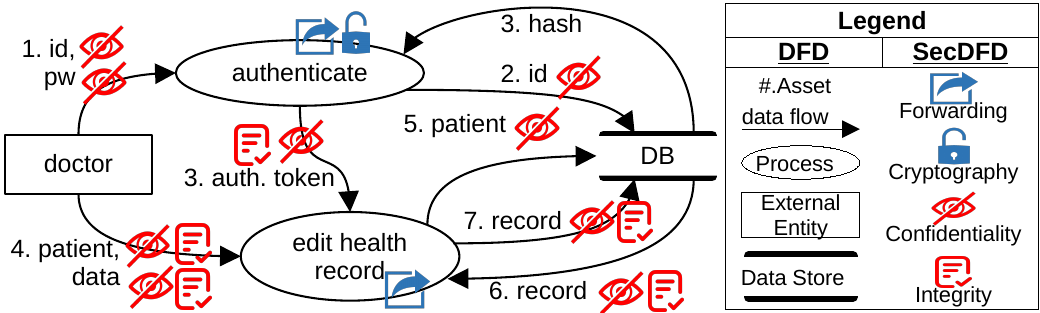}
	\caption{Excerpt of a design model with annotations from the SecDFD security DSL, showing the editing of a health record}
	\label{fig:secdfd}
\end{figure}

As discussed above, it is essential to plan security, i.e., which security features are needed where, early\,\cite{McGraw2004}, especially in software design.
This paradigm of considering security as early as possible is called security by design.
To achieve security by design, security experts can use techniques such as \textsc{STRIDE} threat modeling\,\cite{Shostack2008} (\textsc{STRIDE}: Spoofing, Tampering with Data, Repudiation, Integrity, Denial of Service, and Elevation of Privilege) to systematically identify and reason about possible security threats to the system early.
The identified threats are then used to select appropriate security features to mitigate the identified threats at design time.
To systematize this reasoning, data flow diagrams (DFDs) are used to identify and specify planned data flows in the software architecture (i.e., between external entities, processes, and data stores).
As an example of such a DFD, \cref{fig:secdfd} shows an excerpt of a DFD for iTrust, focusing on the use case of a doctor managing health records.
Intuitively, doctors can edit patients' health records, but must first authenticate to the system; hashed login credentials and patient health records are stored in the same database.
For each of the identified elements, a security expert then reasons how the STRIDE threat categories may impact that element. 
For example, doctors authenticating at the system using their IDs and passwords could be threatened by spoofing, i.e., a third party being able to pretend to be a doctor in the hospital that has access to sensitive information.
Although considering a specification of the system architecture allows for more systematic reasoning about possible threats early in software development, relevant information for this reasoning, such as that the ID and password are sensitive information used to identify a user, is not explicitly captured.
\textit{Modeling security threats for reasoning requires languages to express the system and all relevant security threats, typically called security DSLs.}

\subsection{Design-Time Security DSLs}
Security DSLs allow the specification of, and the reasoning about, design-time security concerns. 
In general, DSLs are languages that are dedicated to a specific context or domain, i.e., security, and facilitate domain-specific tasks by providing task-specific abstractions.
While DSLs are less expressive than programming languages, they are effective by focusing on a particular domain\,\cite{Wasowski2023}, such as security.
Security DSLs are often security-specific extensions of common software modeling languages, e.g., UML models or DFDs as shown in \cref{fig:secdfd}.
Many model-based secure design techniques have been introduced, typically complemented by DSLs for capturing security knowledge and reasoning about this knowledge.
As such, security DSLs typically allow annotating software design models with security requirements and reasoning about them (e.g., about consistency).
For example, SecDFD\,\cite{Tuma2019a} improves the planning of secure data processing in terms of joining, forwarding, or cryptographic operations using DFDs.
\Cref{fig:secdfd} uses SecDFD to specify which assets represent secrets and which security-related processing (e.g., cryptographic hashing) is planned.
Similarly, DSLs such as UMLsec\,\cite{Juerjens2005} support annotating UML models with security requirements---among others, allowing to structure systems into security levels and to plan secure data communication. 
Another example, SecBPMN, allows annotating business process models\,\cite{Salnitri2015}.

Security DSLs have been studied thoroughly\,\cite{Nguyen2015,DenBerghe2017,Uzunov2012} and support security reviews and planning secure software design. 
Unfortunately, no study has investigated the relationship between the security design expressed using these DSLs and the implementation.
To the best of our knowledge, \textit{security DSLs typically provide no means to check the conformance of code to the security models expressed in the DSLs. 
Although many automated static code analyzers are available, they do not enable automatic checking of compliance with the planned security design of the respective software project.}

\subsection{Generative Security Technologies}
\noindent
Although design-time security DSLs allow for relatively detailed planning of secure systems, their implementation remains primarily a manual task.
There are generative techniques that assist at design time, e.g., by generating secure architectures and access control configurations from security requirements models, but actual code generation is limited.
For security protocol DSLs code generators exist that can generate code for executing these protocols, e.g., Java code\,\cite{Modesti2014}, but the generated code has to be manually embedded into the system.
Also, these DSLs are usually not aimed at system development, but at the development of security features such as cryptography, which are usually not written by system developers, but are security features that are reused from libraries\,\cite{Hermann2025}.
Still, there are works that generate code from design models extended with security DSLs, e.g., UML models annotated with UMLsec, but the code generation is limited, e.g., only the general class structure is generated, which facilitates manual identification of the locations where security features are needed, but the security-critical code is not generated\,\cite{Arogundade2021}.

Literature primarily focuses on the generation of secure code, today with a particular focus on generative AI\,\cite{Sai2024}.
However, general code assistants tend to produce insecure code\,\cite{Pearce2022}, which requires a manual review of generated code by security experts.
To counteract these weaknesses, researchers experiment with how to adjust prompts to generate more secure code\,\cite{Nazzal2024}.
On the other side, code generators explicitly focusing only on security have been studied, i.e., a security model for hardening existing code\,\cite{Li2024}.
Secure code generation based on traditional AI uses techniques such as code templates to generate secure usages of cryptographic APIs\,\cite{Krueger2020}.
\textit{While these techniques help developers avoid insecure security feature implementations when implementing a system, they do not help in assessing whether all planned security features have been implemented as expected}---even if an implementation of a security feature is locally secure at the code level, its embedding in the overall system may still be insecure.

\subsection{Tool-supported Security Engineering}
\edit{
Tool-supported security engineering comprises interactive development and verification processes that rely on specialized tools to analyze implementations against formally specified security properties. In contrast to high-level design-time modeling approaches, these tools typically operate at a low level of abstraction, close to the source code or even the generated machine code. Representative examples include B4MSecure\,\cite{b4msecure}, F*\,\cite{fstarpopl}, Frama-C\,\cite{framac_book}, Jasmin\,\cite{jasmin}, and CryptoVerif~\cite{cryptoverif}. These tools enable developers to encode precise security properties—such as memory safety, functional correctness, cryptographic security, or information-flow guarantees—and to incrementally refine high-level security goals into concrete, machine-checkable verification conditions.

Despite their strong assurance guarantees, tool-supported verification approaches incur substantial costs in terms of expertise, specification effort, and verification time. They often require detailed formal models, user interaction with theorem provers, and tight coupling between code and specifications. Consequently, their applicability is largely limited to highly critical systems and components, such as cryptographic libraries, microkernels, or software evaluated at high assurance levels (e.g., EAL~6 or EAL~7 under the Common Criteria\,\cite{CC}). For many practical development contexts, these approaches remain prohibitively heavyweight. Therefore, this paper focuses on more lightweight security analysis techniques that aim to improve scalability and usability while still providing actionable security insights.
}

\subsection{Security Analyzers}
\label{sec:background:implementationcecks}
%
%
At the code level, vulnerabilities can be detected automatically by security analyzers that check code for instances of weaknesses.
To this end, security analyzers provide one or multiple security checks. 
For example, the information flow analysis JOANA\,\cite{Giffhorn2008} or the\textit{OWASP Dependency Check}\,\cite{OWASPDependency} provide a single check each, while analyzers such as \textit{SonarQube}\,\cite{SonarSource,Lenarduzzi2018} or CodeQL\,\cite{Moor2007,CodeQL} provide numerous checks.

The literature classifies analyzers according to various dimensions:
Chess and West\,\cite{Chess2007} distinguish based on the technique employed to check code, considering, among others, \textit{Program verification and property checking}, i.e., techniques that verify code against a specification, and \textit{Bug finding static analysis tools}, i.e, searching for predefined source code patterns. 
Similarly, Lipp et al.\,\cite{Lipp2022} differentiate \textit{syntactic} and \textit{semantic} analyzers.
While \textit{Syntactic analyzers} conform to the definition of \textit{Bug finding static analyzers} by Chess and West, \textit{Semantic analyzers} leverage additional information for analysis, such as control flow or data flow, by lifting the source code to a more abstract representation. 
Another dimension is the code supported, such as Java or C\,\cite{McGraw2004}.
Al Breiki and Mahmoud\,\cite{AlBreiki2014} differentiate analyzers based on the format of the implementation they are analyzing: \textit{source code}, \textit{byte code} or \textit{binary}. 
Nachtigall et al.\,\cite{Nachtigall2022} differentiate tools by groups of usability issues such as \textit{Understandable Warning Messages}, \textit{Fix support}, or \textit{User Feedback}. 
In addition, they categorize analyzers based on the interfaces they provide, such as \textit{Command-Line Tools}, \textit{Standalone}, i.e., with a separate graphical user interface, \textit{IDE tools}, or \textit{Tools with multiple interfaces}. 

All of the above classifications focus on the analysis techniques employed or how developers can use the tools, but not on how they actually relate to security, which is the relevant aspect for this work. 
Besides these classifications, analyzers themselves and literature, e.g., the OWASP Source Code Analysis Tools List\,\cite{OWASPFoundation2022}, commonly classify analyzers according to what they are capable of detecting.
Thereby, they mainly use the \textit{Common Weakness Enumeration (CWE)}, a list of known weaknesses, as a reference.
Although the CWE categorizes a weakness by its influence on high-level security goals, such as \textit{confidentiality}, \textit{integrity}, \textit{availability}, \textit{access control}, or \textit{non-repudiation}, assessment of a detected weakness still requires huge manual effort and considerations since these objectives are very broad.
While this enables weaknesses to provide some degree of abstraction that can help developers, they do not allow the categorization of analyzers due to their large number and still relatively specific nature.
Therefore, taking into account the above classifications and considering further literature, we identified the following categorization of checks that can be provided by security analyzers:


\textit{1) API -- Security API Usage:}
This category includes checks that detect usage patterns of security-related APIs, e.g., cryptographic APIs, that are known to be insecure.
A prominent subclass (e.g., analyzers such as \textit{CogniCrypt}\,\cite{Krueger2017}) focuses on cryptographic APIs, checking for weak encryption algorithms or creating a secret key with hard-coded constants.
This category is mentioned by Zhang et al.\,\cite{Zhang2023} and Kulenovic et al.\,\cite{Kulenovic2014}

\textit{2) IF -- Information Flow:}
Analyzers like \textit{JOANA}\,\cite{Giffhorn2008} examine information flow---that is, data flow and control flow that may carry information implicitly\,\cite{Hough2022}---to identify violations of a given policy, i.e., critical information not flowing to a location.
Taint analyzers, such as \textit{FlowDroid}\,\cite{Arzt2014a}, propagate taint labels along data flows but often only consider explicit flows\,\cite{Balliu2017}.
This class is mentioned by Kulenovic et al.\,\cite{Kulenovic2014}.
This category includes checks with a dedicated purpose of investigating information flows. 
We do not cover checks that use information flow analysis to detect other kinds of issues. 

\looseness=-1
\textit{3) D -- Dependency:}
Analyzers such as \textit{OWASP Dependency Check}\,\cite{OWASPDependency} examine dependencies, e.g., as specified in build systems such as Maven, to determine whether used libraries contain known but unfixed vulnerabilities.
This class of analyzers is mentioned in the survey of Imitiaz et al.\,\cite{Imtiaz2021}.

\looseness=-1
\textit{4) VCP -- Vulnerable Coding Practices:} These checks search for coding practices known to be insecure, such as code prone to buffer overflows\,\cite{Cowan1998}, null pointer dereference\,\cite{Jin2021}, but also using inherently dangerous functions (e.g., \textit{FlawFinder}\,\cite{Wheler2001}).
In contrast to \textit{Security API Usage}, these coding practices result in security vulnerabilities without direct association with security-related functionality, such as encryption or access control.

\textit{5) P -- Permissions:} This category comprises checks that identify problems related to permission management, such as missing or problematic permission checks (e.g., \textit{PaX}\,\cite{Zhang2019}) or more permissions being claimed than necessary (e.g., Tang et al.\,\cite{Tang2017}).
This category also relates to access control vulnerabilities described by Kulenovic et al.\,\cite{Kulenovic2014}.
\smallskip

\begin{figure}
	\includegraphics[width=\columnwidth]{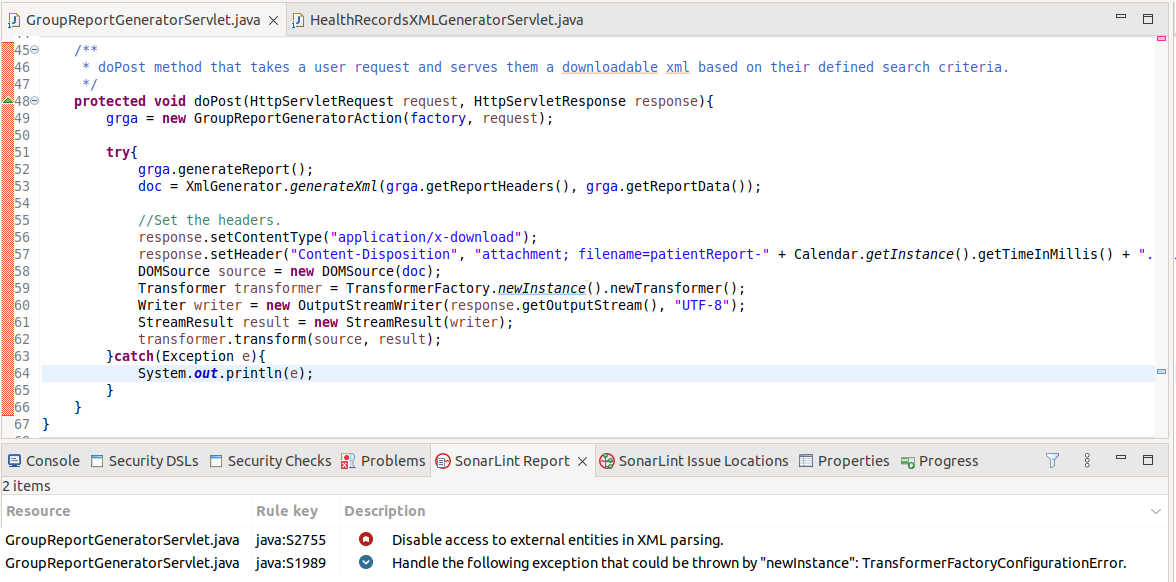}
	\caption{SonarQube check results on iTrust}
	\label{fig:sonarqube}
\end{figure}

While vulnerabilities in all of the categories introduced above can have a severe impact on the security of a software system,
\textit{unfortunately, static analyzers may output many potential security issues that can be challenging to evaluate concerning their impact on the security design of the project at hand.}
We illustrate this issue by applying \textit{SonarQube}\,\cite{SonarSource,Lenarduzzi2018}, which comprises security checks from most categories identified, to iTrust,
where it detects 1,352 issues, 37 of which are explicitly security-related.
The remaining issues could potentially be security-related, because they include reliability issues and references to known weaknesses from the \textit{Common Weakness Enumeration (CWE)}\,\cite{cwe}.

As an example, \cref{fig:sonarqube} shows the source code of a class that implements the functionality of the \textit{edit health report} process in \cref{fig:secdfd}.
This code allows doctors to export their reports in an XML format.
In this code, SonarQube found two security issues.
\edit{The first issue, S2775, points to a potential XXE attack vulnerability for a used XML parser and is associated with CWE-611 (Improper Restriction of XML External Entity Reference), which occurs when an XML parser processes external entities without proper restrictions, allowing attackers to access local files, cause denial-of-service, or execute arbitrary code by referencing malicious URIs.
It is also linked to CWE-827 (Improper Control of Document Type Definition), where the software fails to restrict DTD references, enabling attackers to manipulate XML parsing, consume excessive resources, or expose sensitive data.
The second issue, S1989, identifies that an exception is thrown from a servlet method and is associated with CWE-600 (Uncaught Exception in Servlet).
Thus, the issue allows exposure of sensitive debugging information or crashing the application if exceptions are not properly handled, potentially revealing stack traces or system details to attackers
}

These three weaknesses (CWE-611, CWE-827, and CWE-600) describe seven possible technical impacts, such as bypassing protections, that could compromise either confidentiality, integrity, or availability, access control to the system, or a combination of these four.
However, most of these findings likely have no direct impact on the security of iTrust.
It is up to the user to determine the actual impact and whether any of the system's security objectives are actually compromised.
\textit{Inspecting all reported issues is tedious, and their impact on the security design is difficult to grasp.}

\subsection{Security Compliance Checks}
Some techniques already check code for specific design-time security aspects, such as structural compliance with DFDs\,\cite{Peldszus2019}
or compliance with SecDFD security aspects\,\cite{Tuma2022}.
Existing analyzers such as FlowDroid\,\cite{Arzt2014a} are automatically configured based on security objectives specified using SecDFD\,\cite{Tuma2022}, or are used to check security specifications from design models\,\cite{Geismann2021, Katkalov2013,Toeberg2022,Yurchenko2017}.
%
Other checks utilize call patterns that violate design-time security aspects
\,\cite{Peldszus2021}.
Concepts such as dynamic class loading also require dynamic compliance checks\,\cite{Peldszus2024,Peldszus2022a}.
\textit{While these works successfully demonstrate the feasibility of linking design-time security with code analysis, they are tailored to specific security DSLs and code checks. They provide a foundation for our work, but a systematic and general study that relates design-time security DSLs and code analyzers is lacking.}

%% file: sections/03_methodology.tex
\section{Methodology}
\noindent\looseness=-1
To study the gap between design-time and code-level security, we conducted a systematic review\,\cite{Ralph2023} of state-of-the-art design-time security DSLs and security analyzers as representatives of design-time and code-level security.
As a result of this review, and as an input to enable the systematic analysis of the gap, we synthesized the SecLan model that captures (and enables reasoning about) the relationship between security DSLs and security checks of code-level analyzers.
Based on this model and the concrete security DSLs and checks of security analyzers studied, we quantitatively analyzed the relationship between them.
We conducted interactive interviews\,\cite{Ralph2023} with security experts to qualitatively explore the relationships between design-time security DSLs and security checks to gain further insights.
\Cref{fig:methodology} shows our methodology in detail.

\subsection{Data Sources}
\noindent
To identify relevant security DSLs and code analyzers, we searched for survey papers and supplemented the DSLs and analyzers included therein with additional ones based on expert knowledge.
We provide a list of all selected security DSLs and analyzers in  \cref{tab:dsls,tab:analyzer}, and further details in our replication package\,\cite{Replication}.

\parhead{Security DSLs.}
We identified four surveys\,\cite{Uzunov2012, Nguyen2015, DenBerghe2017, Mashkoor2023} on model-driven security engineering that mention 105 papers, of which 57 provide security DSLs.
Specifically, we considered 37 security DSLs from the survey of Nguyen et al.\,\cite{Nguyen2015}.
The surveys by Van den Berghe et al.\,\cite{DenBerghe2017} and Uzunov et al.\,\cite{Uzunov2012} have a significant overlap with the previous survey and list 9 and 8 additional security DSLs, respectively.
Since Mashkoor et al.\,\cite{Mashkoor2023} took a broader view of model-driven security, their work contributed only three security DSLs relevant to us, while the other papers discuss the general relevance of model-driven design in security engineering or developed secure reference architectures, e.g., for smart cards, without providing DSLs.
In total, 44 papers were referenced in the surveys that did not contain a security DSL, 31 of which were referenced by Mashkoor et al.\,\cite{Mashkoor2023}.
Four additional references may contain security DSLs, but we did not have access to their corresponding publications or preprints.
We added seven DSLs based on our expertise, as some were recently proposed, and two were suggested by experts in the security DSLs derived from the literature.
This resulted in a total of 66 design-time security DSLs for our study (see \cref{tab:dsls}).

\input{sections/tables/dsls}

%

\parhead{Security Analyzers.}
We identified three surveys\,\cite{Liu2023,Nachtigall2022,Zhang2023} on code analysis that mention 72 analyzers.
To get an overview of the analyzers, we classified them according to the categories of checks they provide described in \cref{sec:background:implementationcecks}.
We call analyzers that provide checks of only one category \textit{single-purpose} analyzers, e.g., analyzers that provide only checks that inspect \textit{Information Flow}, such as FlowDroid\,\cite{Arzt2014a}.
In contrast, we classify the analyzers that provide several checks of different categories as \textit{multi-purpose} analyzers, such as SonarQube, providing checks, amongst others, for \textit{API Usage} and \textit{Vulnerable Coding Practices}.
After removing duplicates and non-security-related analyzers, 48 analyzers remained, 25 single-purpose analyzers, i.e., offering only checks from one of the classes described in \Cref{sec:background:implementationcecks}, and 23 multi-purpose analyzers.
Nachtigall et al.\,\cite{Nachtigall2022} list 46 analyzers, of which only 18 provide security checks.
Liu et al.\,\cite{Liu2023} discuss 6 analyzers from which 5 are also listed by Nachtigall et al.\,\cite{Nachtigall2022}.
Zhang et al.\,\cite{Zhang2023} discuss 20 cryptographic API misuse analyzers for Java.
However, 15 analyzers solely focus on \textit{API Usage}, while the 5 remaining ones are \textit{Multi Purpose} tools, such as \textit{SonarQube}\,\cite{SonarSource}.

For our study, we decided to select all identified single-purpose analyzers for an in-depth investigation.
However, the 22 \textit{Multi Purpose} analyzers overlap substantially in the checks they provide.
Find Security Bugs \cite{findsecbugs}, \textit{SonarQube}\,\cite{SonarSource} and \textit{CodeQL}\,\cite{Moor2007}, for example, provide similar checks detecting usage of insecure cryptographic algorithms (\textit{Security API Usage}) or printing the stack trace to the command line (\textit{Vulnerable Coding Practice}).
Consequently, increasing the number of selected \textit{Multi Purpose} analyzers does not increase the number of different checks at some point.
Therefore, we selected a subset of commercial and free-of-use \textit{Multi-Purpose} analyzers mentioned in surveys for different programming languages.
This search results in \textit{SonarQube}\,\cite{SonarSource}, \textit{PMD}\,\cite{pmd}, \textit{Bandit}\,\cite{Bandit}, and \textit{FlawFinder}\,\cite{Wheler2001} (contained in multiple surveys) and the \textit{Clang Static Analyzer}\,\cite{Clang}.
We complement them with \textit{CodeQL}\,\cite{Moor2007,CodeQL} and Frama-C~\cite{framac_book}, among others, because they support \textit{Information Flow} analysis not supported by the other multi-purpose analyzers.
Since the surveys list only a few \textit{Single Purpose} analyzers for some categories, we add additional representatives identified by searching single-purpose analyzers in underrepresented categories.
As \textit{Information Flow} analyzer, we added \textit{FlowDroid}\,\cite{Arzt2014a}, \textit{JOANA}\,\cite{Giffhorn2008}, and \textit{KeY} using a specification of Greiner et al.\,\cite{Greiner2017,Greiner2018_Diss}.
We added \textit{NPDHunter}\,\cite{Jin2021} for \textit{Vulnerable Coding Practices}.
As \textit{Permission} analyzers, we added \textit{PeX}\,\cite{Zhang2019}, which focuses on missing or problematic permission checks, and Tang et al.\,\cite{Tang2017} and Wang et al.\,\cite{Wang2015}, which detect claiming too many permissions.

\edit{We require analyzers that provide checks for \textit{specific} security-related properties as a reasonable basis to establish a relation to security DSLs.
Without this specificity, all DSLs could be related to any of these analyzers.
Consequently, no further information is gained to support developers on \textit{which} property of the DSL can be analyzed.
To provide more detailed insights, we exclude general-purpose analysis and verification frameworks that can be extended to perform security-related checks but do not provide them themselves.
Furthermore, we exclude general approaches such as \textit{fuzzing} unless a specific purpose is described.
For example, frameworks like \textit{Coccinelle}~\cite{lawall18}, \textit{Soot}\,\cite{vallee2010soot} or \textit{KeY}~\cite{ahrendt2016} enable analysis of security-related properties when appropriately extended, e.g., by providing a certain specification.
For example, \textit{FlowDroid}~\cite{Arzt2014a} and the approach of Greiner et al.~\cite{Greiner2017} are based on the frameworks \textit{Soot} or \textit{KeY}, respectively, to analyze information flows.
However, the frameworks themselves cannot be pinpointed to a specific security-related property.
In contrast, we include either the framework itself if \textit{extensions} are closely tied to the framework or the security-related extension directly.
For example, the framework \textit{Frama-C}~\cite{framac_book} specifically lists the Plugins \textit{EVA} and \textit{SECUREFLOW}, which analyze the implementation for vulnerable coding practices and information flow weaknesses.
Thus, we include the \textit{Frama-C} framework.
In contrast, we include \textit{FlowDroid} and the approach Greiner et al.~\cite{Greiner2017} rather than the respective frameworks they extend (i.e., \textit{Soot} or \textit{KeY}).}

\input{sections/tables/analyzer}

In total, we consider \analyzers analyzers that provide \checks security checks (see \cref{tab:analyzer}).
The selected analyzers provide checks for different programming languages, including \edit{243 checks for Java, 155 for C/C++, and 74 for Python.}
Other languages are represented only by a small number of checks each.
Still, the most popular programming languages according to the TIOBE index\,\cite{tiobe} are well covered.

\subsection{Iterative Model Extraction}
\label{sec:method:analysis}
To obtain an understanding of the relationships between security at design time and in the implementation, we iteratively extracted common concepts, such as commonalities between what security DSLs express and what security checks detect, into a \model by studying the identified security DSLs and checks, and extending and generalizing the concepts found.
To this end, we studied the selected design-time security DSLs and code-level security analyzers based on their description in their documentation, scientific papers, and the output of the analyzers themselves.
Specifically, we proceeded as follows:

\begin{figure}
	\centering
	\includegraphics[width=.75\columnwidth]{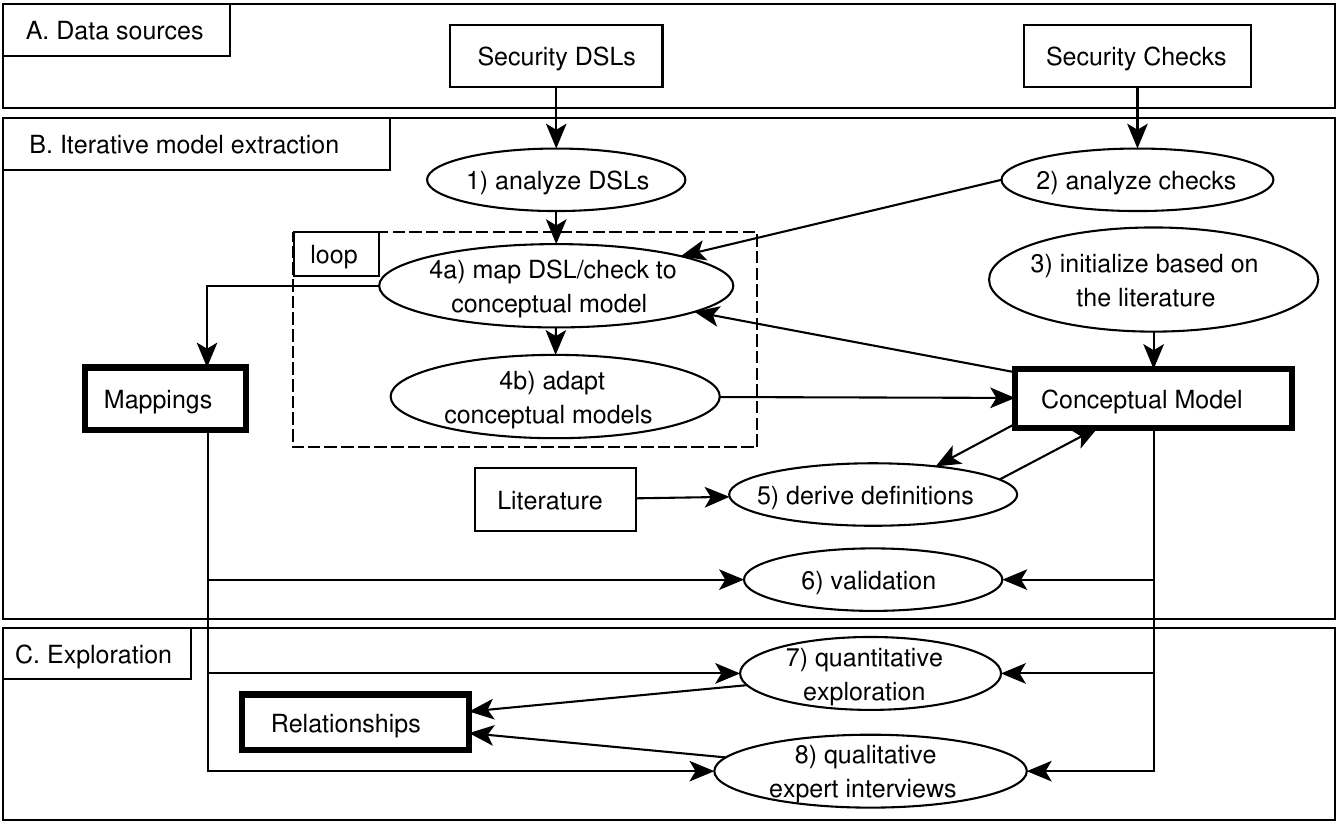}
	\caption{Visualization of the methodology}
	\label{fig:methodology}
\end{figure}

    \textit{1) Analyze DSLs.} First, we analyzed the design-time security DSLs by extracting potentially relevant information, such as the purpose of the DSL, the considered security objectives, relevant elements of the underlying modeling language, and known code-level checks that might be associated with the DSL.
    We started this analysis with a list of relevant aspects, and partially also possible value ranges, known from related publications\,\cite{Uzunov2012, Nguyen2015, DenBerghe2017, Mashkoor2023}, extracted information about the DSLs according to these aspects, and extended this list with further aspects identified during the analysis of the DSLs.

    \textit{2) Analyze Checks.} We then analyzed the code-level security checks provided by the selected analyzers to extract potentially relevant information that encompasses the purpose of each check, such as the vulnerabilities or weaknesses that the check can detect or the implementation elements that are inspected.
    We started with the list of aspects created in step 1) and refined it to include security analyzers and to identify aspects that are common to both design-time security DSLs and code-level analyzers.

    \textit{3) Initialize \model.} After analyzing the security DSLs and checks, we determined which aspects we could collect information on for both design-time security DSLs and code-level analyzers, or which aspects were closely related.
    We then conceptualized these aspects, based on existing literature (cf. \cref{sec:rw}) and our observations from inspecting the security DSLs and checks, and initialized a preliminary version of the \model---focusing primarily on relevant security concerns and a common system representation.

    \textit{4) Map DSLs/Checks and Adapt \model.} We then iteratively classified the security DSLs and analyzers with respect to the preliminary \model, checking whether it was appropriate to describe the DSL or analyzer, starting with the DSLs.
    In case of new or mismatching concepts, we discussed possible adaptations of the \model among the authors until full agreement was reached.
    We then updated the classification of previously mapped DSLs, potentially leading to repeating the adaptation step.
    We applied the same steps to the security analyzers.
    This way, we ensured that the \model is applicable to all security DSLs and analyzers studied.

    \textit{5) Derive Definitions.} Thereafter, we derived concrete definitions of the individual elements identified by systematizing the justifications behind the classifications and adaptations of steps 1--4 and relating them to further literature.
    Based on these definitions, we then iterated over all the elements of the \model and discussed possible relationships between them, again based on further literature and the data collected while studying the DSLs and analyzers.


\textit{6) Validation.}
To further investigate the relationships between security DSLs and analyzers, we had to verify the correctness of the extracted concepts and their application to the studied security DSLs and analyzers.
To this end, we invited all authors of the studied DSLs (186 authors) and analyzers (118 authors) as experts to participate in an online survey.
We found current email addresses for 132 of them, to which we sent the invitation.
\edit{Due to the low response rate from analyzer authors, we proceeded to label additional security analyzers.
Subsequently, we sent a reminder to authors who had not yet responded and extended invitations to 25 authors of the newly labeled analyzers.}

\begin{figure}
    \centering
    \includegraphics[width=0.9\textwidth]{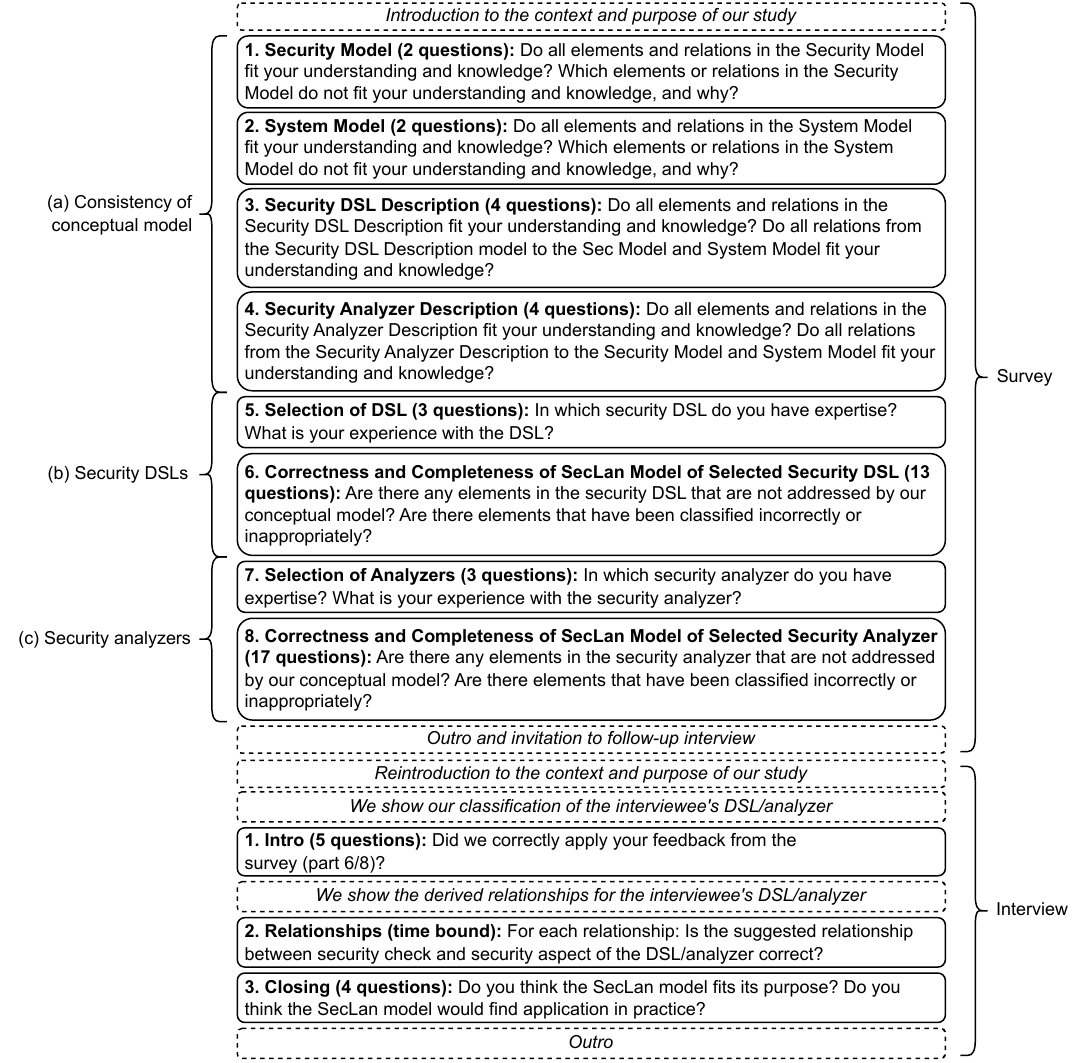}
    \caption{Overview of the survey and interview questions for validation of the \model and exploration of derived relationships}
    \label{fig:questions}
\end{figure}

\edit{\Cref{fig:questions} gives an overview of the survey structure and questions asked.}
The first part of this survey focused on the \model in general, especially in terms of correctness and completeness, and whether it matches the experts' understanding of the domain.
For this purpose, the questionnaire introduces the general idea of SecLan and the extracted concepts using \cref{fig:models,fig:example,fig:SystemElementsConceputalModel}.
Then, the interviewees are iteratively asked for their opinion on the extracted concepts, using multiple-choice questions as well as free text answers to identify concrete improvements.
In the second part, we asked the experts to select their DSL or analyzer from a list of all the DSLs and analyzers we studied, showed how we applied the extracted concepts to the security DSLs and security checks, and asked whether we had classified them correctly with respect to the \model.
As visualization, we use a JavaDoc-like documentation automatically generated from instances of the \model we created (cf. \cref{sec:tool}).
Of the invited experts, \edit{22 participated} in the first part of the survey, and \edit{21 answered} the questions in the second part. 
\edit{\Cref{tab:participants} gives an overview of the participants, their backgrounds, and in which parts of the survey they participated.}

\subsection{Exploration of Relationships Between Security DSLs and Analyzers}
\noindent
Based on the classification of the security DSLs and security checks provided by the analyzers according to their common concepts as captured in the \model, we quantitatively and qualitatively investigated the relationships between the security design as expressed in the security DSLs and what is checked in code.

\textit{7) Quantitative Exploration.} We extracted all relationships between design-time security DSLs and security checks, and quantitatively analyzed the relationships with respect to aspects such as the number of relationships, the strength of the relationships, and characterizing factors (step 7 in \cref{fig:methodology}).

\textit{8) Qualitative Exploration.} We conducted interviews with security experts to explore relationships between security design, as expressed in security DSLs, and code checked by security analyzers (step 8 in \cref{fig:methodology}).
As experts, we invited all respondents to our survey to participate in a follow-up interview.
We recruited seven experts in design-time security DSLs \edit{and two experts in security analyzers}.
In the interviews, we focused on the security DSL or analyzer for which the interviewee is an expert, i.e., one of its authors.
\edit{\Cref{fig:questions} shows the individual steps and questions of these exploratory interviews.}
Before each interview, we made sure to adjust the application of the \model to the interviewee's DSL \edit{or analyzer} based on the feedback given by the interviewee in the questionnaire (step 6) \edit{and validated changes together with the interviewee (interview part 1)}.
We then interactively explored the relationships to security checks of the studied analyzers \edit{or the security aspects of the studied security DSLs (depending on whether the interviewee is an expert in a security DSL or an analyzer) respectively} that we derived in our quantitative exploration (step 6\edit{, interview part 2}).
\edit{If the participant is an expert in a DSL, we asked whether the relationships to security checks made sense to the interviewee and what these checks detected could affect the security aspects of the interviewee's DSL. 
If the participant is an expert in an analyzer, we asked whether the relationships to the security DSLs made sense to the interviewee and whether the checks provided by the interviewee's analyzer could affect the security aspects of the security DSLs.}
\edit{Due to a potentially high number of derived relationships, we time-bound this interview part, spending at most one hour per interview on this part.}
We then engaged in an open discussion about the relationships between security DSLs and code-level security checks \edit{(interview part 3)}.
Our supplementary materials\,\cite{Replication} contain the detailed interview guide.

\begin{table}[]
    \centering
    \caption{Participants in the validation and qualitative exploration}
    \label{tab:participants}
    \scriptsize
    \setlength{\tabcolsep}{3pt}
    \begin{tabular}{rlccc|cccc|c|c}
        \toprule
        \multirow{2}{*}{\textbf{ID}\textsuperscript{*}} & \multirow{2}{*}{\textbf{Analyzer/DSL Name}} & \multirow{2}{*}{\textbf{Type}} & \multirow{2}{*}{\textbf{Role}}& \multirow{2}{*}{\textbf{Expertise}} & \multicolumn{6}{c}{\textbf{Participation}} \\
                    & &                  &      &               & \multicolumn{5}{c}{\smaller \textbf{6) Validation}} & \smaller \textbf{8) Exploration}\\
        \midrule
                    &&&&                                       & \multicolumn{4}{c}{\footnotesize SecLan Model} & \multirow{2}{*}{\rotatebox{90}{\parbox{2.5cm}{\footnotesize SecLan Model Application to DSL/Analyzer}}} & \\
                  &&&  &                                        & \rotatebox{90}{\footnotesize Security Model} & \rotatebox{90}{\footnotesize System Model} & \rotatebox{90}{\footnotesize DSL Description} & \rotatebox{90}{\footnotesize Analyzer Description} & & \\
        \midrule
        I1 & SecBPMN    & DSL & Main Author, Developer/Maintainer & > 5 Years & \checkmark & \checkmark & \checkmark & \checkmark & \checkmark & \checkmark \\
        I2 & SecDFD     & DSL & Main Author, Developer/Maintainer & > 5 Years & \checkmark & \checkmark & \checkmark & \checkmark & \checkmark & \checkmark \\
        I3 & Architectural Dataflow & DSL & Main Author, Developer/Maintainer & 3 -- 5 Years & \checkmark & \checkmark & \checkmark & \checkmark & \checkmark & \checkmark \\
        I4 & UMLsec     & DSL & Co-Author & > 5 Years & \checkmark & \checkmark & \checkmark & \checkmark & \checkmark & \checkmark \\
        I5 & UMLsec     & DSL & Maintainer/Developer & 1 -- 2  Years & \checkmark & \checkmark & \checkmark & \checkmark & \checkmark & \checkmark \\
        I6 & Attack Propagation & DSL & Main Author, Developer/Maintainer & 3 -- 5 Years & \checkmark & \checkmark & \checkmark & \checkmark & \checkmark & \checkmark \\
        I7 & PbSD       & DSL & Co-Author& n/a & \checkmark & \checkmark & \checkmark & \checkmark & \checkmark & \checkmark \\
        I8 & CodeQL & Analyzer & Developer/Maintainer & > 5 Years & \checkmark & \checkmark & \checkmark & \checkmark & \checkmark & \checkmark\\
        I9 & Thuderhorse@ESLint     & Analyzer & Co-Author & > 5 Years & \checkmark & \checkmark & \checkmark & \checkmark & \checkmark & \checkmark \\
        \midrule
        I10 & CARDS      & DSL & Co-Author, Developer/Maintainer & < 1 Year & \checkmark & \checkmark & \checkmark & \checkmark & \checkmark & \texttimes \\
        I11 & SysML-Sec & DSL & Main Author & n/a & \checkmark & \checkmark & \checkmark & \checkmark & \checkmark & \texttimes \\
        I12 & Attack-Fault Trees & DSL & Main Author & > 5 Years & \checkmark & \checkmark & \checkmark & \checkmark & \checkmark & \texttimes \\
        I13 & PMD       & Analyzer & Developer/Maintainer & > 5 Years & \checkmark & \checkmark & \checkmark & \checkmark & \checkmark & \texttimes \\
        I14 & SecureUML & DSL & Main Author & n/a & \checkmark & \checkmark& \checkmark& \checkmark& \checkmark& \texttimes \\
        I15 & CppCheck  & Analyzer & Main Author, Developer/Maintainer & > 5 Years & \checkmark & \checkmark & \texttimes& \texttimes& \texttimes& \texttimes \\
        I16 & n/a   & Analyzer & n/a & n/a & \checkmark & \checkmark & \texttimes& \checkmark & \texttimes& \texttimes \\
        I17 & FlowDroid & Analyzer & Main Author, Developer/Maintainer & > 5 Years & \checkmark & \checkmark & \checkmark & \checkmark & \checkmark & \texttimes \\
        I18 & CCMsec    & DSL & Co-Author & n/a & \checkmark & \checkmark& \checkmark& \checkmark& \texttimes& \texttimes \\
        I19 & UMlsec    & DSL & Developer/Maintainer & 3 -- 5 Years & \checkmark & \checkmark& \checkmark& \checkmark& \checkmark& \texttimes \\
        I20 & n/a   & Analyzer & n/a & n/a & \checkmark & \checkmark & \checkmark & \checkmark & \texttimes & \texttimes\\
        I21 & CogniCrypt & Analyzer & Co-Author & n/a & \checkmark & \checkmark & \checkmark & \checkmark & \checkmark & \texttimes\\
        I22 & PMD    & Analyzer & Developer/Maintainer & > 5 Years & \checkmark & \checkmark & \checkmark & \checkmark & \checkmark & \texttimes\\
        \bottomrule
    \end{tabular}

    \vspace{4pt}
    \scriptsize \textsuperscript{*}The order of the participants is not chronological; participants who participated in the exploration are shown first
\end{table}

Interview participation was voluntary, and did not involve any advantage or disadvantage to potential interviewees.
The circumstances of the interview were explained to participants before they agreed to participate.
All interviews were conducted remotely via Zoom, lasted an average of one hour, and were not recorded.
The first author actively conducted each interview while another author took notes.
All participants were assured that their personal information or interview statements would only be published in aggregated or anonymized form.
For this reason, our supplementary material\,\cite{Replication} includes only the interview guide, but not the transcripts.

%% file: sections/tables/dsls.tex
%
{
    \scriptsize
    \setlength{\tabcolsep}{2pt}
    \begin{xltabular}{\textwidth}{rlXp{1.9cm}r}
        \caption{Overview of the Security DSLs Analyzed}
	\label{tab:dsls}\\
    
            \toprule 
            & \textbf{Name} & \textbf{Description} & \textbf{Category} & \textbf{Ref.}\\
            \midrule 
        \endfirsthead

            \multicolumn{4}{c}%
            {\tablename\ \thetable{} -- continued from previous page} \\
            \toprule
            & \textbf{Name} & \textbf{Description} & \textbf{Category} & \textbf{Ref.}\\
            \midrule 
        \endhead

            \hline \multicolumn{4}{r}{{Continued on next page}} \\ \hline
        \endfoot

            \bottomrule	
        \endlastfoot
        
	1 &		AC-PIM                    & Platform Independent Model for access control, separating policies, access decisions, and enforcement. & ACA & \cite{Burt2003} \\
	2 &		Access Analysis           & Extends the Palladio Component Model with confidentiality-related security. & ACA, TMID, IFIA, DPA, POSA & \cite{Kramer2017} \\
	3 &		ADM-RBAC                  & An extension to ADM that allows modeling role-based access control. & ACA & \cite{Diaz2008} \\
	4 &		Ahn et al.                & The metamodel of a role-based access control model. & ACA & \cite{Ahn2002} \\
	5 &		AMF                       & Assurance Management Framework to ensure security models are fully realized and employed. & ACA & \cite{Ahn2007} \\
	6 &		AoASM                     & An UML extension for aspect-oriented attack scenario models and intrusion detection aspects. & TMID &  \cite{Zhu2008} \\
	7 &		AORDD                     & Models risk-driven aspects to prioritize security trade-offs & TMID, CGA, DPA & \cite{Georg2010} \\
	8 &	    ASM                       & UML profile for aspectual scenario modeling focusing on non-functional requirements. & ACA, DPA, SCSA & \cite{Sanchez2009} \\
	9 &	    Break-Glass RBAC          & Integrates break-glass policies for emergencies into business process models. & ACA & \cite{Schefer-Wenzel2014} \\
	10 &	Busch et al.              & An extension of the UML-based Web Engineering (UWE) language to model security concepts. & ACA, SCSA & \cite{Busch2014} \\
	11 &	CARDS                     & A language for specifying data flow constraints and architecture-level assumptions. & IFIA & \cite{Geismann2021} \\
	12 &	CCMSec                    & An extension of the CORBA Component Model (CCM) with security concepts. & ACA & \cite{Reznik2007} \\
	13 &	DiaAspect/DiaSpec         & A light-weight Architecture Description Language. & CA, ACA & \cite{Jakob2009} \\
	14 &	DRBD-AFT                  & A combined model of dynamic reliability block diagrams (DRBD) and attack-fault trees (AFT). & SCSA & \cite{Kumar2017,Kumar2020} \\
	15 &	FDAF Metamodel            & A UML extension to weave security aspects into a UML architecture design.  & CGA, CA & \cite{Dai2006} \\
	16 &	Gomaa et al.              & Secure software architectures based on secure connectors for intercomponent communication. & CGA, DPA, SCSA & \cite{Gomaa2006} \\
	17 &	Hoisl et al.              & UML extension for secure object flow at the PIM level. & DPA, IFIA & \cite{Hoisl2012} \\
	18 &	INTER-TRUST             & Annotations used for the dynamic monitoring of security policies. & CGA & \cite{Horcas2016} \\
	19 &	Kim et al.                & Describes user assignment to roles and permissions of roles to access objects and operations. & ACA & \cite{Kim2011} \\
	20 &	Koch et al.               & A language for role-permission assignments. & ACA & \cite{Koch2002} \\
	21 &	Menzel et al.             & Enables modeling of service-oriented architecture & ACA, DPA & \cite{Menzel2009} \\
	22 &	Morin et al.              & Specifies access control policies and connections to the software architecture. & ACA & \cite{Morin2010} \\
	23 &	Mouelhi et al.            & Supports specifying, deploying, and testing of access control policies & ACA & \cite{Baudry2008} \\
	24 &	Mouheb et al.             & The Meta-Model for Specifying Security Hardening Plans & POSA, SH & \cite{Mouheb2009} \\
	25 &	Nakamura et al.           & Stereotypes to address business-level security intents. & DPA, SCSA & \cite{Nakamura2005} \\
	26 &	Nguyen et al.             & Allows to specify access control and delegation policies. & ACA & \cite{Nguyen2014} \\
	27 &	Oladimeji et al.          & A lightweight UML extension with authorization and obligation policies. & CGA & \cite{Oladimeji2007} \\
	28 &	OrBAC EMF                 & A realization of the OrBAC model in EMF. & ACA & \cite{Kaddani2014} \\
	29 &	OSL-ISDL                  & Adds references to the Interaction System Design Language to the Obligation Specification Language. & ACA & \cite{Neisse2013} \\
	30 &	Pavlich-Mariscal et al.   & A UML profile for specifying access control diagrams. & ACA & \cite{Pavlich-Mariscal2009} \\
	31 &	PbSD                      & A pattern-based method to describe role-based access control for objects and actors in the system. & ACA & \cite{Abramov2012} \\
	32 &	Pinto et al.              & Models and links security concerns to a system architecture model. & SCSA & \cite{Horcas2015,Pinto2015} \\
	33 &	RBAC DCM                  & A hierarchical design class model to describe RBAC classes and operations. &  ACA & \cite{Yu2005} \\
	34 &    RBML                      & A UML-based Role-Based Metamodeling Language. & ACA & \cite{Kim2006} \\
	35 &	SAL                       & A Security Analysis Language to analyze flows of data objects through distributed systems. & IFIA, DPA, CA, TMID & \cite{Eby2007} \\
	36 &	SAM Security Extension    & Defines in the SAM which information is sensitive to which action/task. & DPA & \cite{Yu2005} \\
	37 &	Satoh et al.              & Models platform information required for creating security policies. & ACA & \cite{Satoh2006} \\
	38 &	SecBPMN                   & Security annotations to represent security aspects of a business process.  & ACA, DPA, CGA, ADSSA & \cite{Salnitri2015} \\
	39 &	SecDSVL                   & Visual modeling of security details in enterprise systems security management processes. & CA & \cite{Almorsy2014} \\
	40 &	SECDW                     & A UML profile for specifying confidentiality constraints in conceptual multidimensional modeling. & DPA, ACA, CGA & \cite{Fernandez-Medina2007} \\
	41 &	SecDFD                    & Annotations to specify data flow contracts on data flow diagrams. & IFIA & \cite{Tuma2019a} \\
	42 &	SecML                     & A requirements metamodel extended with security concepts for defining security requirements. & ACA, DPA & \cite{Sánchez2009} \\
	43 &	SECTET-PL                 & A policy language influenced by Object Constraint Language and interpreted on UML models. & ACA & \cite{Alam2007} \\
	44 &	SecureMDD                 & Modeling of a system, its functional and security requirements, and attackers. & SCSA, CA, DPA, POSA & \cite{Moebius2009} \\
	45 &	Secure Tropos             & A goal-based requirements language for expressing security concerns. & CGA, DPA, TMID & \cite{Mouratidis2007} \\
	46 &	Secure xADL               & Extends xADL with Constructs for describing security characteristics. & ACA & \cite{Ren2005} \\
	47 &	SecureSOA                 & A domain-specific language to specify security configuration patterns. & CGA & \cite{Menzel2010} \\
	48 &	SecureUML                 & UML-based language security in distributed systems. & ACA & \cite{Lodderstedt2002} \\
	49 &	Security@Runtime          & The metamodel for the Security@Runtime approach. & CGA & \cite{Elrakaiby2014} \\
	50 &	Security Primitives Lib.  & A library of security building blocks for SPACE. & SCSA, CA & \cite{Gunawan2009,Gunawan2011} \\
	51 &	Security Model            & Allows modeling of security dysfunctional behavior. & TMID & \cite{Brunel2014,Brunel2014} \\
	52 &	Secure SaaS               & Integrates security patterns into a more comprehensive pattern for handling more threats. & ADSSA & \cite{MoralGarcia2014} \\
	53 &	Seifermann et al.         & Annotations to specify data flow-related security properties in the Palladio Component Model. & IFIA & \cite{Seifermann2021} \\
	54 &	Sensitivities             & Extends mechatronic UML with sensitivities of data received or sent from ports in components. & IFIA & \cite{Gerking2019,Gerking2018} \\
        55 &	Shin and Gomaa            & Allows to describe security-relevant elements while separating system and security concerns. & SCSA, SH, DPA & \cite{Shin2004} \\
	56 &	SSD RBAC                  & A metamodel for Static Separation of Duty (SSD) and Role-Based Access Control (RBAC). & ACA & \cite{Ray2003} \\
	57 &	SysML-Sec                 & A security extension to SysML. & DPA, ACA & \cite{Apvrille2013} \\
	58 &	UMLS                      & Extends the UML with annotations to address confidentiality throughout the development process  & DPA, ACA & \cite{Heldal2003} \\
	59 &	UMLsec                    & UML profile defining security annotations for dependencies, information flow, and data processing. & ACA, IFIA, DPA, SCSA, POSA & \cite{Juerjens2005} \\
	60 &	VBAC-PIM                  & The Meta Object Facility model for the  View-based Access Control Model. & ACA & \cite{Fink2006} \\
	61 &	Visual RBAC               & A visual language to specify access and security policies according to role-based access control & ACA & \cite{Giordano2010} \\
	62 &	Wada et al.               & An UML profile with key model elements for service-oriented applications. & SCSA & \cite{Wada2006} \\
	63 &	Walter et al.             & An ADL extension for modeling access control policies, attacker capabilities, and system vulnerabilities. & ACA,  DPA,TMID, SCSA & \cite{Walter2022} \\
	64 &	Wolter et al.             & A security policy model based on a simplistic view of a service-oriented architecture. & CGA, DPA, SCSA & \cite{Wolter2008} \\
	65 &    Xiao et al.               & A RBAC-based metamodel for adaptive and secure multi-agent systems. & ACA & \cite{Xiao2008} \\
	66 &    Xu et al.                 & The metamodel for aspect-oriented modeling of security threats. & TMID & \cite{Xu2006} \\
\end{xltabular}
    \vspace{-17.2pt}
    \noindent
    \begin{center}
        \scriptsize 
        Categories: 
        ACA -- Access Control Aspects; 
        CGA -- Compliance and Governance Aspects; 
        CA -- Cryptographic Aspects;
        DPA -- Data Protection Aspects;
        ADSSA -- Application-Domain-Specific Security Aspects; 
        IFIA -- Information Flow and Isolation Aspects;
        POSA -- Physical and Operational Security Aspects; 
        SCSA -- System and Communication Security Aspects;
        SH -- Security Hardening
        TMID -- Threat Modeling and Intrusion Detection; 
    \end{center}
}

%% file: sections/tables/analyzer.tex
\begin{table}
	\caption{Overview of the security analyzers investigated with their \textit{Name}, a \textit{Description}, the number of checks they provide (\textit{\#Checks}), the categories (see \cref{sec:background:implementationcecks}) that the provided checks cover (\textit{Category}), and references to relevant publications (\textit{Ref.}).}
	\label{tab:analyzer}
    \setlength{\tabcolsep}{3pt}
    \scriptsize
	\begin{tabularx}{\textwidth}{rlXrlr}
		\toprule		
            & \textbf{Name} & \textbf{Description}  & \textbf{\#Checks} & \textbf{Category} & \textbf{Ref.}\\
		\midrule
			1 &   Amandroid & Inter-component data flow analysis framework for security vetting of Android apps. & 5 & API, IF & \cite{Wei2018} \\
			2 &   Bandit & Comprises checks to find common security issues in Python code. & 40 & API, VCP, P & \cite{Bandit} \\
			3 &   BinSight  & Static program analysis to attribute API misuse to its source. & 1 & API & \cite{Muslukhov2018} \\
			4 &   CDRep & Repairs cryptographic misuse defects in Android apps. & 1 & API & \cite{Ma2016} \\
			5 &   Clang Static Analyzer &  Security-related checkers of the Clang Static Analyzer.  & 15 & VCP & \cite{Clang} \\
			6 &   CMA & Crypto misuse analyzer based on analyzing runtime information. & 1 & API & \cite{Shuai2014} \\
			7 &   CodeQL & A static analysis framework for discovering vulnerabilities across a codebase. & 196 & API, VCP, IF, P & \cite{CodeQL} \\
			8 &   CogniCrypt & CogniCrypt is an API analysis framework for Java. & 1 & API & \cite{Krueger2017,CogniCrypt} \\
			9 &   CppCheck & Focuses on detecting undefined behavior and dangerous coding constructs. & 2 & VCP & \cite{CppCheck} \\
			10 &   CryptoChecker & Elicit security rules by clustering commonalities among semantic code changes & 1 & API & \cite{Paletov2018} \\
			11 &   CryptoGuard & Scalable detection of cryptographic API misuse vulnerabilities in Java projects. & 1 & API  & \cite{Rahaman2019} \\
            12 &   CryptoLint & Analysis of crypto-related information flows. & 1 & API & \cite{Egele2013} \\
			13 &   CryptoTutor & Detection of cryptographic misuses and suggestion of possible repairs. & 1 & API & \cite{Singleton2020} \\
			14 &   Dr. Checker & A bug-finding tool for Linux kernel drivers. & 2 & VCP & \cite{Machiry2017} \\
			15 &   ErrorProne & Static analysis tool for Java that catches common programming mistakes. & 2 & API, VCP & \cite{ErrorProne} \\
			16 &   Flawfinder & Lexical analysis for vulnerabilities based on function names. & 49 & VCP & \cite{Wheler2001} \\
			17 &   FlowDroid & Context-, flow-, field-, object-sensitive and lifecycle-aware taint analysis for Android. & 1 & IF & \cite{Arzt2014a} \\
            18 &   Frama-C & Analyzer for analyzing C-Code extended by various security related Plug-Ins & 15 & IF, VCP & \cite{framac_book}  \\
            19 & GitHub Dependabot & Automated dependency updates & 1 & D & \cite{dependabot} \\
			20 &   Hotfixer  & Crypto API misuse hotfixing at Java application runtime. & 1 & API & \cite{Newbury2020} \\
            21 &   Infer & A static analyzer provided by Facebook to detect bugs in Java and C/C++ code & 2 & VCP & \cite{calcagno15} \\  
			22 &   JOANA & Java analyzer for information flow control-based security leaks. & 1 & IF & \cite{JOANA,Snelting2014} \\
			23 &   MalloDroid & Detects potential vulnerabilities against MITM attacks. & 1 & API & \cite{Fahl2012} \\
			24 &   NPDHunter & Binary analysis for Null Pointer Dereference vulnerabilities. & 1 & VCP & \cite{Jin2021} \\
			25 &   OWASP Dependency Check & Software composition analysis tool for detecting vulnerable dependencies. & 1 & D & \cite{OWASPDependency} \\
			26 &   PeX & Static permission check error detector for Linux. & 1 & P & \cite{Zhang2019} \\
			27 &   PMD & Rule subset that flags potential security flaws. & 17 & API, VCP, P & \cite{pmd} \\
			28 &   Pyre/Pysa & A tool for performing taint analysis on Python code. & 1 & IF & \cite{Pyre} \\
            29 &   Rafnsson et al. ESLint rules & ESLint rules for certain classes of cross-site scripting, SQL injection, and misconfiguration vulnerabilities in JavaScript. & 3 & API, IF & \cite{Rafnsson2020} \\
            30 &   Semgrep & Lightweight static analysis for many languages based on patterns that look like source code. & 100\textsuperscript{*} & API, IF, VCP, P & \cite{semgrep} \\
			31 &   Snyk Open Source & Vulnerability detection based on known security issues found in open-source libraries. & 1 & D & \cite{Snyk} \\
			32 &   SonarQube & Subset of the Java rules classified according to the CWE, OWASP Top 10, or SANS Top 25. & 57 & API, VCP, P& \cite{SonarSource} \\
			33 &   Tang et al. & Detects over-claims of permissions in Android apps. & 1 & P & \cite{Tang2017} \\
			34 &   Vulvet & Multiple program analysis techniques for detecting vulnerabilities in Android apps. & 1 & API, VCP & \cite{Gajrani2020} \\
			35 &   Wang et al. & Static and dynamic analysis to detect permission gaps in Android applications. & 1 & P & \cite{Wang2015} \\
			36 &   Xu et al. & Analyzes crypto API usages in Android applications based on probabilistic models. & 1 & P & \cite{Xu2020} \\ 
		\bottomrule	
        \end{tabularx}
        
        \scriptsize Categories:  1) API -- Security API Usage; 2) IF -- Information Flow; 3) D -- Dependency; 4) VCP -- Vulnerable Coding Practices; 5) P -- Permissions

        \textsuperscript{*}\,Randomly drawn subset
    \end{table}

%% file: sections/04_0_conceptual_model.tex
\section{SecLan Model}
\label{sec:model}
\begin{figure}
	\center
	\includegraphics[width=.8\columnwidth]{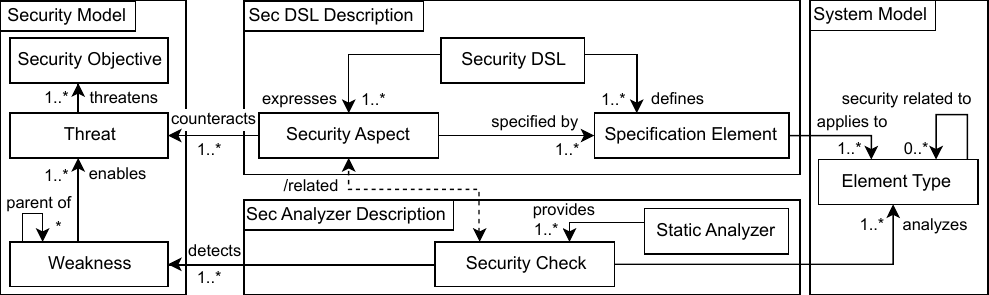}
	\caption{The SecLan model shows the common security concepts and system elements related to security DSLs and security checks}
	\label{fig:models}
\end{figure}

\noindent\looseness=-1
By analyzing \dsls design-time security DSLs and \checks security checks provided by \analyzers security analyzers, we identified security concepts and the relevant elements of the subject system as the only common concepts to establish relationships between all of them.
We captured these common concepts in the \model shown in \cref{fig:models}.
SecLan allows the systematic representation of relationships between security checks of code and the security design provided by security DSLs.

SecLan consists of four sub-models.
\textit{Sec DSL Description}, shown in the center of \cref{fig:models}, captures the common elements of security DSLs relevant for the relation to code-level checks.
\textit{Sec Analyzer Description} captures security checks offered by static analyzers.
Common security concepts (\textit{Security Model}) and security-relevant system elements (\textit{System Model}), shown on the left and right of \cref{fig:models} respectively, establish a relation between \textit{Sec Analyzer Description} and \textit{Sec DSL Description}.
Additional aspects may be necessary to fully describe a design-time security DSL or security analyzer, but in this work, we focus only on those concepts that are suitable for relating DSLs and analyzers that are common to all of them.

\subsection{Security Model}
\label{sec:model:secmodel}
\begin{figure}
    \centering
    \includegraphics[width=.9\columnwidth]{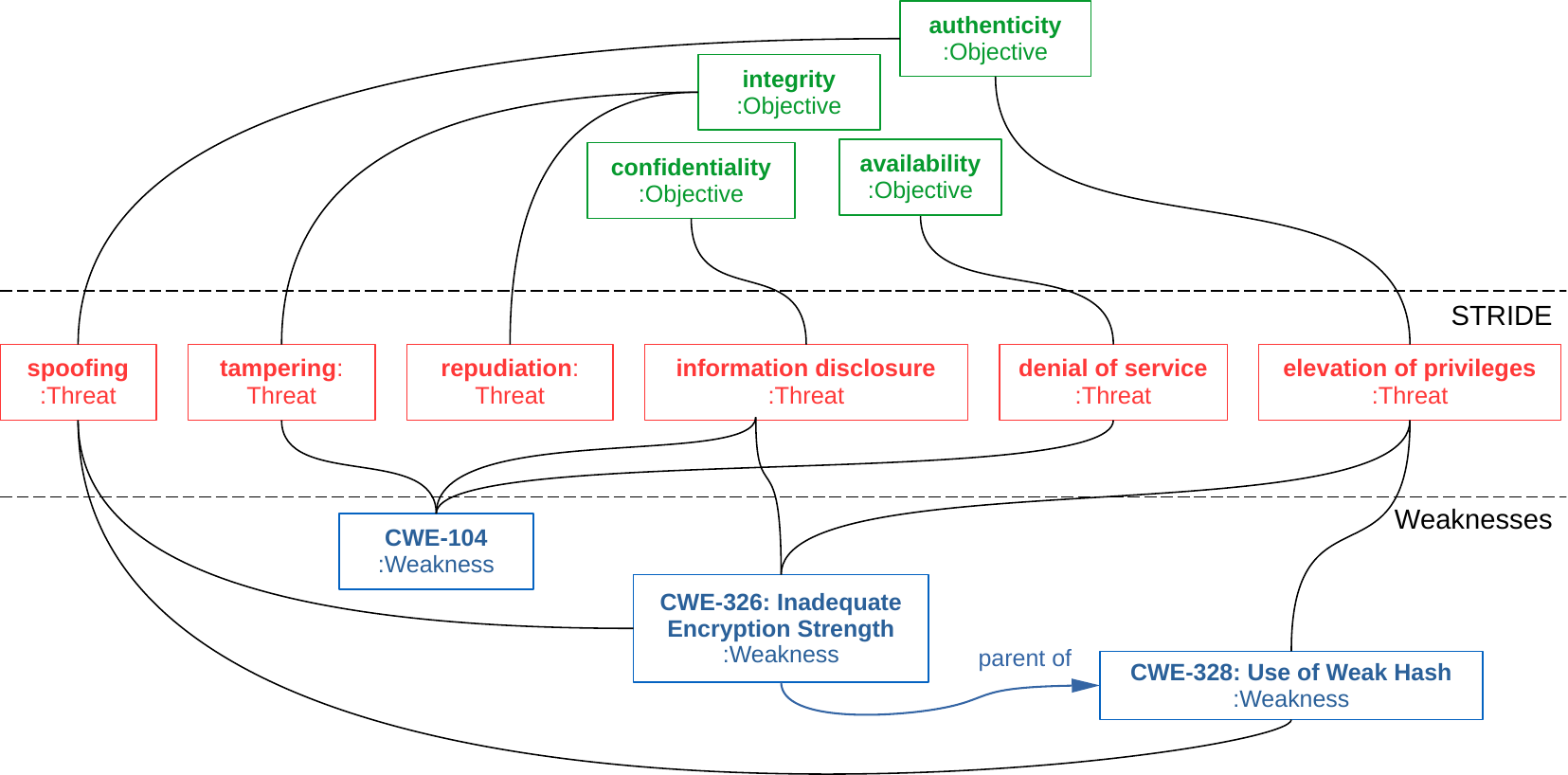}
    \caption{Example for an instance of the security model based on the CIA, STRIDE, and exemplary security weaknesses}
    \label{fig:example}
\end{figure}

We identified three related security concepts of security DSLs and analyzers that are common to all DSLs and analyzers, respectively, and need to be considered: security objectives, threats, and weaknesses.
The \textit{Security Model}, whose instance is shown in \cref{fig:example}, captures these concepts to provide a fundamental view of software security.

In addition to security objectives, threats, and weaknesses, countermeasures were often explicitly considered by security DSLs (62\%), e.g., in planning needed cryptography (16\%), as in SecDFD.
A significant number of the security DSLs (48\%) target the planning of access control.
Concrete countermeasures were also analyzed in many security checks, primarily to verify the secure use of cryptographic APIs (18\%), such as CogniCrypt\,\cite{Krueger2017}.
However, many of the DSLs and checks examined do not have such an explicit relationship.
We observed that encryption was the only countermeasure considered by DSLs and analyzers.
Other countermeasures, e.g., access control, are only considered in either DSLs or analyzers.
Accordingly, since countermeasures do not apply to all DSLs and checks, we did not explicitly include them as a separate element in the \model.
Nevertheless, the focus of security DSLs on concrete countermeasures can be captured via the security aspects they express, i.e., by having a security aspect whose purpose is to represent a specific countermeasure.

We now present the common elements we have identified that apply to all security DSLs and all security analyzers:

\parhead{Security objective.} All security checks and DSLs, as well as the CWE and other literature, refer to high-level security objectives, such as the commonly used CIA triad of confidentiality, integrity, and availability.
However, these terms are used with different abstractions and, therefore, cannot be used to relate security checks to DSLs.
Still, they are useful abstractions for reasoning about system security. 
In line with the literature\,\cite{Tuma2019a,Swanson2006,Seifermann2022}, we describe \textit{security objectives} (called security property in ISO 27001 and 27002) as high-level goals that must be achieved to ensure the security of a system. 
The coarse granularity of \textit{security objectives} makes it challenging to reason about the relation between security DSLs and security checks with only a few possible distinctions, e.g., confidentiality, integrity, or availability.
For example, only stating that SecDFD in our running example is related to CogniCrypt\,\cite{Krueger2020} because they are related through confidentiality does not provide any insight into which security flaws could potentially occur, violating the security of the system described with SecDFD. 
For a software engineer, the insight that selecting an encryption algorithm with inadequate encryption strength would violate the security contracts defined by SecDFD to secure information disclosure better details the relation. 
Consequently, we capture \textit{threats} and \textit{weaknesses} in our \model as a more detailed representation, which is commonly applied for more fine-granular reasoning on how \textit{security objectives} are addressed or violated.


\parhead{Threat.} Security objectives are threatened by \textit{threats} that may cause the system to violate its objectives or allow an attacker to do so.
The NIST\,\cite{SP800-30} defines a threat as a circumstance or event that could adversely affect assets or operations, such as information disclosure due to the exploitation of a vulnerability.
Design-time security engineering, such as threat modeling\,\cite{Shostack2014}, explicitly focuses on identifying threats and planning appropriate security measures to mitigate them.

Common threat categories are provided in threat modeling frameworks, such as STRIDE\,\cite{Shostack2008}.
For example, \textit{Information Disclosure} covers exposing sensitive information, such as a medical record from \cref{fig:secdfd}.
In the \model shown in \cref{fig:example}, we focus on security threats as specified in STRIDE
Still, further threats related to other aspects, such as privacy, could be included by extending the \model, i.e., with LINDDUN\,\cite{RoblesGonzalez2020}. 

\parhead{Weakness.}
\looseness=-1
Threats are potentially enabled by \textit{weaknesses} of a system.
The NIST\,\cite{SP800-160} defines a weakness as a flaw or characteristic that can lead to undesirable behavior.
For example, a weak cryptographic algorithm could allow an adversary to read data without permission, such as the sensitive information defined using SecDFD in \cref{fig:secdfd}.
A weakness in the system design can manifest in the system's implementation as security-critical bugs, such as implementation errors or incorrect configurations, impacting the assumed security properties.
Security checks of analyzers typically detect vulnerabilities, which are concrete instances of weaknesses (see \cref{sec:background:implementationcecks}).
The Common Weakness Enumeration (CWE)\,\cite{cwe} provides an exhaustive list of weaknesses from which software-intensive systems may suffer.
A single weakness may be further detailed by one or more weaknesses, represented in the \model as a \texttt{parent of} relationship.
For example, the weakness CWE-328 discussed above is a more detailed instance of the weakness CWE-326.
In this paper, we have populated the instance of the security model with the entire CWE in version 4.12\,\cite{cwe}.
\smallskip

\validation{Security Model}{
\edit{In the validation with 22 security experts}, all of them agreed with the security objectives, threats, and weaknesses of the \textit{Security Model} as well as their relations.
}




\subsection{System Model}
\label{sec:model:systemmodel}
On the level of the system described using a security DSL or analyzed by a security analyzer, we have to consider different types of elements, relationships between them, and implications if an element of a specific type is vulnerable.
We first introduce the common element types that we identified and their semantic relationships, and then, what security-related implications are there for other element types if an element has been compromised.

\subsubsection{Element Types and Semantic Relationships.}
\begin{figure}
	\center
	\includegraphics[width=.7\columnwidth]{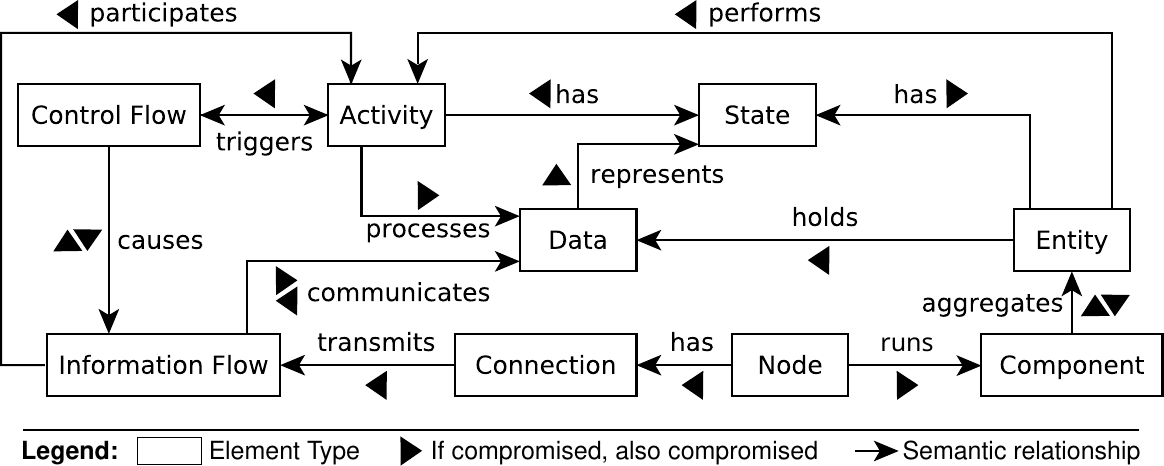}
	\caption{System element types relevant in the context of security
}
	\label{fig:SystemElementsConceputalModel}
\end{figure}

Design-time models and code both describe systems, but with different abstractions of the elements that make up a software system, such as components, objects, or functions.
We have identified common system element types relevant to security that map to both design and code, allowing us to relate design-time security DSLs and code-level checks.
The \textit{System Model} (right side of \cref{fig:models}) describes these types of system elements (\textit{Element Type}), with the concrete identified types and their semantic relations shown in \cref{fig:SystemElementsConceputalModel}.


\textit{Data.}
A central element type in software security is \textit{Data}, which conveys information through a collection of values.
While 67\,\% of the DSLs analyzed address \textit{Data}, e.g., by labeling communicated information with security objectives (cf. to \textit{SecDFD} in \cref{fig:secdfd}).
\textit{Data} is directly analyzed by 35\,\% of the checks.

\textit{Entity.}
\textit{Data} can be held by \textit{Entities}, which can be a physical actor, software object, or external system such as a database.
For example, a class in code is an \textit{Entity} that holds \textit{Data} via its attributes.
Only 11\,\% of the checks analyze entities, but 85\,\% of the DSLs specify security-related information on them.

\looseness=-1
\textit{Activity.}
To realize a system's behavior, an \textit{Entity} performs an \textit{Activity} that processes \textit{Data} or communicates with other \textit{Activities}.
The processes of the DFD in \cref{fig:secdfd} are instances of an \textit{Activity} as well as the operations in code that implement these.
Most security checks are performed on this element type (83\,\%), and 67\,\% of the DSLs refer to it.

\textit{State.}
\textit{Entities} and \textit{Activities} can have a \textit{State} which is generally represented by \textit{Data}, i.e., by the semantics of its values.
Examples of \textit{State} include the values of object attributes and the roles assigned to an adversary in a system.
\textit{States} are targeted by 40\,\% of the security DSLs and 9\,\% of the checks.

\looseness=-1
\textit{Control Flow.}
The execution of \textit{Activities} is orchestrated by \textit{Control Flow}.
DFDs contain it only implicitly, but design models such as UML activity diagrams express it explicitly, and 8\,\% of the DSLs refer to it.
In code, operation calls represent control flow, and 18\,\% of the security checks analyze them.

\textit{Information Flow.}
When \textit{Data} is exchanged between \textit{Activities} (indirectly also between \textit{Entities}, i.e., receiving data can be seen as an \textit{Activity}), this can form an \textit{Information Flow}.
The literature\,\cite{Balliu2017} distinguishes between information flow and data flow.
While data flow considers any flow of data in the system, information flow focuses on security-critical data, either explicitly communicated or indirectly communicated by control flow.
Since the security-relevant consideration is that transmitted data conveys information, we subsume both under the term information flow.
While 27\,\% of security DSLs allow for planning secure information flow, 38\,\% of security checks address the leakage of sensitive information via techniques such as taint analysis\,\cite{Hough2022}.

\textit{Component.}
Multiple software \textit{Entities} can be combined into a \textit{Component} that encapsulates specific functionality.
Components have different granularity, ranging from an entire system to semantically related classes.
\edit{Components do not necessarily have to be part of the system itself or their internal need to be fully accessible, but can also be third-party libraries used by a system.}
While only 1\,\% of checks are at the component level, this is 34\,\% of security DSLs.

\textit{Node.}
A \textit{Component} can be deployed on a \textit{Node}, which is a physical device executing software.
While 21\,\% of the security DSLs analyzed address \textit{Nodes}, no security checks do so.

\textit{Connection.}
Communication between \textit{Activities} or \textit{Entities} is established either through physical \textit{Connections} between \textit{Nodes}\edit{, e.g., via LAN, the Internet, or local sockets,} or internally within a component.
Security DSLs refer more often to connections (25\,\%) than security analyzers\,(2\,\%).

\subsubsection{\securityRelation}
Attacks typically propagate through a system\,\cite{Walter2022,ATTCK}; once an attacker compromises the first element of a system, such as by exploiting a vulnerability, the attacker gains control of everything dominated by that element and attempts to expand its influence from there by exploiting additional vulnerabilities.
Accordingly, such propagation has to be considered when assessing the impact of a possible vulnerability.
To support this assessment, the \model captures the potential of immediate propagation without the need to exploit a different vulnerability with the \textit{\securityRelation} relationship, providing a basis for reasoning about the specifics of concrete cases.

This reasoning is not limited to code, but spans the gap between design-time and implementation-level security.
As an example, consider an insecure \textit{Connection}, e.g., discovered by an analyzer.
While a \textit{Connection} is outside the scope of a DFD, a compromised connection can allow an attacker to obtain or alter the \textit{Information Flow} over it that is considered in SecDFD.
However, this does not mean that all security objectives are always compromised, and there may be concrete cases where nothing else is compromised.
For example, if the information communicated by the \textit{Information Flow} over the compromised \textit{Connection} is encrypted, the confidentiality and integrity of the information is still preserved. 
However, in this case, an attacker can still compromise the availability of the data being transmitted.

We systematically derived the \textit{\securityRelation} relationships by triangulating the observations of our study with related literature (cf. \cref{sec:method:analysis}).
In the following, we give an overview of the relationships identified and the rationale behind them.

\textit{Data $\rightleftarrows$ Information Flow $\rightleftarrows$ Control Flow.}
According to Balliu et al.\,\cite{Balliu2017}, information flow analysis defines an information flow as a combination of \textit{Control Flow} and \textit{Data Flow}.
If the \textit{Data} in an information flow is compromised, e.g., because it can be manipulated by an attacker, then the information flow transporting this data is also compromised. 
Taint trackers like FlowDroid\,\cite{Arzt2014a} or JOANA\,\cite{Giffhorn2008} flag other \textit{Data} as tainted if they are affected by an already tainted information flow.  
Consequently, we describe a bi-directional \textit{\securityRelation} relation between \textit{Information Flow} and \textit{Data}. 
If data is used to determine the control flow, observing the program behavior allows information about the information flow to be obtained.
As discussed above, an information flow can be controlled if the \textit{Control Flow} can be illegally modified. 
Therefore, we also define a bi-directional \textit{\securityRelation} relation between \textit{Information Flow} and \textit{Control Flow}. 

\textit{Activity $\rightarrow$ Control Flow, Activity $\rightarrow$ Information Flow.}
From the above reasoning, it follows that if an \textit{Activity} is compromised, then the \textit{Control Flow} it triggers, or the \textit{Information Flow} in which it participates, can be manipulated.
Consequently, there is an \textit{\securityRelation} relation between \textit{Activity} and \textit{Control Flow} and between \textit{Activity} and \textit{Information Flow}, respectively. 

\textit{Connection $\rightarrow$ Information Flow.}
The analysis by Kramer et al.\,\cite{Kramer2017}, UMLSec\,\cite{Juerjens2005}, or Almorsy et al.\,\cite{Almorsy2013} all evaluate whether communicated information is secure. 
They flag communicated information as insecure if the \textit{Connection} over which the information flows is not correctly secured, e.g., by not employing encryption.  
Therefore, we describe a \textit{\securityRelation} relationship between \textit{Connection} and \textit{Information Flow}.
A relevant type of attack illustrating this relation is man-in-the-middle attacks: If an attacker can access the communication, the communicated information can be altered or read. 

\textit{Node $\rightarrow$ Connection.}
In network security, attacks like \textit{Spoofing} attacks, \textit{Denial of Service} attacks, or \textit{Wormhole}\,\cite{Hu2003} threaten the network itself by compromising a network \textit{Node} to change network topography, flooding the network with messages, or modifying routing paths through the network\,\cite{Pawar2015}.
Consequently, a network as a whole, i.e., its \textit{Connections}, can be compromised if one of its \textit{Nodes} is compromised.
This is why we consider a \textit{\securityRelation} relationship between \textit{Node} and \textit{Connection}.

\textit{Node $\rightarrow$ Component.}
The analyses of Walter et al.\,\cite{Walter2022} and Hahner\,\cite{Hahner2024} define that an attacker has access to all \textit{Components} running on a \textit{Node} that the attacker compromised.
This is why we define a \textit{\securityRelation} relation from \textit{Node} to \textit{Component}. 

\textit{Component $\rightleftarrows$ Entity.}
Similarly to \textit{Nodes}, if a \textit{Component} is affected by a security issue, it is likely that also the composed \textit{Entities} are affected.  
For example, in the case of a maliciously exchanged library, any \textit{Entities} it contains, i.e., classes that provide functionality, must be considered compromised.
\edit{
Conversely, a compromised \textit{Entity} can also affect the \textit{Component} it is part of.
For example, a virtual machine (\textit{Entity}) which contains the VM escape vulnerability can be exploited to compromise the host system (\textit{Component})~\cite{Alnaim19_VMescape}.
}
Therefore, we define a \textit{\securityRelation} relationship between \textit{Component} and \textit{Entity}.

\textit{Entity $\rightarrow$ Data, Entity $\rightarrow$ Activity.}
Due to the differences in what can be expressed using \textit{Entities}, we have to consider its relationships from two perspectives, an abstract perspective as considered in some security DSLs and a more narrow, low-level perspective on \textit{Entities} in code.
In social engineering, attackers compromise users (typically considered in design-time security DSLs by elements relating to the concept of an \textit{Entity}) to obtain or manipulate \textit{Data} like passwords\,\cite{Wang2020}. 
Similarly, if a database (\textit{Entity}) is accessible by an attacker, e.g., due to poor permission management, the database content can also be obtained or altered\,\cite{Mousa2020}.
Attackers can also use social engineering to convince users (\textit{Entity}) to perform certain \textit{Activities}, e.g., triggering a process\,\cite{Wang2020}. 
On the code level, if a class, e.g., via exploiting a deserialization vulnerability\,\cite{Peldszus2024}, has been compromised, none of the \textit{Activities} provided by this \textit{Entity} can be trusted. 
Consequently, we define a \textit{\securityRelation} relation between \textit{Entity} and \textit{Data}.
Also, we define a \textit{\securityRelation} relation between \textit{Entity} and \textit{Activity}. 

\textit{Data $\rightarrow$ State.}
As the \textit{State} is defined by \textit{Data}, logically, if the \textit{Data} can be modified by an attacker, the \textit{State} can be equally modified. 
Consequently, we define a \textit{\securityRelation} relation between \textit{Data} and \textit{State}. 

\textit{State $\rightarrow$ Activity, State $\rightarrow$ Entity.}
Related to these two relationships, \textit{CWE-642 (External Control of Critical State Data)} explicitly states that modification of a state can be used to either perform ``unauthorized actions (\textit{Activity}) or access unexpected resources (e.g, \textit{Data} held by \textit{Entities}).''
In a program, if executing an action depends on some state variable and an attacker can modify this variable, the attacker can decide whether the method can be executed. 
This is why we define a \textit{\securityRelation} relation between \textit{State} and \textit{Activity} and \textit{State} and \textit{Entity}. 
Please note that a compromised \textit{Activity} can also lead to a compromised state via compromised \textit{Data} modified by the \textit{Activity} that represents the \textit{State}.

\validation{System Elements}{
\edit{While 19 experts} fully agreed with the extracted system element types, two experts reported two minor concerns.
One shows concerns about the relation between \textit{Information Flow} and \textit{Entity}, since there may be cases where an information leak may not be due to a single \textit{Entity}.
However, the participant stated that this \textit{Entity} is what you want to have to pinpoint the locations to fix information flow violations---this confirms the fact that there is a relationship to be considered, which is the only aspect we wanted to express.
The other perceived the semantic relation between \textit{Control Flow} and \textit{Information Flow} as correct, but discussed whether it could also be realized transitively over \textit{Activity}.
No expert found misconceptions in the \securityRelation relations.
\edit{One expert in analyzers pointed out that the derived abstraction is not perfectly capable of representing the corresponding source code elements, a stretch that stems from also having to address the need to represent DSLs.}
}

\subsection{Security DSL Description}
\begin{figure}
	\center
	\includegraphics[width=\textwidth]{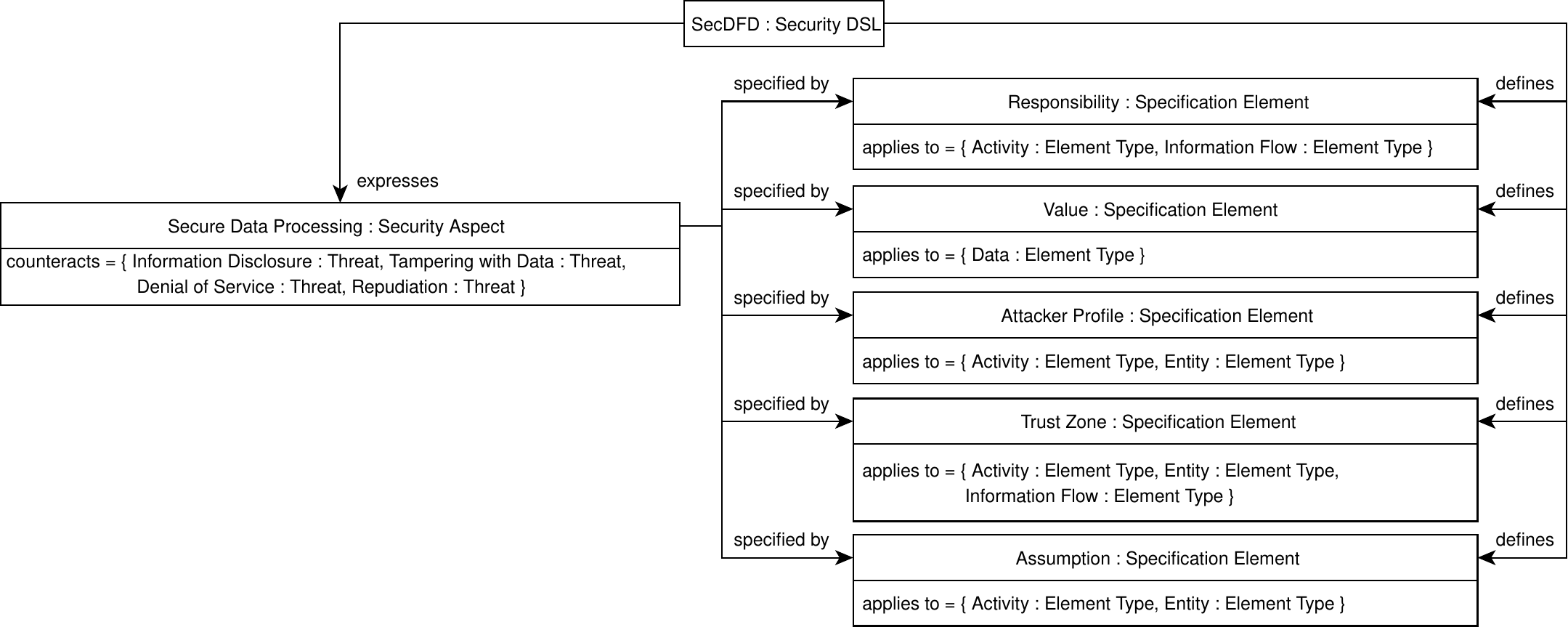}
	\caption{Application of the SecLan model to SecDFD (Associations with elements from \cref{fig:SystemElementsConceputalModel,fig:example} are shown as attributes)}
	\label{fig:application:secdfd}
\end{figure}

\noindent
A \textit{Security DSL} is used to design software systems to achieve one or more security objectives. 
To this end, all DSLs express the security aspects they target using the specification elements defined in the DSL, i.e., the set of specification elements defines the syntax of the security DSL, and the security aspects define the semantics to be expressed.
Security aspects and specification elements are realized differently in each security DSL.
While some related work maps them to the intended security objective, e.g., \textit{integrity},
other works\,\cite{Geismann2021,Tuma2019a,Katkalov2013,Kramer2017} structure security aspects of specification elements with finer granularity, e.g., to provide aspects for secure information flow or secure physical access.
In our analysis of 66 security DSLs, we identified counteracted threats as the abstraction applicable to all of them, e.g., supporting the abstraction of security objectives by describing a security DSL via the threats that threaten these security objectives.
\Cref{fig:application:secdfd} shows an exemplary description of SecDFD, which will be explained in detail below.

\textit{Security Aspect.} 
DSLs define \textit{Security Aspects} to counteract potential \textit{Threats} from the security model (cf. \cref{fig:example}).
For example, SecDFD defines the security aspect of \enquote{Secure Data Processing}, which aims to counteract the threats of \textit{Information Disclosure}, \textit{Tampering with Data}, \textit{Denial of Service}, and \textit{Repudiation}, e.g., due to data leaks or illegal modifications caused by insecure data processing.

\textit{Specification Element.}
The syntactic means to specify \textit{Security Aspects} are \textit{Specification Elements}, which provide security-related information by being applied to instances of the system element types of \cref{fig:SystemElementsConceputalModel}.
For example, SecDFD defines the specification elements \textit{Responsibilities} to label \textit{Activities} (processes in DFDs) with data processing contracts, such as a cryptographic operation or whether data will be joined.
Conceptually, SecDFD uses the same specification element for all possible processing contracts and details their responsibilities, i.e., joining data in the element's properties.
The specification element \textit{Value} allows for labeling \textit{Data} (assets in DFDs) with security objectives.
While these security objectives correspond to those from \cref{fig:example}, this is the only specification element in SecDFD with such a relationship, and considering the specification elements of all security DSLs, such a relationship to security objectives has only a fraction of all elements.
Besides these two specification elements, \cref{fig:application:secdfd} shows the specification elements \textit{Attacker Profile}, \textit{Trust Zone}, and \textit{Assumption}.

\validation{Sec DSL Description}{Except for \edit{three experts, the other 19 experts} fully agreed with the precise terminology for the \textit{Security Aspect}.
The first noted that in the goal modeling community, \textit{Confidentiality} and \textit{Integrity} should be enforced by a security DSL, and therefore, should be \textit{Security Aspects}, whereas we have subsumed it under \textit{Security Objectives} to which he still agrees as an indirect but appropriate solution.
The other expert asked why a \textit{Security Aspect} cannot immediately remedy a \textit{Weakness}, which we again only cover indirectly via \textit{Threats} that enable \textit{Weaknesses}.
\edit{The third expert considered the definition of specification element as too abstract.}
\edit{Except for one expert,} all experts fully agreed with the relationships between \textit{Sec DSL Description} and \textit{System Model}.}

\subsection{Security Analyzer Description}
\begin{figure}
    \begin{subfigure}{.495\textwidth}
	\center
	\includegraphics[width=\textwidth]{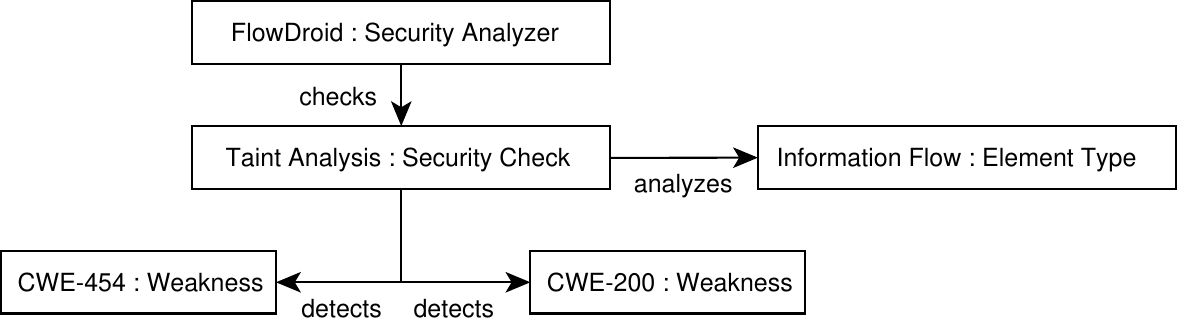}
        \caption{FlowDroid}
	\label{fig:application:flowdroid}
    \end{subfigure}
    \hfill
    \begin{subfigure}{.495\textwidth}
	\center
	\includegraphics[width=\textwidth]{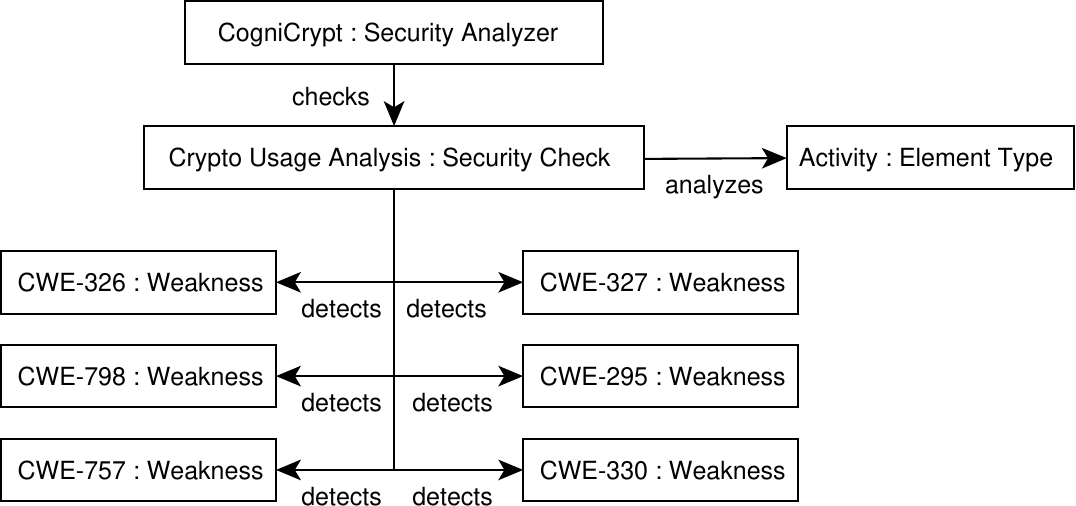}
        \caption{CogniCrypt}
	\label{fig:application:cognicrypt}
    \end{subfigure}
    \caption{Application of the SecLan model to security analyzers}
    \label{fig:application:checks}
\end{figure}

A \textit{Security Check} is usually provided as part of an \textit{Analyzer}.
While analyzers such as SonarQube or CodeQL implement multiple security checks, others like FlowDroid (see \cref{fig:application:flowdroid}) implement exactly one check.
A \textit{Security Check} inspects certain code elements, described by their \textit{Element Type}, for errors related to predefined \textit{Weaknesses}.
For example, FlowDroid analyzes \textit{Information Flow}.
The \textit{Security Check} then reports detected instances of the \textit{Weakness}, i.e., CWE-200 and CWE-454 in the case of FlowDroid.
Relevant weaknesses are either documented directly by analyzer developers, e.g., in SonarQube check descriptions, or identified in academic papers, e.g., by Zhang et al.\,\cite{Zhang2023} for cryptographic \textit{API Usage} analyzers.
For analyzers where no weaknesses were documented, we searched the CWE\,\cite{cwe} for relevant weaknesses.
For the security analyzer VuRLE, we had to specify a wide variety of weaknesses since it is ML-based and has been trained on six entirely different types of vulnerabilities\,\cite{Ma2017}.

While it is relatively obvious in the case of FlowDroid that it provides a single check that can be executed in multiple contexts, such as detecting potential exposure of sensitive information (CWE-200) or external initialization of variables (CWE-454) by reversing what is considered a source or sink, this is not so obvious in other cases.
For example, CogniCrypt provides the CrySL language, a DSL that allows specifying the correct use of an API, which itself is outside the scope of this work.
Although this language allows the specification of multiple security checks, each concerning a different API, the underlying principle is the same, and, therefore, we resorted to considering CogniCrypt to provide only one check (see \cref{fig:application:cognicrypt}).
However, due to the flexibility of CogniCrypt, its security check can address a wide range of weaknesses, with the cryptography-related weaknesses being the focus of the initial version of CogniCrypt\,\cite{Krueger2017}.

\validation{Sec Analyzer Description}{\edit{The 20 of the 21 experts} who answered the questions on the \textit{Sec Analyzer Description} (one did not answer these questions) agreed with the identified concepts and their relationships to the \textit{Security Model} and \textit{System~Model}.
\edit{The one expert not agreeing acknowledged that the relationships make sense at a high level, but was missing low-level details specific to analyzers, which unfortunately do not generalize across analyzers according to our analysis of \analyzers security analyzers.}
}

\subsection{Application of SecLan's to DSLs and Analyzers}
\noindent
We successfully applied the SecLan conceptual model to the \dsls security DSLs, and \analyzers analyzers analyzed.
\Cref{fig:application:secdfd,fig:application:checks} show the application of the SecLan conceptual model to the SecDFD security DSL from our running example and the security analyzers FlowDroid and CogniCrypt.
We particularly show the analyzer \textit{FlowDroid} as it has already been used in combination with SecDFD to investigate compliance between the design and implementation\,\cite{Peldszus2019,Tuma2022}.

\edit{
    \Cref{tab:description-stats} shows statistics on the concrete applications in terms of the size of the description files created.
    On average, a description has 158 lines of JSON code and contains 4,062 characters, primarily for description texts.
    Particularly, security analyzer descriptions can get relatively large due to the huge number of checks provided by multi-purpose analyzers, reaching up to 2,438 lines of JSON code and 59,850 characters.
}
The instances with respect to the \model created for all the security DSLs and analyzers we have studied can be found in our SecLan repository\,\cite{Replication}.

\begin{table}
    \caption{Statistics on the sizes of the created descriptions of \analyzers security analyzers and \dsls security DSLs}
    \label{tab:description-stats}
    \centering
    \begin{tabular}{llcc}
        \toprule
         	  &            & \textbf{LoC}       & \textbf{Chars} \\
        \midrule
        Min     & all       & 29 & 495 \\ 
                & DSLs      & 58 & 1,417 \\ 
                & Analyzers & 29 & 495 \\ 
        \midrule 
        Average & all       & 158.31 & 4,062.08 \\ 
                & DSLs      & 122.41 & 3,346.39 \\ 
                & Analyzers & 224.14 & 5,374.17 \\ 
        \midrule 
        Max     & all       & 2,438 & 59,850 \\ 
                & DSLs      & 378 & 10,658 \\ 
               & Analyzers & 2,438 & 59,850 \\ 
        \bottomrule
    \end{tabular}
\end{table}

As intended in step 6 of our methodology, we asked experts not only about the general concepts we extracted, but also about how we addressed their security DSLs or analyzers.
In particular, we needed to verify the correctness of the models we created to ensure that our qualitative and quantitative investigation was based on a valid database. 
To this end, in the second part of our survey, we asked the experts to verify our description of their DSL or analyzer according to the \model. 
Among others, we asked whether there are elements from the security DSL or the security analyzer that are not addressed by our conceptual model to determine whether our conceptual model is incomplete. 
Another question we asked was whether elements of their respective DSL or analyzer are classified incorrectly or inappropriately. 
A question especially targeting \textit{Security Analyzers} was whether there are incorrectly or inappropriately assigned CWEs or even missing ones. 

We received \edit{16 responses} to the second part of our questionnaire that covers 10 DSLs (CARDS\,\cite{Geismann2021}, Attack-Fault Trees\,\cite{Kumar2017}, PbSD\,\cite{Abramov2012}, SecBPMN\,\cite{Salnitri2015}, SecDFD\,\cite{Tuma2019a}, SecureUML\,\cite{Lodderstedt2002}, SysML-Sec\,\cite{Apvrille2013}, UMLsec\,\cite{Juerjens2005}, the Architectural Dataflow DSL\,\cite{Seifermann2022} and the Attack Propagation DSL\,\cite{Walter2022}) and \edit{five analyzers} (FlowDroid\,\cite{Arzt2014a}, PMD\,\cite{pmd}, \edit{CogniCrypt\,\cite{CogniCrypt}, CodeQL\,\cite{CodeQL}, and an ESLint plugin\,\cite{Rafnsson2020}}).
These cover a wide range of security aspects and system representations, including business process models, dataflow diagrams, component-based architectures, and UML diagrams.
Security aspects addressed are confidentiality, integrity, authenticity of data, accountability, auditability, secure dataflow, access control, secure dependencies, and communication security.

Based on the feedback provided, we adjusted the models, whereby the most adjustments were necessary for SecBPMN.
For example, one participant explained that the SecBPMN specification element \textit{Integrity} has two semantics: to label security requirements of \textit{Data} as we considered it and to express integrity requirements of a task.
Therefore, we added a link to the system element \textit{Activity}.
In addition, the participant stated that an additional element applies to a \textit{Security Aspect} of SecBPMN.
One participant added \textit{Information Flow} to a \textit{Specification Element} in SecDFD.
Another participant removed \textit{Information Flow} from a \textit{Specification Element} in the architectural dataflow DSL.
One participant stated that specification elements related to a \textit{Security Aspect} of UMLsec (\textit{Secure Dependency}) should only apply to \textit{Information Flow}, while we also specified \textit{Control Flow}, since the specification of the UMLsec security aspect explicitly considers the potential for an insecure dependency to be a security violation\,\cite{Juerjens2005}.
Another participant validating UMLsec did not raise such a concern.
After adjusting the models regarding these minor comments from the experts, we had a valid basis for further qualitative and quantitative investigations. 

\validation{Application of SecLan}{
No expert found major inconsistencies in the application of SecLan to their DSLs or \edit{analyzers, but some identified adjustments to individual DSL elements or security checks.}
We encountered challenges mainly for SecBPMN due to different semantics of specification elements, which we resolved based on the expert's feedback.
Adjustments to the specifications of all other DSLs and analyzers were minor, such as adding or deleting a single system element type to \textit{Specification Elements}.
For FlowDroid, we did not cover all possible application scenarios, i.e., detected weaknesses, since it only offers a building block for a security check.
\edit{For PMD, one of two experts suggested adjustments to the system model elements analyzed for individual checks.}
}

\section{Relating Security Checks and Security DSLs}
\noindent
%
%
Security DSLs and checks of security analyzers are on opposite sides of the gap between design-time and code-level security.
\edit{
For instance, using the SecDFD DSL (see \cref{fig:secdfd}), security experts specify what confidential information in iTrust is and what the expected data processing is, e.g., what data flows are planned and where data is expected to be encrypted.
As shown in \cref{fig:application:secdfd}, the security aspect expressed by SecDFD aims to counteract \textit{Information Disclosure}, \textit{Tampering with Data}, \textit{Denial of Service}, and \textit{Repudiation}.
Therefore, it specifies security assumptions with respect to \textit{Activities}, \textit{Information Flow}, \textit{Data}, and \textit{Entities}, making use of five different \textit{Specification Elements}.
The question is which vulnerabilities in the implementation, as detected by the checks, are suitable for invalidating the security aspects specified using DSLs on the other side of the gap.
}

\edit{
For example, consider CogniCrypt as shown in \cref{fig:application:cognicrypt}. 
Its crypto usage analysis analyzes system elements of the type \textit{Activity}, onto which SecDFD expresses security aspects.
The check is capable of detecting instances of the weaknesses \textit{CWE-326: Inadequate Encryption Strength}, \textit{CWE-327: Use of a Broken or Risky Cryptographic Algorithm}, \textit{CWE-798: Use of Hard-coded Credentials}, \textit{CWE-295: Improper Certificate Validation}, \textit{CWE-330: Use of Insufficiently Random Values}, and \textit{CWE-757: Algorithm Downgrade}.
All of them are potentially threatening the proper implementation of cryptography planned using SecDFD.
This relationship is visible in corresponding instances of the weaknesses posing threats of \textit{Information Disclosure}, \textit{Tampering with Data}, \textit{Repudiation}, and \textit{Spoofing}, most of which are meant to be counteracted by the security aspect specified using SecDFD.
}

\edit{
As CogniCrypt is a relatively obvious example, with a very directly visible relationship to the security aspect of SecDFD, let us consider a more general, less obvious example, such as a check for potential buffer overflows, such as the \textit{OverflowBuffer} check provided by CodeQL.
Buffer overflows are known to enable a variety of exploits, and the question to answer is whether, in principle, there is the possibility of the secure data flow aspect expressed by SecDFD being violated, and what are the elements of the relationship that would need to be inspected to assess whether a concrete vulnerability affects the security design.
The check \textit{OverflowBuffer} analyzes code elements corresponding to system elements of types \textit{Entity}, \textit{Activity}, \textit{Information Flow}, and \textit{Data} to detect buffers, i.e., an \textit{Entity} in the program that holds \textit{Data}, that can overflow. 
For a buffer overflow to be possible, the buffer must be accessed out of range, where the range is concrete \textit{Data} values.
To detect such cases, \textit{Information Flow} within the implementation of an \textit{Activity}, e.g., its statements, is analyzed.
Since SecDFD is used to express security aspects primarily on \textit{Data}, e.g., security labels, and corresponding \textit{Information Flow} between processes within a DFD (i.e., \textit{Activities}), there is a direct overlap with the security check on the side of the system model.
The question is whether the security objective, e.g., confidentiality and integrity of data, aimed to be achieved by counteracting corresponding threats, e.g., \textit{Information Disclosure} and \textit{Tampering with Data}, can be enabled by the weakness that the check can detect.
In the case of \textit{OverflowBuffer}, this is \textit{CWE-119: Improper Restriction of Operations within the Bounds of a Memory Buffer}, which has the potential to enable threats with respect to \textit{Information Disclosure}, \textit{Tampering with Data}, and \textit{Denial of Service}.
Therefore, we must conclude that a buffer overflow, in principle, can prevent the planned data processing from occurring as intended, e.g., information labeld as sensitive using SecDFD could be accessed by an attacker due to an out of range read explointing the overflow.
This is visible in direct connections between SecDFD's security aspect and CodeQL's security check over both the system model, e.g., both being related to \textit{Data}, and the security model, e.g., via the directly enabled threat of \textit{Tampering with Disclosure}.
}

\edit{As shown above, we have to consider security checks to be related to security aspects} of a DSL via their relationship to both the \textit{Security Model} and the \textit{System Model}.
Weaknesses detected by security checks enable threats, while security aspects aim to counter them.
We exploit this relationship to relate security checks to security DSLs in combination with their relationships to the types of system elements that are either analyzed or enriched with security-related information.
\textit{A check is related to a security aspect if and only if at least one detected weakness enables a threat countered by the security aspect, and the targeted system elements are identical or form a security relationship.}

\looseness=-1
Since instances of \textit{Sec DSL Description} and \textit{Sec Analyzer Description}, as well as the instances of the \textit{Security Model} (cf. \cref{fig:example}) and the \textit{System Model} (cf. \cref{fig:SystemElementsConceputalModel}), comprise a graph consisting of typed nodes and typed edges, we can formalize this relationship as follows.
Consider a graph consisting of a set of typed nodes $n$$\in$$N$ with a type $type_n$ denoted as $n:type_n$ and a set of directed, typed edges $e$$\in$$E$ between these nodes.
An edge $e$ connects two nodes $n_1$ and $n_2$ and has a type $type_e$, denoted as $n_1${\scriptsize$\xrightarrow{\text{type}_e}$}$n_2$.
The node types $type_n$ and edge types $type_e$ are defined in the \model in \cref{fig:models}.

A \textit{Security Aspect} $a$ and \textit{Security Check} $b$ are related if there exist two paths from $a$ to $b$ that are specified by the sequence of edge types on them, as in the function \textit{related}:

\begin{small}
\begin{center}
	$related:$ $a{:}SecurityAspect, b{:}SecurityCheck$ $|$
	$\exists path(a \xrightarrow[]{\text{counteracts}}\xleftarrow[]{\text{enables}} 
	\xrightarrow{\text{(parent of)}^*}
	\xleftarrow{\text{detects}} b)$
	$\land$ $\exists path(a \xrightarrow{\text{specified by}} \xrightarrow{\text{applies to}}
	\xleftarrow{\text{(\securityRelation)}^*}
	\xleftarrow{\text{analyzes}} b)$
\end{center}
\end{small}

\noindent
We use a regular path expression\,\cite{Ebert2010} $path(a$$\rightarrow$$b)$ between nodes $a$ and $b$ that describes the sequence of edge types of this path as a regular expression.
Please note that only acyclic paths are allowed, i.e., each edge appears at most once on a path.
Ebert et al.\,\cite{Ebert2010} explain this notation in detail.
At the given position, an edge type can be expected to appear exactly once {\scriptsize($\xrightarrow{\text{type}_e}$)} or an arbitrary number of times ({\scriptsize$\xrightarrow{(\text{type}_e)^*}$}) that also includes not appearing on the path.
An edge pointing to the right ($\rightarrow$) is navigated in the direction specified in the \model in \cref{fig:models} and an edge pointing to the left ($\leftarrow$) is navigated in the opposite direction.

\looseness=-1
To explore the relationship between a security aspect and a security check in detail, we can compute all possible paths between them.
This allows us to explore further potential relationships that may be relevant to consider when relating a discovered vulnerability to the security design.
For the security aspect of a DSL and a security check to be related, there must be at least 6 edges between them. 
On the path through the security model, there must always be a $counteracts$, an $enables$, and a $detects$ edge.
On the path through the system model, there is always a $specified$ $by$, a $applies$ $to$, and an $analyzes$ edge.
However, both paths can be longer by including additional edges in addition to these required edges.

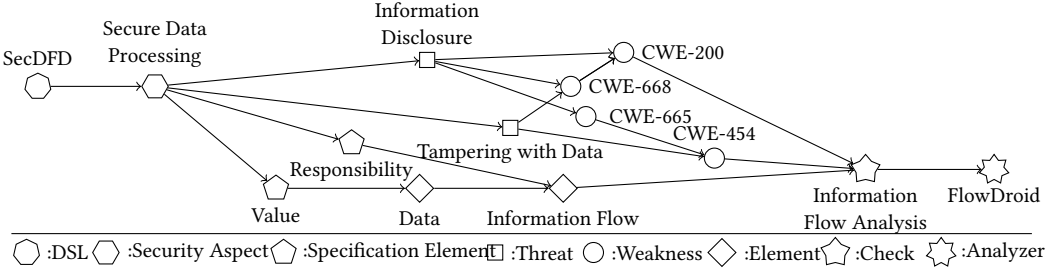
\begin{figure}
	\center
	\begin{footnotesize}
	\begin{tikzpicture}[every label/.style={align=center}]
		\node[shape=regular polygon,regular polygon sides=7,draw=black,minimum size=.35cm] (secdfd) [label=above:{SecDFD}] at (-4.15,-0.35) {};
        
		\node[shape=regular polygon,regular polygon sides=6,draw=black,minimum size=.35cm] (aspect) [label=above:{Secure Data\\ Processing}] at (-2.6,-0.35) {};

        \node[shape=regular polygon,regular polygon sides=5,draw=black,minimum size=.35cm] (value) [label=below:{Value}] at (-1,-1.7) {};
        \node[shape=regular polygon,regular polygon sides=5,draw=black,minimum size=.35cm] (responsibility) [label=below:{Responsibility}] at (0,-1.1) {};

		\node[shape=rectangle,draw=black,minimum size=.2cm] (confidentiality) [label=above:{Information\\ Disclosure}] at (1,0) {};
		\node[shape=rectangle,draw=black,minimum size=.2cm] (integrity) [label=below:{Tampering with Data}] at (2.1,-0.9) {};

		\node[shape=circle,draw=black] (cwe200) [label=right:{CWE-200}] at (3.6,0.1) {};
		\node[shape=circle,draw=black] (cwe454) [label=above:{CWE-454}] at (4.8,-1.3) {};
		\node[shape=circle,draw=black] (cwe668) [label=right:{CWE-668}] at (2.9,-0.35) {};
		\node[shape=circle,draw=black] (cwe665) [label=right:{~CWE-665}] at (3.1,-0.75) {};

		\node[shape=star,draw=black] (check) [label=below:{Information\\ Flow Analysis}] at (6.8,-1.45) {};
		\node[shape=star,star points=7,draw=black] (analyzer) [label=below:{FlowDroid}] at (8.5,-1.45) {};

		\node[shape=diamond,draw=black] (data) [label=below:{Data}] at (0.9,-1.7) {};
		\node[shape=diamond,draw=black] (flow) [label=below:{Information Flow}] at (2.8,-1.7) {};

		\path [->] (secdfd) edge (aspect);
		\path [->] (aspect) edge (confidentiality);
		\path [->] (confidentiality) edge (cwe200);
		\path [->] (confidentiality) edge (cwe665);
		\path [->] (cwe665) edge (cwe454);
		\path [->] (confidentiality) edge (cwe668);
		\path [->] (cwe668) edge (cwe200);
        
		\path [->] (aspect) edge (value);
		\path [->] (aspect) edge (responsibility);

		\path [->] (aspect) edge (integrity);
		\path [->] (integrity) edge (cwe454);
		\path [->] (integrity) edge (cwe668);
		\path [->] (cwe668) edge (cwe200);

		\path [->] (cwe454) edge (check);
		\path [->] (cwe200) edge (check);

		\path [->] (value) edge (data);
		\path [->] (responsibility) edge (flow);
		\path [->] (data) edge (flow);
		\path [->] (flow) edge (check);
		\path [->] (check) edge (analyzer);

		\path (-4.5,-2.3) edge (9.0,-2.3);
		\node[shape=regular polygon,regular polygon sides=7,draw=black,minimum size=.35cm] [label=right:{:DSL}] at (-4.3,-2.55) {};
		\node[shape=regular polygon,regular polygon sides=6,draw=black,minimum size=.35cm] [label=right:{:Security Aspect}] at (-3.25,-2.55) {};
		\node[shape=regular polygon,regular polygon sides=5,draw=black,minimum size=.35cm] [label=right:{:Specification Element}] at (-0.9,-2.55) {};
		\node[shape=rectangle,draw=black,minimum size=.2cm] [label=right:{:Threat}] at (1.9,-2.55) {};
		\node[shape=circle,draw=black] [label=right:{:Weakness}] at (3.2,-2.55) {};
		\node[shape=diamond,draw=black] [label=right:{:Element}] at (4.9,-2.55) {};
		\node[shape=star,draw=black] [label=right:{:Check}] at (6.4,-2.55) {};
		\node[shape=star, star points=7,draw=black] [label=right:{:Analyzer}] at (7.8,-2.55) {};

	\end{tikzpicture}
\end{footnotesize}
\caption{Relationship between SecDFD and FlowDroid}
\label{fig:paths}
\end{figure}

When we apply $related$ to the SecDFD example and the taint analyzer FlowDroid, which was used by Tuma et al.\,\cite{Tuma2022} in a security compliance check, then we get the relationships shown in \cref{fig:paths}.
FlowDroid is immediately related with SecDFD over \textit{Data} and \textit{Information Flow} and detects the weaknesses \textit{CWE-200 (Exposure of Sensitive Information)} and \textit{CWE-454 (External Initialization of Trusted Variables)} that directly enable the threats \textit{Information Disclosure} and \textit{Tampering with Data} that should be counteracted by SecDFD.
Further relationships exist over the parent-child relationships of the weaknesses.
For example, {\textit{CWE-668  (Exposure of Resource)}} is a parent of CWE-200, which could result in further security objectives being indirectly threatened.

%% file: sections/05_evaluation.tex
\section{Exploration of Relationships between Security DSLs and Security Checks} 
\noindent\looseness=-1
We aim to gain deeper insight into the relationships between design-time security DSLs and code-level security checks as a first step toward bridging the gap between security design and code.
For this, we analyzed their relationships quantitatively using SecLan and conducted exploratory interviews with experts in security DSLs and security checks to gain qualitative insights.

\input{sections/04_1_recommendations.tex}

\subsection{Quantitative Exploration.}
\noindent
\looseness=-1
To characterize the gap between security DSLs and analyzers, we examined which of the identified aspects of system security are covered by design-time security DSLs and checks of code-level security analyzers, and the relationship between DSLs and checks.

\subsubsection{Elements Covered by DSLs and Analyzers.}
\Cref{tab:statistics} lists how often the elements of the system model are covered by security DSLs and checks of security analyzers, which threats DSLs counteract most often, and which threats are enabled by the weaknesses that security checks can detect.

\begin{table}
	\caption{Frequency of SecLan Model Elements Observed in Security DSLs and Checks of Security Analyzers}
	\label{tab:statistics}
	\center
	\setlength\tabcolsep{2pt}
	\scriptsize
	\begin{tabular}{l|ccccccccc|ccccccc}
		\toprule
		& \multicolumn{9}{c}{\textbf{System Model}} 	& \multicolumn{6}{c}{\textbf{Security Model}} \\

		&
		\vspace{-2pt}\rotatebox{90}{Activity} &
		\rotatebox{90}{Component} &
		\rotatebox{90}{Connection} &
		\rotatebox{90}{Control Flow} &
		\rotatebox{90}{Data} &
		\rotatebox{90}{Entity} &
		\rotatebox{90}{\parbox{1.2cm}{Information Flow}} &
		\rotatebox{90}{Node} &
		\rotatebox{90}{State} &

		\rotatebox{90}{Spoofing} &
		\rotatebox{90}{\parbox{1.1cm}{Tampering\newline with Data\vspace{3pt}}} &
		\rotatebox{90}{Repudiation} &
		\rotatebox{90}{\parbox{1.6cm}{Information Disclosure\vspace{5pt}}} &
		\rotatebox{90}{\parbox{1.3cm}{\raggedright Denial of Service\vspace{5pt}}} &
		\rotatebox{90}{\parbox{1.6cm}{\raggedright Elevation of Privilege}} \\

		\midrule

		Security DSLs & 
        71.2\,\% & 
        36.4\,\% & 
        27.3\,\% & 
        9.1\,\%  & 
        71.2\,\% & 
        90.9\,\% & 
        28.8\% & 
        22.7\,\% & 
        42.4\,\%

		& 52\,\%
		& 67\,\%
		& 38\,\%
		& 77\,\%
		& 26\,\%
		& 28\,\% \\

		Security Analyzers & 
        80.6\,\% &  
        13.9\,\% & 
        19.4\,\%  &
        33.3\,\% &  
        38.9\,\% &  
        36.1\,\% &  
        52.8\,\% & 
        2.8\,\%  & 
        30.6\,\%

		& 80.6\,\%
		& 100.0\,\%
		& 58.3\,\%
		& 100.0\,\%
		& 88.9\,\%
		& 80.6\,\% \\

        Security Checks & 
        83.2\,\% &  
        1.7\,\% & 
        3.0\,\%  &
        17.9\,\% &  
        33.6\,\% &  
        10.7\,\% &  
        38.7\,\% & 
        0.2\,\%  & 
        8.0\,\%

		& 42.8\,\%
		& 81.4\,\%
		& 20.4\,\%
		& 81.0\,\%
		& 58.2\,\%
		& 41.9\,\% \\
        \midrule
        
        On Shortest Path & 
        41.9\,\% &  
        0.3\,\% & 
        0.5\,\%  &
        1.3\,\% &  
        17.2\,\% &  
        7.1\,\% &  
        8.4\,\% & 
        0.1\,\%  & 
        2.8\,\%

		& 12.1\,\%
		& 42.2\,\%
		& 4.0\,\%
		& 58.8\,\%
		& 10.3\,\%
		& 10.9\,\% \\ 
		\bottomrule
	\end{tabular}
\end{table}

\parhead{Covered System Model Elements.}
Security checks and DSLs target specific types of system elements to describe or analyze a system (see \cref{tab:statistics}).
The gap between security design and code is, among others, visible in the differences in element types they target.
Security checks primarily analyze concrete code statements, and therefore, the element type \textit{Activity} (83.2\,\%), and \textit{Information Flow} within or between \textit{Activities} (38.7\,\%).
\textit{Activity} is also one of the most targeted elements in security DSLs (71.2\,\%).
\textit{Still, in contrast to security checks, design-time security DSLs also focus on coarser-grained system elements}, primarily \textit{Entities}, i.e., classes and actors of a system.
Security DSLs exclusively cover the coarsest system elements, i.e., \textit{Nodes}, \textit{Connections}, and \textit{Components}.
In addition, security DSLs often target \textit{Data} (71.2\,\%), mainly to specify which data is critical.

\looseness=-1
Since checks typically analyze code written in specific programming languages, their characteristics may affect which conceptual language elements are checked.
One observation is that \textit{Information Flow} is primarily checked by Java and C/C++ analyzers, but rarely in Python.
Primarily, C/C++ analyzers focus on checking code elements related to \textit{Data}, mainly examining specifics such as low-level memory operations (e.g., array accesses or memory allocation).
Overall, the differences are minimal and unlikely to have a systematic impact on the gap between security design and code.

\begin{table}
	\caption{Weaknesses Detected the Most by the Checks}
	\label{tab:statistics:weaknesses}
	\center
	\smaller
	\setlength\tabcolsep{1pt}
	\begin{tabularx}{\columnwidth}{X c c c c}
		\toprule
		 {\textbf{CWE}} & {\textbf{All}} & {\textbf{C/C++}} & {\textbf{Java}} & {\textbf{Python}} \\
		\midrule
CWE-327: Use of a Broken or Risky Cryptographic Algorithm & 50 (31\%) & 21 (23\%) & 25 (21\%) & 6 (16\%) \\ 
CWE-20: Improper Input Validation & 41 (25\%) & 16 (17\%) & 25 (21\%) & 1 (3\%) \\ 
CWE-295: Improper Certificate Validation & 28 (17\%) & 4 (4\%) & 24 (20\%) & 2 (5\%) \\ 
CWE-78: Improper Neutralization of Special Elements used in an OS Command ('OS Command Injection') & 26 (16\%) & 6 (7\%) & 9 (8\%) & 14 (38\%) \\
CWE-798: Use of Hard-coded Credentials & 24 (15\%) & 0 (0\%) & 22 (19\%) & 1 (3\%) \\ 
CWE-326: Inadequate Encryption Strength & 21 (13\%) & 1 (1\%) & 17 (14\%) & 3 (8\%) \\ 
CWE-200: Exposure of Sensitive Information to an Unauthorized Actor & 21 (13\%) & 6 (7\%) & 14 (12\%) & 1 (3\%) \\ 
CWE-120: Buffer Copy without Checking Size of Input ('Classic Buffer Overflow') & 21 (13\%) & 21 (23\%) & 0 (0\%) & 0 (0\%) \\
CWE-89: Improper Neutralization of Special Elements used in an SQL Command ('SQL Injection') & 20 (12\%) & 4 (4\%) & 9 (8\%) & 7 (19\%) \\
CWE-330: Use of Insufficiently Random Values & 20 (12\%) & 2 (2\%) & 19 (16\%) & 0 (0\%) \\
\bottomrule
	\end{tabularx}
\end{table}

\parhead{Covered Security Model Elements.}
While security aspects of DSLs directly map to threats, security checks map to weaknesses they can detect.
We assigned 164 different CWEs to \checks security checks (on average 1.7 CWEs per check). 
The most assigned weaknesses are shown in \cref{tab:statistics:weaknesses}.
\edit{
These weaknesses align with our check categories and include insecure API usage, especially with regard to cryptography, as well as issues with permissions and unsanitized information flows.
However, we observed differences across programming languages. For instance, C/C++ analyzers target language-specific issues, such as insecure coding practices like CWE-120 (classic buffer overflow), the top-ranked issue for C/C++ but absent from languages like Java and Python.
Python has a strong focus on improper neutralization of external information (CWE-78 and CWE-89), which are less prevalent in Java and C/C++. However, the more general CWE-20 (improper input validation) ranks first in Java.
In Java and C/C++, broken or risky cryptographic algorithms (CWE-327) share first place with CWE-20 and CWE-120, respectively.
}

\edit{
To further explore how these weaknesses are addressed by security analyzers, \cref{tab:check-categories} categorizes the distribution of security checks across the five analysis categories introduced in \cref{sec:background:implementationcecks}. 
The API category (\textit{Security API Usage}) is most prevalent with 19 analyzers, highlighting the focus on detecting insecure usage of security-critical APIs (e.g., cryptographic or network-related APIs). 
This aligns with tools like CogniCrypt\,\cite{Krueger2017} and the broader literature\,\cite{Zhang2023,Kulenovic2014}, which emphasize the importance of API misuse as a common source of vulnerabilities.
\textit{Vulnerable Coding Practices (VCP)} follows closely, with 14 analyzers targeting coding patterns prone to security risks, such as buffer overflows or null pointer dereferences. 
Consequently, the API and VCP categories underscore a focus on local code-level issues, such as individual insecure API calls or coding practices.
In contrast, checks with a system-wide scope, such as \textit{Information Flow (IF)} checks or checks related to \textit{Permissions (P)}, are less prevalent, with only 8 and 9 analyzers providing such checks.
\textit{Dependency (D)} checks are the least represented, with just 3 analyzers.
However, this category includes widely used tools like GitHub Dependabot, suggesting that the sample may underrepresent dependency-focused checks due to their integration into broader platforms rather than standalone analyzers.
}

\begin{figure}
\begin{minipage}[t]{.4\textwidth}  
    \captionof{table}{Check categories covered by the inspected security analyzers}
    \label{tab:check-categories}
    \centering
    \small
    \begin{tabular}{cc}
        \toprule
        \textbf{Category} & \textbf{Number of analyzers} \\
        \midrule
        P   & 9 \\
        VCP & 14 \\
        D   & 3 \\
        API & 19 \\
        IF  & 8 \\
        \bottomrule
    \end{tabular}
\end{minipage}\qquad
\begin{minipage}[t]{.40\textwidth}
    \captionof{table}{Security aspect categories covered by the inspected security DSLs}
    \label{tab:dsl-categories}
    \centering
    \vspace{-9pt}
    \small
    \begin{tabular}{cc}
        \toprule
        \textbf{Category} & \textbf{Number of DSLs} \\
        \midrule
        ACA   & 37 \\
        DPA & 20 \\
        ADSSA   & 2 \\
        CA & 6 \\
        CGA & 11 \\
        IFIA  & 8 \\
        POSA & 4 \\
        SCSA  & 13 \\
        SH  & 2 \\
        TMID & 8 \\
        \bottomrule
    \end{tabular}
\end{minipage}
\end{figure}

\edit{
When examining the security aspects defined in the 66 security DSLs, we identified recurring themes that reflect both fundamental security objectives (such as confidentiality, integrity, and availability) and specialized mechanisms like access control, cryptographic operations, and information flow management. 
Our observations yielded the following categories, which provide a structured lens for understanding, comparing, and designing security DSLs.
\Cref{tab:dsl-categories} gives an overview of the frequencies of the observed categories, while \cref{tab:dsls} shows which categories each DSL addresses.

\textit{ACA - Access Control Aspects:}
56\,\% of all security DSLs, including SecureUML and Visual RBAC, focus on regulating access to specific resources under defined conditions. Access control aspects include, among others, role-based permissions, authorization rules, and accountability.

\textit{DPA - Data Protection Aspects:}
These aspects introduce information that safeguards data against unauthorized disclosure or tampering. These aspects align with the CIA triad and privacy regulations, such as the GDPR. 30\,\% of all DSLs, such as SysML-Sec, prioritize this category by providing respective annotations.

\textit{SCSA - System and Communication Security Aspects:}
20\,\% of the DSLs provide security aspects that address secure communication channels, network protections, and dynamic threat responses to mitigate risks such as eavesdropping or denial-of-service attacks.

\textit{CGA - Compliance and Governance Aspects:}
These aspects are used to align system designs with organizational and legal requirements. 
17\,\% of the DSLs, e.g., Secure Tropos, provide aspects belogning to this category to formalize security policies, constraints, and regulatory compliance.

\textit{IFIA - Information Flow and Isolation Aspects:}
12\,\% of the DSLs, such as the Security Analysis Language (SAL) and CARDS, provide aspects to ensure that data flows only through authorized channels to prevent unauthorized disclosure.

\textit{TMID - Threat Modeling and Intrusion Detection:}
The aspects in this category focus on proactive threat analysis and dynamic response mechanisms. 
Security aspects in this category are found in 12\,\% of DSLs. One example is the Aspect-Oriented Attack Scenario Model (AoASM).

\textit{CA - Cryptographic Aspects:}
 8\,\% of the DSLs provide security aspects that are used to describe encryption algorithms and validate them against known vulnerabilities. Examples include DiaSpec and SAL, which validate encryption algorithms against known vulnerabilities.

\textit{POSA - Physical and Operational Security Aspects:}
This category captures security aspects that introduce concepts to handle physical access control and operational resilience.
Although they are less represented, 6\,\% of the DSLs introduce aspects of this category, such as trust zones and emergency modes.

\textit{ADSSA - Application-Domain-Specific Security Aspects:}
3\,\% of the DSLs provide aspects to address specific needs of a specialized application domain.
Examples include aspects that describe the needs of IoT device authentication or healthcare privacy.

\textit{SH - Security Hardening:}
This category contains security aspects that provide frameworks with information to reduce attack surfaces and mitigate vulnerabilities through secure configurations.
This category is less commonly represented, with only 3\,\% of the DSLs providing related security aspects.

The investigation of these categories reveals well-covered areas, such as access control and data protection, as well as underrepresented domains, such as physical security and security hardening.
}

\looseness=-1
To examine the gap between what security checks can detect and what security DSLs attempt to counteract, \cref{tab:statistics} shows the threats that are enabled by the weaknesses detected by the analyzed security checks and those that DSLs counteract.
Both security DSLs and checks primarily focus on \textit{Information Disclosure} (77\% and 79\%) and \textit{Tampering with Data} (67\% and 76\%).
\textit{Denial of Service} is the third most relevant threat for security checks (52\% of all checks) but the least considered in security DSLs (only by 26\%).
Since most weaknesses can threaten a system's availability, e.g., due to exceptions, extensive coverage of \textit{Denial of Service} by checks is expected.
In contrast to the significant differences in CWEs addressed by checks in the context of different programming languages (see \cref{tab:statistics:weaknesses}), we observed less significant differences in programming languages for the analyzed system elements.
\textit{Spoofing} and \textit{Elevation of Privileges} are considered slightly more often in Java than in other languages, while \textit{Repudiation} is addressed above average in Python.

\subsubsection{Relationship between Security DSLs and Checks.}
Using the security DSL and analyzer descriptions, we calculated all relationships across the gap between the \checks security checks and the 112 security aspects of the DSLs, \edit{resulting in 56,631 relationships that relate 554 of the specified 559 (99.1\,\%) security checks to at least one security aspect.}
Our analysis reveals that security aspects and checks are generally broadly related.
When allowing unlimited navigation along the edges \textit{\securityRelation} and \textit{parent of} (see \cref{fig:models,fig:example,fig:SystemElementsConceputalModel}), on average \edit{506 (90\,\% of all checks)} security checks are related per aspect.
Still, there are aspects to which no check is related, e.g., privacy of SecBPMN, which is rather coarse-grained.
The highest number of related checks is 554 of the \checks checks, which is the case for \edit{23 security aspects that either consider general security requirements (15 security aspects), such as in \textit{SecML}, general access control security policies (5 security aspects) that could be bypassed, and 3 more general security aspects of \textit{UWE} on Secure Connections, Crossite Scripting, and SQL Injections.}

\begin{figure}
    \centering
    \includegraphics[width=0.7\linewidth]{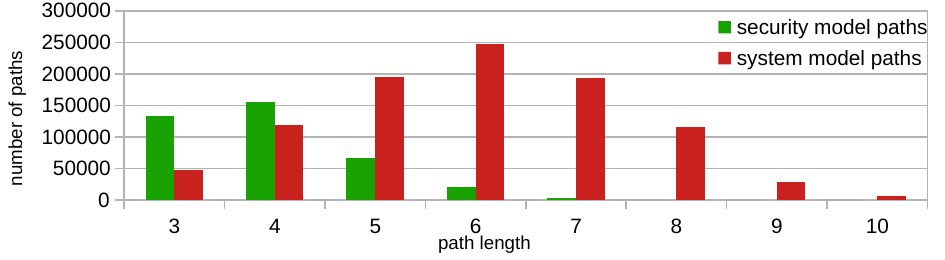}
    \caption{Number of paths between security aspects and checks with a given path length.}
    \label{fig:path-lengths}
\end{figure}

For a security aspect and check to be related, the paths between them, one over the system model and one over the security model, together must be at least 6 edges long.
Note that, as visible in \cref{fig:paths,fig:path-lengths}, the minimum length of a path over the system model is 3 edges over each model.
\edit{We identified two dimensions that allow us to assess the degree of relationship between a security aspect and a check based on the paths.}

\edit{
\begin{description}
    \item[\textbf{Path Length:}] First, the number of elements between a security aspect and a check could be considered a relevant metric, indicating whether a security check finding can immediately invalidate a design-time security assumption or whether it needs propagation through the system.
Shorter paths between a security aspect and a check suggest a more direct and immediate impact of a finding on the security aspect. For example, a check that invalidates a design-time security assumption with minimal intermediate steps is more critical than one requiring longer propagation through the system. 
Conversely, longer paths, while still relevant, are harder to trace manually due to the complexity of the context involved.
SecLan provides a way to quantitatively explore the gap by automatically generating such hard-to-notice paths.

    \item[Path Quantity:] Second, a higher number of paths between a security aspect and a check increases the likelihood that at least one concrete, exploitable path exists. 
    Although multiple paths might appear redundant, they actually expand the potential attack vectors, indicating a stronger and more critical relationship between the security aspect and the check.
    As a result, a greater number of paths not only broadens the attack surface but also heightens the probability that at least one path remains vulnerable in practice.
\end{description}
}

\edit{
    \Cref{fig:path-lengths} gives an overview of the length of all paths detected.
}
On average, we observed 16.9 paths between each pair of related security aspect and check over the system model, and 6.7 paths over the security model.
The paths over the system model have an average length of 6 edges, while the shortest path has an average of 3.5 edges.
Over the security model, there are on average 4 edges per path and 3.1 edges on the shortest paths.
\edit{
    While there seem to be primarily shorter paths over the security model, i.e., a security aspect being connected to a check over one weakness or two (path lengths of 3 or 4), most detected paths over the system model seem to be longer.
}

\begin{figure}
\begin{subfigure}[t]{\columnwidth}
	\center
	\includegraphics[width=.75\columnwidth]{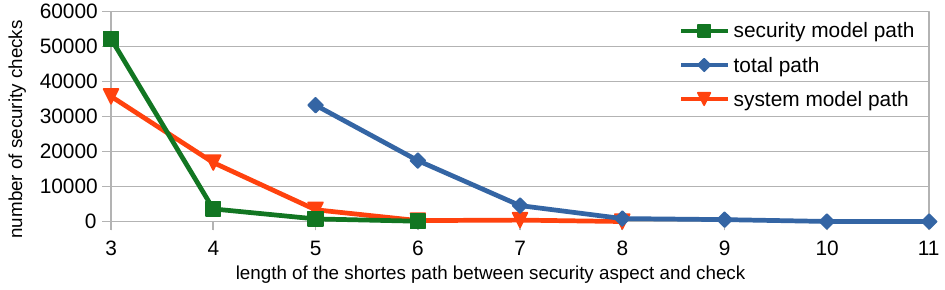}
	\caption{Number of security checks related to a security aspect at a given length of the shortest paths between them}
	\label{fig:statistics:checks-path}
\end{subfigure}

\begin{subfigure}[t]{\columnwidth}
	\center
	\includegraphics[width=.75\columnwidth]{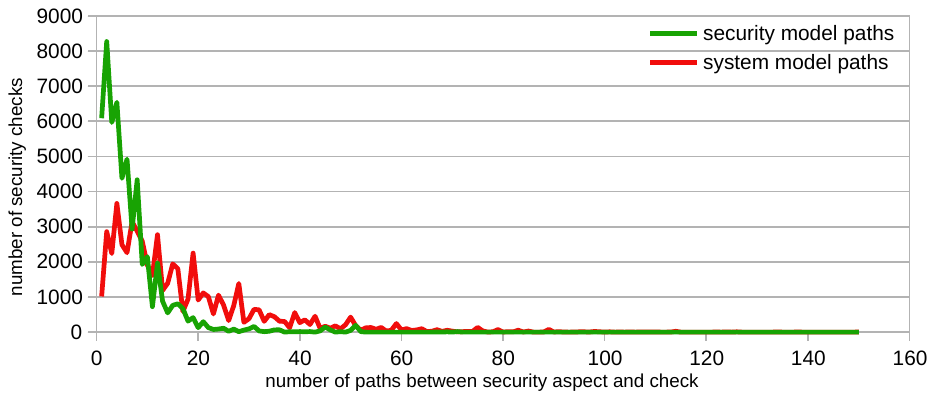}
	\caption{Number of security checks related to a security aspect at a given number of paths between them}
	\label{fig:statistics:checks-path-width}
\end{subfigure}
	\caption{Lengths of paths vs number of paths}

\end{figure}

To get a better understanding of how directly security checks and aspects are related, we investigate how many relationships are derived for different path lengths in \cref{fig:statistics:checks-path}.
The majority of the relationships (58.6\,\%) stem from the shortest possible paths for the security and system model (a total path length of 6 edges).
However, the number of minimum-length paths for both the system model paths and the security model paths is greater than the total number of paths where both are minimal.
Another 37.9\,\% of the relationships result from paths where one of the two paths has an extra edge.
While longer paths over the security model are rare, the paths over the system model can get quite long, even reaching a length where each system model element type is visited once.
However, only 3.5\,\% of the relationships stem from paths with more than seven edges.

\edit{
To further investigate the path quantity as a metric for the relationship between security aspects and checks, we counted the distinct paths connecting them across the system and security models, as shown in \cref{fig:statistics:checks-path-width}. 
On average, the relationship between a security aspect and a check is supported by \edit{16.9 paths over the system model and 6.7} paths over the security model.
More than 95\,\% of all relationships have paths over the security model with less than 18 paths.
In contrast to this, only 64.2\,\% of the relationships have less than 18 paths over the system model.
To cover 95\,\% of relationships, up to 49 paths must be considered for the system model.

While most relationships involve a rather small number of paths, the maximum number of paths between a security aspect and a check reaches \edit{150 paths over the system model (for 1 relationship) and 71 over the security model (for 15 relationships).} 
These cases suggest that certain security aspects are highly interconnected with checks, potentially offering multiple avenues for exploitation.

Despite these extremes, the generally moderate number of paths in most cases suggests a manageable but meaningful attack surface, where the likelihood of an exploitable path remains significant.
This highlights the importance of considering path quantity to understand not only the direct relationships but also the broader context in which vulnerabilities may emerge.

A higher quantity of paths between a security aspect and a check increases the probability that at least one exploitable path exists, even if others are not applicable in the concrete case. 
This highlights the necessity of comprehensive path exploration in security assessments, as a narrow focus on individual paths may overlook critical vulnerabilities. 
Not only the shortest paths are relevant to reason about a relationship but all paths that provide additional information.
Tools like SecLan can help systematically explore these relationships, ensuring that both possible direct and possible indirect connections are considered when assessing a detected issue.}

\begin{figure}
    \begin{subfigure}[t]{\columnwidth}
	   \centering
      \includegraphics[width=.6\columnwidth]{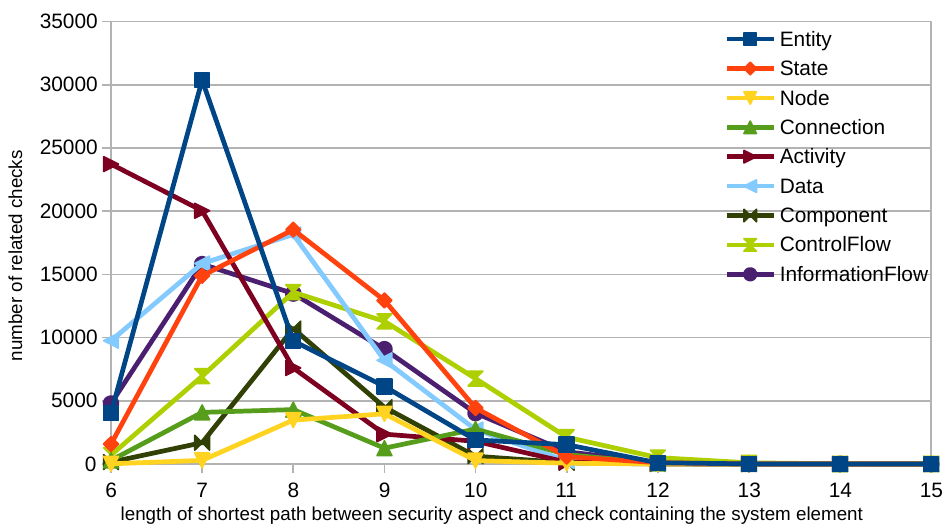}
	   \caption{Shortest path at which a system element appears}
	   \label{fig:paths-per-systemelement}
    \end{subfigure}
    
    \begin{subfigure}[t]{\columnwidth}
	   \centering
      \includegraphics[width=.6\columnwidth]{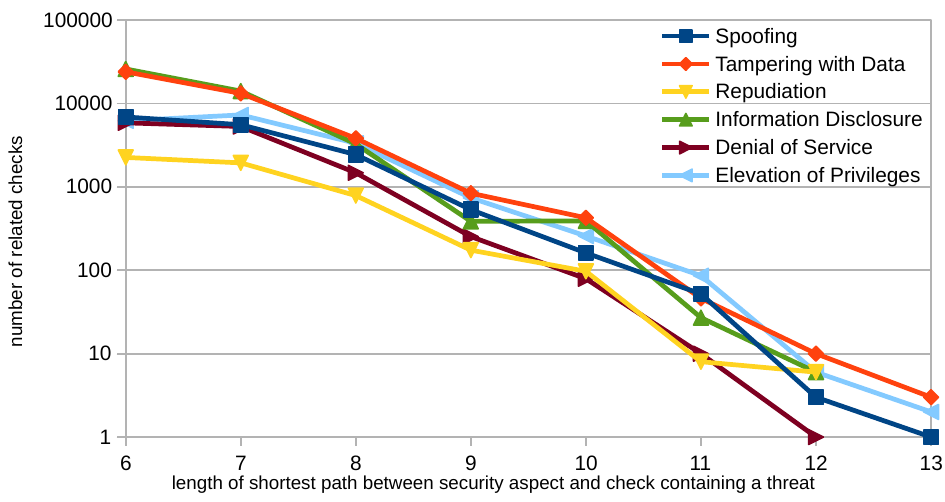}
	   \caption{Shortest path at which a threat appears}
	   \label{fig:paths-per-threat}
    \end{subfigure}
    \caption{The length of the shortest path at which an element from the SecLan model appears on the path between a security aspect and a security check}
\end{figure}

\edit{
In our general analysis of the paths expressing relationships between security aspects and security checks, we have seen differences between the path lenghts and quantities depending on the system model and seurity model.
To further investigate the impact of the security model and what are the most relevant system model elements with respect to the relationships, we counted how often each system model element appears on the shortest path between related security aspects and security checks.
As visible in \cref{tab:description-stats}, following the prevalence in DSL and check descriptions, \textit{Activities} are by far the most represented on the shortest paths between security aspects and checks.
While \textit{Data} and \textit{Entity} follow according to the high presence in descriptions, even though \textit{Entities} are only covered by few checks, this is the opposite for other system elements.
Particularly, \textit{Infomration Flow} is on fewer relationships than the percentage of DSLs and checks addressing \textit{Information Flow}.

To better understand these variations, \Cref{fig:paths-per-systemelement} shows at which path length which of the system model element appears how often.
Particularly, \textit{Activities} appear primarily on the shortes possible paths, unlike all other system element types. 
This reflects the high number of security aspects and checks immediately addressing \textit{Activities} and therefore being related directly.
All other element types have their peak at first or second additional element on a path, i.e., lenghts 4 and 5.
This is also true for \textit{Information Flow}, which is why it does not appear often on the shortest paths, as there are shorter paths with other elements, e.g., an immediate mapping based on \textit{Activity}.
Considering \textit{Information Flow} as an implicit element in the communications of \textit{Activies}, this seems to be expectable.
}

While our system model defines a clear chain of security relations between the system model elements, the relationships in the security model are less clear.
Threats provide further security-related information about the relationships.
Therefore, we investigated the relationships regarding the threats they involve, counting for each threat the shortest paths that contain this threat as shown in \cref{fig:paths-per-threat}.
Furthermore, this allows us to study in more detail how prominent the different threats are in the relationships.
\textit{Information Disclosure} and \textit{Tampering with Data} are the most frequent threats on the relationships which follows their dominance in the security DSLs and what threats the weaknesses detected by checks enable (cf. \cref{tab:statistics}).
Also, these two threats are the only ones for which nearly 60\,\% or more of their relationships are present for the minimal path length of 6 edges, while this is below 50\,\% for all other threats.
\textit{Repudiation} is more present on longer paths relative to its total number of occurrences than any other threat, \edit{particularly not significantly dropping a path length of 12 edges}, while for all other threats the number of relationships decreases steadily with path length.
While security checks rarely explicitly address repudiation, side-effects of weaknesses they detect may lead to repudiation issues.
\edit{In contrast, \textit{Denial of Service} is mainly present on shorter paths, i.e., there is only one path with a length of 12.
While most threats decrease steadily with the path length, \textit{Information Disclosure} has a local maximum at a total path length 10 edges.}
This is however negligible compared to the number of threats appearing for the first time on shorter paths.
Overall, there are only minor but no significant differences between the individual threats.

\subsection{Qualitative Exploration.}
%
\edit{
We examined the relationships that SecLan derived for security DSLs and checks through interactive interviews with seven security DSL experts and two security analyzer experts.
We recruited all of the experts through the survey in which they validated SecLan's application to their DSL or analyzer.
}
\edit{For the security DSLs, the} interviewees were experts in Sec\-BPMN\,\cite{Salnitri2015}\,\interview{1}, Sec\-DFD\,\cite{Tuma2019a}\,\interview{2}, the Architectural Dataflow DSL\,\cite{Seifermann2022}\,\interview{3}, UMLsec\,\cite{Juerjens2005}\,\interview{4,5}, the Attack Propagation DSL\,\cite{Walter2022}\,\interview{6}, and PbSD\,\cite{Abramov2012}\,\interview{7}.
\edit{For the analyzers, the interviewees were experts in CodeQL\,\cite{CodeQL}\,\interview{8} from GitHub and the plugin Thunderhorse for ESLint\,\cite{Rafnsson2020}\,\interview{9}}.
All of the experts interviewed held a Ph.D. or were Ph.D. students.
\edit{Three} interviewees went to industry after gaining a Ph.D., while the other \edit{six} interviewees are working in academia (four faculty positions, one postdoctoral researcher, and one research associate).
While all experts together examined hundreds of relationships, we identified only \edit{seven} false positives \interview{1,3,8}.
Six false positives occurred because SonarQube checks point to relatively general weaknesses that threaten more security objectives than what the vulnerability the check detects actually enables.
This was mentioned by \interview{3}, stating that \textit{CWE 20 is too broad. SonarQube rules should not relate to broad weaknesses}. 
\edit{For one false positive, the expert in \interview{8} stated that the logging of sensitive information should only affect the confidentiality of the system.
However, SecLan also reported a relationship with the security aspect \textit{No Upflow} of \textit{UMLsec}, which addresses system integrity.
The expert concluded that SecLan is not expressive enough in this regard because this issue arises at the instance level rather than at the type level, which is SecLan's current focus.} 
In seven of the nine interviews, the interviewed expert initially flagged some checks as false positives, but upon following the reasoning provided by SecLan, possibly looking at the CWEs and check descriptions (to which SecLan provides pointers that the interviewees leveraged) \interview{2,3,4}, they acknowledged that the security check was or could be related to the security aspect.

\edit{
Since the security analyzer experts had no experience with security DSLs, we investigated UMLsec, which addresses multiple security concerns, such as access control, secure communication, and information flow. This enabled us to discuss the calculated relationships more precisely, particularly with regard to false positives and false negatives.
}

\edit{
In two cases, the experts identified a false negative while exploring the relationships for their analyzer \interview{8,9}.
For example, I\textsubscript{8} mentioned that the security check detects vulnerabilities related to \textit{time of creation, time of use} (CWE-367) that should result threaten access control, but no relation to UMLsec's role-based access control (rbac) was reported by SecLan. 
Upon investigation, we found that the discrepancy stemmed from an outdated local CWE version (4.12), which did not yet associate CWE-362 (the parent of CWE-367) with confidentiality or access control, unlike the current version (4.19.1).
Similarly, I\textsubscript{9} stated that their check for detecting \textit{unknown object injection} should affect almost everything, while SecLan again did not report a relationship with rbac.
}

\edit{
For all other relationships, the experts confirmed their alignment with the security aspects modeled by the DSLs. 
These findings highlight how the accuracy of SecLan’s relationship detection depends on the completeness and currency of the underlying security model data, rather than limitations in the technique itself.
}


Due to the broad scope of security DSLs, there are many relationships between security checks and DSLs.
I\textsubscript{3} mentioned that this may be the case because many checks and corresponding CWEs are aimed at a broad objective, such as potential \textit{Information Disclosure}.
I\textsubscript{6} stated that \textit{all means that surround the [access control] engine \dots is a threat to access control} and thus anything detected by the checks that could lead to bypassing an access control engine running PbSD is related, which, in the end, is \textit{true for [almost] every check}.
\edit{The experts of \interview{8,9} also support this by stating that \textit{the CWEs are definitely too broad}.
In particular, the expert in \interview{9} stated that too many CWEs are created and that they are being broadened further if a reported vulnerability cannot be classified.
Furthermore, the expert notes that the high-level objectives in the CWE, such as confidentiality, integrity, and availability, are too coarse-grained for practical usage.
The expert argued that the impact of a weakness should be classified in more detail, for example, with information \textit{what an attacker can do with the affected part of the program}.}

Four out of nine experts \interview{1,2,3,5} stated that the relationships, although being correct, are quite numerous and that it would be beneficial to reduce the number of checks---issue prioritization is a well-known challenge, but is out of the scope of this work.
In \interview{1}, the expert \textit{was wondering whether some security relations can be removed}.
This expert also stated that research into prioritization could be beneficial, e.g., by \textit{considering the level of propagation, e.g., level zero (no step)\dots}.     
In \interview{2}, the expert stated that \textit{the number of warnings [relations] should be reduced as everything is connected right now}. 
The expert in \interview{5} stated that \textit{the number of checks recommended for secure links [security aspect in UMLsec] is a lot} and that \textit{less recommended checks would be ok}. 

The experts in \interview{1,3} acknowledged that most of the relationships are clear, e.g., the expert of \interview{3} stated that \textit{most of the things [checks] recommended are pretty clear}. \interview{1,2,3,4,6} showed that the reasoning paths provided by SecLan helped the experts to understand the relationship between the security DSL and the security check.
Three experts explicitly stated that the relationship became particularly clear when they more closely examined the CWEs and security check descriptions referenced by SecLan \interview{2,3,4}.
In \interview{2}, the expert was not sure \textit{how it [SonarQube rule RSPEC-5679] is related to SecDFD}. Then the expert stated \textit{first looking at the description, it was hard to understand, but looking at the explanation [the relation], it's clear} after showing the expert the relation and description of \textit{CWE287} and \textit{CWE347}. 
In \interview{3}, the expert stated that the relation is not always immediately clear, \textit{but after looking it [the relation and descriptions of the weaknesses] up, it gets clear why it is related}. 
In \interview{4}, the expert inspected \textit{SonarQube rule RSPEC-1989} and stated that \textit{after looking into the reasoning and check description, the check is related}.


\textit{In summary, SecLan successfully identifies relationships and helps the security experts to systematically explore them, thereby revealing relationships between security aspects of DSLs and security checks that would otherwise be neglected.}

\smallskip




\section{Threats to Validity}
\parhead{External validity.}
The generalizability of the derived \model could be limited by the reviewed security DSLs and security analyzers. 
Missed security analyzers or checks could have led to missing classes in the SecLan model, inappropriate definitions, or commonalities.
To avoid this, we primarily based our selection of DSLs and analyzers on surveys that summarize the state of the art of the research area.
In summary, we derived security DSLs from four surveys and 30 analyzers from three surveys.

One aspect to consider is that security analyzers or DSLs may not be representative.
To mitigate this threat, we extended our review for underrepresented categories of checks, e.g., single-purpose analyzers targeting information flow, by searching for additional analyzers in the literature.
Similarly, we added missing DSLs based on the authors' knowledge.
In addition, we asked in the questionnaire about security DSLs and analyzers known to the participants, adding two additional DSLs to the security DSLs considered.
This limits the threats to generalizability.

Another aspect is that the programming languages supported by analyzers could have an influence, making our findings not generally applicable.
However, our analysis includes numerous checks for each of the most popular programming languages according to the TIOBE index\,\cite{tiobe}, but we did not observe significant effects on the gap studied.

The perspective of the security experts who volunteered to participate in the survey and interview may not be representative, as they are primarily DSL experts.
However, although both security DSLs and security checks are widely used in practice\,\cite{Clavel2008,Juerjens2008,Shostack2014,Vassallo2018}, commercial tools such as SonarQube have made static security analyzers more widely available and they are now part of many CI pipelines.
For this reason, we expect that security experts in general will be more familiar with static security analyzers than with security DSLs.
Therefore, it is essential that participants have expertise in security DSLs, which is the case in our study.
Furthermore, the experts cover different modeling languages, levels of abstraction, and security aspects, which mitigates threats on the generalizability and applicability of the concepts extracted in our study.
\edit{
We observed no noticeable differences when comparing the responses of security DSL experts and security analyzer experts.
This is another indication of a low impact on the validity of our study.
Additionally, the relatively small number of participants could threaten the generalizability of our study.
However, the small differences in participants' survey and interview responses indicate a limited threat.
}

The validity of the explorative interviews could be threatened by factors, such as subject expectancy bias and selection bias.
Especially, subject expectancy bias could have affected the interviews, as the SecLan was proposed as a solution to bridge the gap between security DSLs and checks.
This could have lead to the insights only being applicable to the concrete context of the interviews.
We mitigated these threats by inquiring about the applicability of the concepts and including multiple interviewers.

\parhead{Internal validity.}
The analysis of security DSLs and analyzers could be affected by selection bias due to possible author preferences for particular DSLs or analyzers.
To mitigate this threat, as described above, we based the selection on surveys and asked external experts for further relevant security DSLs and analyzers.

Another threat to validity is observer bias of the authors of this study.
To mitigate this threat, for the analysis of security DSLs and analyzers, we conducted the analysis in an iterative process and in a team of four researchers, involving discussions until full agreement was achieved.
This agreement has additionally been validated by the authors of the security DSLs and analyzers that have been studied.
Of the 304 authors, we were able to contact 132 authors of which 22 validated the SecLan model and 18 confirmed the classification of their security analyzer or DSL. 

Since one interviewer conducted all interviews for the qualitative exploration, potential effects of differences in interviews should be limited.
However, this increases the likelihood of observer bias, which we mitigated by always including at least one other author in the interviews and discussing the interview with all authors.
Therefore, observer bias should have been limited.

%% file: sections/04_1_recommendations.tex
\subsection{Tool Support for the Exploration of Relationships}
\label{sec:tool}
\begin{figure}
	\includegraphics[width=\columnwidth]{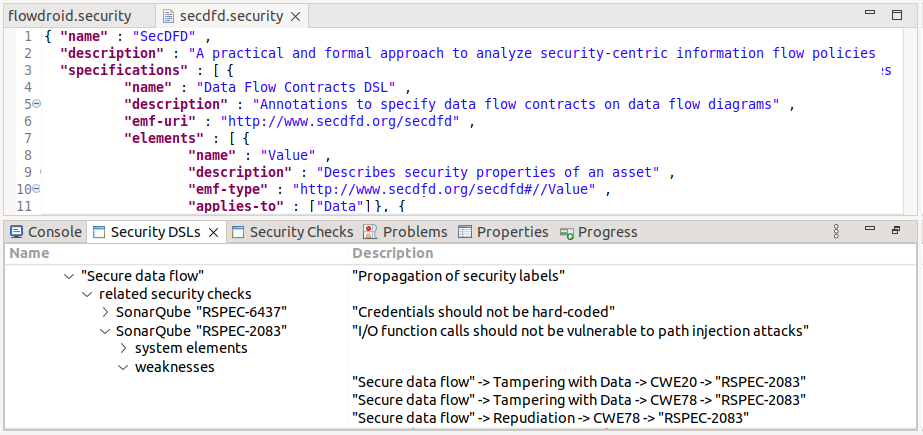}
	\vspace{-16pt}
	\caption{Screenshot showing an excerpt from the description of SecDFD and an explorable view of identified relationships}
	\label{fig:tool}
	\vspace{-3pt}
\end{figure}

\noindent\looseness=-1
For the exploratory analysis of relationships in our study (steps 7 and 8 in \cref{fig:methodology}), but also for reuse by researchers and practitioners, we created tooling for describing security DSLs and analyzers, computing the relationships between them, and providing an interactive UI for exploring them (lower half of \cref{fig:tool}).
We used this tooling to specify the descriptions of security DSLs and analyzers used in the validation of SecLan, and to form the basis for our initial repository of security DSL and security check descriptions.
The entire tooling as well as the descriptions are contained in our supplementary materials\,\cite{Replication}.

The SecLan tooling provides a JSON-compliant textual syntax for describing security DSLs and analyzers according to the \model in \cref{fig:models}.
For this purpose, we use the language engineering toolkit Xtext\,\cite{xtext}, which allows us to provide editing support over an IDE-independent language server\,\cite{Barros2022}.
The description of SecDFD is shown in the upper half of \cref{fig:tool}.
To provide a human-readable version, we can perform an automated transformation of this presentation into JavaDoc-like HTML pages, which serve as a human-readable interface to the SecLan repository of security DSLs and security analyzers.
The generated website is accessible online\footnote{SecLan Repository: \url{https://gravity-tool.org/seclan-repository/}} and archived in our replication package\,\cite{Replication}.
To lower the effort of applying the SecLan tooling in further studies focusing on other DSLs, we allow semi-automatic importing of EMF metamodels\,\cite{Steinberg2008} and UML profiles\,\cite{uml}, which are the most commonly used languages for specifying security DSLs.
The tooling also supports linking of the SecLan model elements to elements of concrete security DSLs, in case they are based on EMF and/or UML Profiles.

To facilitate replication and further studies that may need refinements of the models derived in this work, we provide a pre-defined security model and system model, both of which are extensible, to relate descriptions defined in this language to them.
The security model is based on STRIDE as captured in \cref{fig:example}.
We provide weaknesses extracted from the MITRE CWE\,\cite{cwe} database and link them to the threats of STRIDE\,\cite{Shostack2008}.
The security objectives are confidentiality, integrity, availability, and authenticity.
The common system elements, e.g., \textit{Activity} or \textit{Entity}, and the security model can be referenced from the descriptions of security DSLs and analyzers.

Using the descriptions of security analyzers and DSLs, the tooling allows the calculation of the relationships between security DSLs and checks automatically and presents them in an interactive view as shown at the bottom of \cref{fig:tool}. 
For each relationship, all discovered paths can be displayed if desired.

%% file: sections/05_1_discussion.tex
\section{Discussion}
\label{sec:discussion}
\noindent
Based on the identified relationships between design-time security DSLs and code-level security checks, as well as their exploration with experts, we identified five key findings that are relevant to researchers and practitioners for bridging the gap between security design and code.

\begin{observation}
	\textbf{There are only few commonalities between design-time and implementation-level security.}
	\textit{Only abstractions of the parts of a system to which security considerations apply, and the high-level threats and corresponding weaknesses, are commonalities between design-time security DSLs and code-level analyzers. In addition, many aspects that apply only to subsets of security must be considered to relate design-time and implementation-level security.}
\end{observation}

\noindent
Our first finding highlights the abstraction gap between design-time security DSLs and code-level analyzers. 
Relating analyzers to security DSLs requires investigating potential transitive relationships of system elements that are not directly captured in the DSLs. 
In doing so, few commonalities between design-time security, represented in this study by security DSLs, and implementation-level security, represented by security analyzers, can be exploited.
Instead, it is necessary to consider different subsets of security in detail, as well as potential interactions between these subsets, to effectively establish the relationships between security DSLs and implementation-level checks. 
While this finding is not surprising, our study provides empirical evidence for this gap and the empirically validated \model that allows for further exploration of this relationship. 

\recommendation{
    In order to assess the impact of a vulnerability on the security design, it is important to be aware of the many potential relationships and to consider details of each security aspect to narrow down what needs to be considered.
    However, such considerations should not be limited to details in isolation, e.g. only cryptography, but must also consider the interactions between different security aspects.
    These interactions are not yet well understood and should be the subject of further research.
}

\begin{observation}
\textbf{Affected security objectives depend on the location context of a weakness in the system.}
\textit{Without information about the local context in which a code-level weaknesses was detected, i.e., ways to restrict what other types of system elements are reachable, it has the potential to compromise most security aspects of a security design.}
\end{observation}

\noindent\looseness=-1
Weakness descriptions in the CWE\,\cite{cwe} typically list several conditions under which security objectives may be compromised.
Therefore, a single vulnerability has the potential to compromise most of the fundamental security objectives, and it must be determined which ones actually affect the security aspects of the security DSLs specifying the security design.
Developers must carefully consider the context, that is, what part of the system an identified vulnerability applies to and how it is used, to determine its impact.
This observation led to the request for prioritization in our interviews, for which the \model and derived paths provide a basis for developing prioritization techniques, e.g. considering discussed metrics such as number of paths and their length.
SecLan helps developers to bridge the gap by providing an overview of the relationships to be examined and prevents effects from going unnoticed.
It also provides a framework to systematically investigate the impact of context. 

\recommendation{
In a review, security experts should systematically evaluate all possible relationships and use the context of the concrete instances to eliminate inappropriate relationships.
Further research is needed on how to automatically infer such contextual information to properly assess a vulnerability.
}

\begin{observation}
\textbf{Local weaknesses can propagate to~the entire system scope.}
\textit{
	Weaknesses not only threaten security objectives applicable to the type of element on which they are detected, but their effects can propagate to arbitrary types of elements throughout the system.
}
\end{observation}
\noindent\looseness=-1
Detected weaknesses are mostly not capable of directly enabling a threat against a security objective, but there is usually a longer attack path through the system starting at the affected system element.
Looking at the paths over the system model extracted in this paper, we found on average 16.9 different paths between system elements analyzed by a security check and a potentially affected security objective.
These paths span an average of 4 system elements, and even if only the shortest path is considered for each pair of check and security aspect, it is an average of 1.4 element types.
While these numbers are high already, it is important to keep in mind that the effect of the weakness could propagate over an arbitrary number of instances of these identified system elements, which results in a wide gap to consider.
SecLan provides a first step toward a better understanding of how the effects of vulnerabilities can propagate throughout a system.

\recommendation{
It is essential also to consider non-obvious relationships within the system when assessing the potential impact of a vulnerability.
This study particularly shows how far such relationships can propagate, therefore calling for further research on proper criteria from when to stop following the propagation of security issues through the system.
}

\begin{observation}
\textbf{Even security experts are overwhelmed.}
\textit{Even security experts are not aware of all relationships between what security analyzers detect and the security objectives specified in a security design.}
\end{observation}
\noindent
While it is well known that most developers are not security experts, although sometimes assumed otherwise\,\cite{Ryan2023}, even security experts can easily be overwhelmed by the enormous complexity involved in bridging the gap, i.e., relating weaknesses to a security design.
In the validation of this work, security experts interviewed consistently showed that there are cases where they did not expect a relationship between a weakness that can be detected by a check and aspects of the security design.
This means that in situations where we need to prioritize security tasks---which is the norm in practice\,\cite{Mazurek2022}---even security experts need support in making informed decisions that take into account the impact on the overall security design.
As visible in our user study, when both, security DSLs and analyzers, are described according to their commonalities, i.e., the SecLan conceptual model, analyzer findings can be automatically related to security aspects of DSLs, enabling a systematic and easy assessment of analyzer findings.
However, this work can only be the beginning of the journey toward secure software systems in practice.
More research on effectively presenting relationships to developers is necessary. 

\recommendation{
To address the large gap between design and implementation level security in assessing the potential impact of vulnerabilities, as this study shows, developers should use impact assessment tools, such as available issue prioritization tools. 
However, since available tools often do not take into account contextual information about the security design, we need to facilitate future research, i.e. based on SecLan.
}

\begin{observation}
\textbf{Security checks are described by too broad weaknesses.}
\textit{Particularly, security checks of commercial analyzers point to unnecessarily broad weaknesses, making it difficult to assess their impact on the security~design.}
\end{observation}
\noindent
While most academic security checks do not explicitly refer to weaknesses, commercial analyzers often do so but mostly point to broad weaknesses.
For example, SonarQube rule \textit{S2631 (Regular expressions should not be vulnerable to Denial of Service attacks)} that was often identified as part of false positives in our interviews, mainly refers to broad weaknesses such as \textit{CWE-20 (Improper Input Validation)} that due to their general nature can threaten all three security objectives of the CIA.
However, when inspecting the SonarQube rule in detail, one notices that the real issue lies in the backtracking that regular expression engines use and therefore in how the expression is specified, e.g., an expression whose complexity is exponential to the input size.
The detection rule checks two things, first, whether a backtracking engine is used, and second, if inputs (external but also internal ones) to engines are sanitized.
Weaknesses assigned to security checks being too broad increases the effort required to understand findings. Our quantitative exploration quantifies the effects of this and explicitly shows the consequences.
Providing pointers to less general weaknesses for each of these two checks would not only improve the quality of the relationships established in this work, and thus facilitate bridging the gap between design-time and code-level security, but would also improve the understandability of detected issues.
In the CWE, such more specific weaknesses are called base weaknesses, and even the CWE explicitly recommends that specific vulnerabilities, e.g., detected by security analyzers, should only be mapped to such base weaknesses\,\cite{cwe}.
Therefore, we call to provide more fine-grained weaknesses in a common format and database (e.g., CWE but not on the granularity of base weaknesses).

\recommendation{
Not only to support analyses such as this study, but also to help developers in their day-to-day tasks, such as assessing the impact of analyzer results, a low-effort but highly beneficial change would be to describe analyzers by more precise weaknesses.
While CWE already requires that real-world vulnerabilities, i.e., those found by analyzers, be mapped to specific weaknesses rather than abstract categories of weaknesses, this study highlights the consequences of not doing so.
}

%% file: sections/06_related.tex
\section{Related Work}
\label{sec:rw}
\noindent
Most closely related to SecLan are surveys of model-driven development processes.
Mashkoor et al.~\cite{Mashkoor2023}, from whose work we extracted security DSLs for our investigation, discuss model-driven security engineering in detail.
Similarly, Mohammed et al.~\cite{Mohammed2017} and Kahn et al.~\cite{Khan2021} discuss how security approaches are addressed throughout the software development lifecycle.
Baker et al.\,\cite{Baker2012} propose to embed design-level security concerns in the implementation explicitly.
However, they change the way code has to be written and embed specific measures in aspects.
Establishing relationships between security DSLs and code remains a manual task, usually embedded in a systematic process but based on expert knowledge that we study and conceptualize in SecLan.

\looseness=-1
\parhead{System Representations.}
Artifacts such as code and design models describe systems using different abstractions.
In particular, the systematic representation of systems has been studied extensively in the area of system architecture.
Most design-time modeling languages allow for the representation of a domain-oriented system.
Particularly, the Unified Modeling Language (UML)\,\cite{uml} supports multiple abstractions that can be combined.
In general, the ISO/IEC/IEEE 42010\,\cite{ISO42010} defines requirements for the description of software systems.
However, due to the wide variety, it is a huge manual effort to relate design languages, and even worse to relate them to the implementation.
Often, systematic representations differ in the type of system aspect to be represented, e.g., behavior or structure\,\cite{uml}.
Similar to our work, it is often necessary to extract domain-specific aspects, e.g., the granularity at which variability mechanisms are realized\,\cite{Berger2014}.

\parhead{Security Knowledge.}
\noindent
Several works capture security concerns in ontologies, typically dividing them into high-level security properties, mid-level domain knowledge, and low-level system knowledge.
Herzog et al.\,\cite{Herzog2007} capture 88 threat, 79 asset, and 133 countermeasure classes, and their relationships.
The terms presented vary in granularity, ranging from high-level threats such as spoofing to specific low-level threats such as IP splicing.
SecWAO\,\cite{Busch2015} models similar aspects for web applications. 
These ontologies serve as a reference for our work, but are tailored to specific use cases.


\parhead{Analyzer Exchange Formats.}
There are several formats for a systematic and interchangeable representation of analyzer results, such as the ASSUME Static Code Analysis Tool Exchange Format\,\cite{Kern2019} or the OASIS Static Analysis Results Interchange Format\,\cite{SARIF}.
The ASSUME Static Code Analysis tool exchange format\,\cite{Kern2019} aims at facilitating the comparability of static code analysis results.
Similarly, the OASIS Static Analysis Results Interchange Format (SARIF)\,\cite{SARIF} is intended to allow the integration of static analyzer findings into different environments, e.g., managing the findings of multiple analyzers in a single IDE view.
Among others, the static analyzer CogniCrypt, which we considered in this work, has been adapted to OASIS-SARIF\,\cite{Kummita2019}.
Compared to SecLan, such formats focus only on analyzers and not on relating their results to security DSLs, so they should be extended by our concepts.

\parhead{Weaknesses and Vulnerabilities.}
Known weaknesses and vulnerabilities are collected in various databases, whereby the common weakness enumeration (CWE) and common vulnerability enumeration (CVE) are the most prominent ones.
Although the CWE partially maps CWE entries to design aspects or concepts, SecLan is an addition to the CWE and does not compete with it. 
We combine information from the CWE with other sources, i.e., STRIDE, in the security model.
In combination with information about the system, i.e., the system model, this allows for automatically relating security checks and DSLs.
While the CWE contains references to potential mitigations of weaknesses at views such as code or architecture, it remains a manual task to assess their concrete impact.

%% file: sections/07_conclusion.tex
\section{Conclusion}
\noindent
To better understand the gap between design-time and implementation security, we investigated the relationship between security DSLs and code-level security checks by examining \dsls security DSLs and \checks code-level checks from \analyzers static analyzers concerning their common concepts.
We extracted the \model that relates security DSLs and checks to their common security and system concepts (C1) and used it to classify all DSLs and analyzers (C2).
Based on this, we compute relationships between security DSLs and code-level checks, thereby crossing the gap between them, and provide reasoning about their relationship (C3).
We validated the \model in a questionnaire and explored the relationships quantitatively (C4) and qualitatively in interviews with seven security experts (C5).
We learned that even security experts do not have a complete overview of the relationship between security DSLs and security checks.
One reason is security checks being described by broad CWEs that relate them to too many threats and security objectives.
However, our automatically derived relationships based on the \model helped the experts to understand the relationships thus to bridge the gap.
We therefore call for more research on possible tool support to help developers bridge the gap between security design and implementation.
\edit{To support these efforts, we provide a full replication package\,\cite{Replication}.}
As a short-term, low-effort improvement with high impact, tool developers should relate their security checks to more fine-grained CWEs.
In addition, we found that the effect of weaknesses can propagate throughout the system scope and provide first numbers to help understand this propagation.
Overall, our results show that the gap between design-time security and code-level security is due to an exhaustive number of relationships, many of which are often indirect and transitive, systematically captured for the first time in this work.

%% file: bibliography.bib
@Article{Xiao2008,
  author  = {Liang Xiao},
  journal = {Information and Software Technology (IST)},
  title   = {{An Adaptive Security Model using Agent-Oriented MDA}},
  year    = {2008},
}

@Article{Zhu2008,
  author  = {Zhi Jian Zhu and Mohammad Zulkernine},
  journal = {Information and Software Technology (IST)},
  title   = {{A model-based aspect-oriented framework for building intrusion-aware software systems}},
  year    = {2008},
}

@InProceedings{Gunawan2009,
author = {Linda Ariani Gunawan and Peter Herrmann and Frank Alexander Kraemer},
title = {{Towards the Integration of Security Aspects into System Development Using Collaboration-Oriented Models}},
booktitle = {International Conference on Security Technology},
year = {2009},
}

@InProceedings{Eby2007,
author = {Matthew Eby and Jan Werner and Gabor Karsai and Akos Ledeczi},
title = {{Integrating Security Modeling into Embedded System Design}},
booktitle = {International Conference and Workshops on the Engineering of Computer-Based Systems (ECBS)},
year = {2007},
}

@InProceedings{Moebius2009,
author = {Nina Moebius and Kurt Stenzel and Holger Grandy and Wolfgang Reif},
title = {{SecureMDD: A Model-Driven Development Method for Secure Smart Card Applications}},
booktitle = {International Conference on Availability, Reliability and Security (ARES)},
year = {2009},
}

@InProceedings{Baudry2008,
  author    = {Benoit Baudry and Franck Fleurey and Tejeddine Mouelhi and Yves Le Traon},
  booktitle = {International Conference on Model Driven Engineering Languages and Systems (MODELS)},
  title     = {{A Model-Based Framework for Security Policy Specification, Deployment and Testing}},
  year      = {2008},
}

@InProceedings{Burt2003,
author = {Carol C. Burt and Barrett R. Bryant and Rajeev R. Raje and Andrew Olson and Mikhail Auguston},
title = {{Model Driven Security: Unification of Authorization Models for Fine-Grain Access Control}},
booktitle = {International Enterprise Distributed Object Computing},
year = {2003},
}

@InBook{Busch2014,
  author    = {Marianne Busch and Nora Koch and Santiago Suppan},
  title     = {{Modeling Security Features of Web Applications}},
  year      = {2014},
  booktitle = {Engineering Secure Future Internet Services and Systems},
}

@Article{Schefer-Wenzel2014,
  author  = {Sigrid Schefer-Wenzel and Mark Strembeck},
  journal = {Information and Software Technology (IST)},
  title   = {{Model-driven specification and enforcement of RBAC break-glass policies for process-aware information systems}},
  year    = {2014},
}

@Article{Alam2007,
  author  = {Masoom Alam and Michael Hafner and Ruth Breu},
  journal = {Journal of Software},
  title   = {{Model-Driven Security Engineering for Trust Management in SECTET}},
  year    = {2007},
}

@inproceedings{b4msecure,
  author    = {Akram Idani and Yves Ledru},
  title     = {B for Modeling Secure Information Systems -- The B4MSecure Platform},
  booktitle = {International Conference on Formal Engineering Methods (ICFEM)},
  pages     = {312--318},
  year      = {2015},
  doi       = {10.1007/978-3-319-25423-4_20}
}

@inproceedings{fstarpopl,
  author    = {Jean-Karim Zinzindohoué and Karthikeyan Bhargavan and Jonathan Protzenko and Benjamin Beurdouche},
  title     = {HACL*: A Verified Modern Cryptographic Library
},
  booktitle = {Conference on Computer and Communications Security (CCS)},
  pages     = {1789--1806},
  year      = {2017},
  doi       = {10.1145/3133956.3134043}
}

@book{framac_book,
  title = {Guide to Software Verification with Frama-C: Core Components,  Usages,  and Applications},
  editor = {Nikolai Kosmatov and Virgile Prevosto and Julien Signoles},
  ISBN = {9783031556081},
  ISSN = {2731-5762},
  url = {http://dx.doi.org/10.1007/978-3-031-55608-1},
  DOI = {10.1007/978-3-031-55608-1},
  journal = {Computer Science Foundations and Applied Logic},
  publisher = {Springer International Publishing},
  year = {2024}
}

@inproceedings{jasmin,
  author    = {Almeida, Jos{\'e} Bacelar and Manuel Barbosa and Gilles Barthe and Arthur Blot and Benjamin Grégoire and Vincent Laporte and Tiago Oliveira and Hugo Pacheco and Benedikt Schmidt and Pierre-Yves Strub},
  title     = {Jasmin: High-Assurance and High-Speed Cryptography},
  booktitle = {Conference on Computer and Communications Security (CCS)},
  pages     = {1807--1823},
  year      = {2017},
  doi       = {10.1145/3133956.3134078}
}

@inproceedings{cryptoverif,
  author    = {Benjamin Lipp and Bruno Blanchet and Karthikeyan Bhargavan},
  title     = {A Mechanised Cryptographic Proof of the WireGuard Virtual Private Network Protocol},
  booktitle = {European Symposium on Security and Privacy (EuroS\&P)},
  pages     = {231--246},
  year      = {2019},
  doi       = {10.1109/EuroSP.2019.00026}
}

@InProceedings{Morin2010,
author = {Brice Morin and Tejeddine Mouelhi and Franck Fleurey and Yves Le Traon and Olivier Barais and Jean-Marc Jézéquel},
title = {{Security-driven model-based dynamic adaptation}},
booktitle = {International Conference on Automated Software Engineering (ASE)},
year = {2010},
}

@Misc{dependabot,
    title = {{Dependabot}},
    author = {GitHub},
    howpublisched = {\url{https://github.com/dependabot}},
    year = 2025,
}

@Article{Hoisl2012,
  author  = {Bernhard Hoisl and Stefan Sobernig and Mark Strembeck},
  journal = {Software \& Systems Modeling},
  title   = {{Modeling and enforcing secure object flows in process-driven SOAs: an integrated model-driven approach}},
  year    = {2012},
}

@incollection{vallee2010soot,
  title={{Soot: A Java Bytecode Optimization Framework}},
  author={Vall{\'e}e-Rai, Raja and Co, Phong and Gagnon, Etienne and Hendren, Laurie and Lam, Patrick and Sundaresan, Vijay},
  booktitle={CASCON First Decade High Impact Papers},
  pages={214--224},
  year={2010}
}

@Article{Horcas2016,
  author  = {Jose-Miguel Horcas and Mónica Pinto and Lidia Fuentes and Wissam Mallouli and Edgardo Montes de Oca},
  journal = {Computers \& Security},
  title   = {{An approach for deploying and monitoring dynamic security policies}},
  year    = {2016},
}

@InProceedings{Yu2005,
author = {Huiqun Yu and Dongmei Liu and Xudong He and Li Yang and Shu Gao},
title = {{Secure Software Architectures Design by Aspect Orientation}},
booktitle = {International Conference on Engineering of Complex Computer Systems (ICECCS)},
year = {2005},
}

@InProceedings{Almorsy2014,
author = {Mohamed Almorsy and John Grundy},
title = {{SecDSVL: A Domain-Specific Visual Language to Support Enterprise Security Modelling}},
booktitle = {Australian Software Engineering Conference},
year = {2014},
}

@InProceedings{Muslukhov2018,
author = {Ildar Muslukhov and Yazan Boshmaf and Konstantin Beznosov},
title = {{Source Attribution of Cryptographic API Misuse in Android Applications}},
booktitle = {Asia Conference on Computer and Communications Security},
year = {2018},
}

@InProceedings{Menzel2010,
author = {Michael Menzel and Robert Warschofsky and Christoph Meinel},
title = {{A Pattern-Driven Generation of Security Policies for Service-Oriented Architectures}},
booktitle = {International Conference on Web Services},
year = {2010},
}

@InProceedings{Fahl2012,
author = {Sascha Fahl and Marian Harbach and Thomas Muders and Lars Baumgärtner and Bernd Freisleben and Matthew Smith},
title = {{Why eve and mallory love android: an analysis of android SSL (in)security}},
booktitle = {Conference on Computer and Communications Security (CCS)},
year = {2012},
}

@InProceedings{Ren2005,
  author    = {Jie Ren and Richard N. Taylor},
  booktitle = {Workshop on Software Security Assurance Tools, Techniques, and Metrics},
  title     = {{A Secure Software Architecture Description Language}},
  year      = {2005},
}

@Article{Kim2011,
  author  = {Sangsig Kim and Dae-Kyoo Kim and Lunjin Lu and Suntae Kim and Sooyong Park},
  title   = {{A feature-based approach for modeling role-based access control systems}},
  year    = {2011},
  volume  = {84},
  issue   = {12},
  jounral = {Journal of Systems and Software},
}

@Article{Mouratidis2007,
  author  = {Haralambos Mouratidis and Paolo Giorgini},
  journal = {International Journal of Software Engineering and Knowledge Engineering},
  title   = {{Secure Tropos: A Security-Oriented Extension of the Tropos Methodology}},
  year    = {2007},
  number  = {2},
  pages   = {285--309},
  volume  = {17},
}

@InProceedings{Paletov2018,
  author    = {Rumen Paletov and Petar Tsankov and Veselin Raychev and Martin Vechev},
  booktitle = {Conference on Programming Language Design and Implementation (PLDI)},
  title     = {{Inferring Crypto API Rrules from Code Changes}},
  year      = {2018},
}

@Article{Reznik2007,
  author  = {Julia Reznik and Tom Ritter and Rudolf Schreiner and Ulrich Lang},
  journal = {Electronic Notes in Theoretical Computer Science},
  title   = {{Model Driven Development of Security Aspects}},
  year    = {2007},
  number  = {2},
  pages   = {65--79},
  volume  = {162},
}

@InProceedings{Kaddani2014,
author = {Aziz Kaddani and Amine Baina and Loubna Echabbi},
title = {{Towards a Model Driven Security for critical infrastructures using OrBAC}},
booktitle = {International Conference on Multimedia Computing and Systems (ICMCS)},
year = {2014},
}

@Article{Giordano2010,
  author  = {Massimiliano Giordano and Giuseppe Polese and Giuseppe Scanniello and Genoveffa Tortora},
  journal = {Journal of Visual Languages \& Computing},
  title   = {{A system for visual role-based policy modelling}},
  year    = {2010},
}

@InProceedings{Kim2006,
author = {Dae-Kyoo Kim and Priya Gokhale},
title = {{A Pattern-Based Technique for Developing UML Models of Access Control Systems}},
booktitle = {Computer Software and Applications Conference (COMPSAC)},
year = {2006},
}

@InProceedings{Jakob2009,
author = {Henner Jakob and Nicolas Loriant and Charles Consel},
title = {{An aspect-oriented approach to securing distributed systems}},
booktitle = {International conference on Pervasive services (ICPS)},
year = {2009},
}

@Article{Ahn2002,
  author  = {Gail-Joon Ahn and Seung-Phil Hong and Michael E Shin},
  journal = {Information and Software Technology (IST)},
  title   = {{Reconstructing a formal security model}},
  year    = {2002},
}

@Misc{Pyre,
  title = {{Pyre/Pysa Website}},
    author = {{Meta Platforms}},
    year = {2025},
    URL = {https://pyre-check.org/docs/pysa-basics/},
    note         = {last accessed February 10\textsuperscript{th}, 2025},
}

@Article{Mouheb2009,
  author  = {Djedjiga Mouheb and Chamseddine Talhi and Azzam Mourad and Vitor Lima},
  journal = {Frontiers in Artificial Intelligence and Applications},
  title   = {{An Aspect-Oriented Approach for Software Security Hardening: from Design to Implementation}},
  year    = {2009},
  volume  = {199},
}

@Article{Nguyen2014,
  author  = {Phu H. Nguyen and Gregory Nain and Phu H.Nguyen and Tejeddine Mouelhi and Yves Le Traon},
  journal = {Transactions on Aspect-Oriented Software Development},
  title   = {{Modularity and Dynamic Adaptation of Flexibly Secure Systems: Model-Driven Adaptive Delegation in Access Control Management}},
  year    = {2014},
}

@Article{MoralGarcia2014,
  author  = {Santiago Moral-García and Santiago Moral-Rubio and Eduardo B. Fernández and Eduardo Fernández-Medina},
  journal = {Computer Standards \& Interfaces},
  title   = {{Enterprise security pattern: A model-driven architecture instance}},
  year    = {2014},
  volume  = {36},
  issue   = {4},
}

@article{Gomaa2006,
author = {Hassan Gomaa and Michael Eonsuk Shin},
title = {{Software Requirements and Architecture Modeling for Evolving Non-secure Applications into Secure Applications}},
journal = {Science of Computer Programming},
year = {2006},
Volume = 66, 
Issue = 1,
pages = {60--70},
}

@InProceedings{Shin2004,
author = {Michael Eonsuk Shin and Hassan Gomaa},
title = {{Modelling Complex Systems by Separating Application and Security Concerns}},
booktitle = {International Conference on Engineering of Complex Computer Systems},
year = {2004},
}

@InProceedings{Ma2016,
author = {Siqi Ma and David Lo and Teng Li and Robert H. Deng},
title = {{CDRep: Automatic Repair of Cryptographic Misuses in Android Applications}},
booktitle = {Asia Conference on Computer and Communications Security},
year = {2016},
}

@InProceedings{Fink2006,
author = {Torsten Fink and Manuel Koch and Karl Pauls},
title = {{An MDA approach to Access Control Specifications Using MOF and UML Profiles}},
booktitle = {Electronic Notes in Theoretical Computer Science},
year = {2006},
}

@Article{Horcas2015,
  author  = {Jose-Miguel Horcas and Mónica Pinto and Lidia Fuentes},
  journal = {Journal of Systems and Software},
  title   = {{An Automatic Process for Weaving Functional Quality Attributes Using a Software Product Line Approach}},
  year    = {2015},
  volume  = {112},
}

@InProceedings{Kumar2020,
author = {Rajesh Kumar},
title = {{A Model-Based Safety-Security Risk Analysis Framework for Interconnected Critical Infrastructures}},
booktitle = {International Conference on Critical Infrastructure Protection},
year = {2020},
}

@InProceedings{Gerking2019,
author = {Christopher Gerking and David Schubert},
title = {{Component-Based Refinement and Verification of Information-Flow Security Policies for Cyber-Physical Microservice Architectures}},
booktitle = {International Conference on Software Architecture (ICSA)},
year = {2019},
}

@Article{Fernandez-Medina2007,
  author  = {Eduardo Fernandez-Medina and Juan Trujillo and Rodolfo Villarroel and Mario Piattini},
  journal = {Information Systems},
  title   = {{Developing secure data warehouses with a UML extension}},
  year    = {2007},
  pages   = {826--856},
  volume  = {32},
  issue   = {6},
}

@InProceedings{Brunel2014,
  author    = {Julien Brunel and David Chemouil and Laurent Rioux and Mohamed Bakkali and Frederique Vallee},
  booktitle = {Workshop on Model-Driven Engineering, Verification and Validation},
  title     = {{A Viewpoint-Based Approach for Formal Safety \& Security Assessment of System Architectures}},
  year      = {2014},
}

@InProceedings{Gerking2018,
author = {Christopher Gerking and David Schubert and Eric Bodden},
title = {{Model Checking the Information Flow Security of Real-Time Systems}},
booktitle = { Engineering Secure Software and Systems (ESSoS)},
year = {2018},
}

@InProceedings{Nakamura2005,
author = {Yuichi Nakamura and Michiaki Tatsubori and T. Imamura and Kohichi Ono},
title = {{Model-driven security based on a Web services security architecture}},
booktitle = {International Conference on Services Computing (SCC)},
year = {2005},
}

@InProceedings{Neisse2013,
  author    = {Ricardo Neisse and Joerg Doer},
  booktitle = {Annual Conference on Privacy, Security and Trust},
  title     = {{Model-based specification and refinement of usage control policies}},
  year      = {2013},
}

@InProceedings{Rahaman2019,
author = {Sazzadur Rahaman and Ya Xiao and Sharmin Afrose and Fahad Shaon and Ke Tian and Miles Frantz and Murat Kantarcioglu and Danfeng (Daphne) Yao},
title = {{CryptoGuard: High Precision Detection of Cryptographic Vulnerabilities in Massive-sized Java Projects}},
booktitle = {Conference on Computer and Communications Security (CCS)},
year = {2019},
}

@Misc{CppCheck,
    title = {{CppCheck Feature Overview}},
    author = {Daniel Marjamäki and others},
    URL = {https://cppcheck.sourceforge.io/},
    year = {2025},
    note         = {last accessed February 10\textsuperscript{th}, 2025},
}

@article{ATTCK,
title = {{ATT\&CK-based Advanced Persistent Threat Attacks Risk Propagation Assessment Model for Zero Trust Networks}},
journal = {Computer Networks},
volume = {245},
year = {2024},
doi = {10.1016/j.comnet.2024.110376},
author = {Jingci Zhang and Jun Zheng and Zheng Zhang and Tian Chen and Yu-an Tan and Quanxin Zhang and Yuanzhang Li},
}

@InProceedings{Gunawan2011,
author = {Linda Ariani Gunawan and Frank Alexander Kraemer and Peter Herrmann},
title = {{A Tool-Supported Method for the Design and Implementation of Secure Distributed Applications}},
booktitle = {International Symposium on Engineering Secure Software and Systems (ESSoS)},
year = {2011},
}

@Article{Sanchez2009,
  author  = {Pablo Sanchez and Ana Moreira and Lidia Fuentes and Joao Araujo and Jose Magno},
  journal = {Information and Software Technology (IST)},
  title   = {{Model-driven Development for Early Aspects}},
  year    = {2009},
}

@Article{Dai2006,
  author  = {Lirong Dai and Kendra Cooper},
  journal = {Science of Computer Programming},
  title   = {{Modeling and performance analysis for security aspects}},
  year    = {2006},
  number  = {1},
  pages   = {58--71},
  volume  = {61},
}

@InProceedings{Wada2006,
  author    = {Hiroshi Wada and Junichi Suzuki and Katsuya Oba},
  booktitle = {International Conference on Autonomic and Autonomous Systems (ICAS)},
  title     = {{A Model-Driven Development Framework for Non-Functional Aspects in Service Oriented Grids}},
  year      = {2006},
}

@InProceedings{Singleton2020,
author = {Larry Singleton and Rui Zhao and Myoungkyu Song and Harvey Siy},
title = {{CryptoTutor: Teaching Secure Coding Practices through Misuse Pattern Detection}},
booktitle = {Annual Conference on Information Technology Education},
year = {2020},
}

@InProceedings{Diaz2008,
author = {Paloma Diaz and Ignacio Aedo and Daniel Sanz and Alessio Malizia},
title = {{A model-driven approach for the visual specification of Role-Based Access Control policies in web systems}},
booktitle = {Symposium on Visual Languages and Human-Centric Computing},
year = {2008},
}

@Article{Koch2002,
  author  = {Manuel Koch and Luigi V. Mancini and Francesco Parisi-Presicce},
  journal = {Transactions on Information and System Security},
  title   = {{A graph-based formalism for RBAC}},
  year    = {2002},
  number  = {3},
  volume  = {5},
}

@MastersThesis{Newbury2020,
  author = {Kristen L. Newbury},
  school = {University of Alberta},
  title  = {{Automated Hotfixes for Misuses of Crypto APIs}},
  year   = {2020},
}

@Article{Georg2010,
  author  = {Geri Georg and Kyriakos Anastasakis and Behzad Bordbar and Siv Hilde Houmb and Indrakshi Ray and Manachai Toahchoodee},
  journal = {Transactions on Software Engineering (TSE)},
  title   = {{Verification and Trade-Off Analysis of Security Properties in UML System Models}},
  year    = {2010},
}

@InProceedings{Heldal2003,
  author    = {Rogardt Heldal and Fredrik Hultin},
  booktitle = {European Symposium on Research in Computer Security (ESORICS)},
  title     = {{Bridging Model-Based and Language-Based Security}},
  year      = {2003},
}

@InProceedings{Egele2013,
  author    = {Manuel Egele and David Brumley and Yanick Fratantonio and Christopher Kruegel},
  booktitle = {Conference on Computer \& Communications Security (CCS)},
  title     = {{An empirical study of cryptographic misuse in android applications}},
  year      = {2013},
}

@Misc{JOANA,
    title = {{JOANA (Java Object-sensitive ANAlysis) Project Page}},
    author = {Simon Bischof},
    URL = {https://pp.ipd.kit.edu/projects/joana/},
    year = {2022},
    note         = {last accessed February 10\textsuperscript{th}, 2025},
}

@InProceedings{Ahn2007,
author = {Gail-Joon Ahn and Hongxin Hu},
title = {{Towards realizing a formal RBAC model in real systems}},
booktitle = {Symposium on Access Control Models and Technologies},
year = {2007},
}

@InProceedings{Satoh2006,
author = {Fumiko Satoh and Yuichi Nakamura and Koichi Ono},
title = {{Adding Authentication to Model Driven Security}},
booktitle = {International Conference on Web Services (ICWS)},
year = {2006},
}

@InProceedings{Shuai2014,
author = {Shao Shuai and Dong Guowei and Guo Tao and Yang Tianchang and Shi Chenjie},
title = {{Modelling Analysis and Auto-detection of Cryptographic Misuse in Android Applications}},
booktitle = {International Conference on Dependable, Autonomic and Secure Computing},
year = {2014},
}

@Article{Wei2018,
  author  = {Fengguo Wei and Sankardas Roy and Xinming Ou and Robby},
  journal = {Transactions on Privacy and Security},
  title   = {{Amandroid: A Precise and General Inter-component Data Flow Analysis Framework for Security Vetting of Android Apps}},
  year    = {2018},
}

@InProceedings{Elrakaiby2014,
author = {Yehia Elrakaiby and Moussa Amrani and Yves Le Traon},
title = {{Security@Runtime: A Flexible MDE Approach to Enforce Fine-grained Security Policies}},
booktitle = {International Symposium on Engineering Secure Software and Systems (ESSoS)},
year = {2014},
}

@InProceedings{Seifermann2021,
author = {Stephan Seifermann and Robert Heinrich and Dominik Werle and Ralf Reussner},
title = {{A Unified Model to Detect Information Flow and Access Control Violations in Software Architectures}},
booktitle = {International Conference on Security and Cryptography (SECRYPT)},
year = {2021},
}

@Misc{CogniCrypt,
    title = {{CogniCrypt Website}},
    author = {{Eclipse Foundation}},
    URL = {https://eclipse.dev/cognicrypt/},
    year = {2024},
  note         = {last accessed February 10\textsuperscript{th}, 2025},
}

@InProceedings{Oladimeji2007,
  author    = {Ebenezer Oladimeji and Lawrence Chung and Sam Supakkul},
  booktitle = {International Conference on Software Engineering \& Knowledge Engineering},
  title     = {{A Model-driven Approach to Architecting Secure Software}},
  year      = {2007},
}

@Article{Wolter2008,
  author  = {Christian Wolter and Michael Menzel and Andreas Schaad and Philip Miseldine and Christoph Meinel},
  journal = {Journal of Systems Architecture},
  title   = {{Model-driven business process security requirement specification}},
  year    = {2008},
}

@Article{Pavlich-Mariscal2009,
  author  = {Jaime A. Pavlich-Mariscal and Steven A. Demurjian and Laurent D. Michel},
  journal = {Computers \& Security},
  title   = {{A Framework of Composable Access Control Features: Preserving Separation of Access Control Concerns from Models to Code}},
  year    = {2009},
  number  = {3},
  pages   = {350--379},
  volume  = {23},
}

@Article{Xu2006,
  author  = {Diangxiang Xu and Kendall Nygard},
  journal = {Transactions on Software Engineering (TSE)},
  title   = {{Threat-driven modeling and verification of secure software using aspect-oriented Petri nets}},
  year    = {2006},
  number  = {4},
  volume  = {32},
}

@Article{Ray2003,
  author  = {Idrakshi Ray and Robert France and Na Li and Geri Georg},
  journal = {Information and Software Technology (IST)},
  title   = {{An aspect-based approach to modeling access control concerns}},
  year    = {2003},
  number  = {9},
  volume  = {46},
}

@Article{Sánchez2009,
  author  = {Óscar Sánchez and Fernando Molina and Jesús García-Molina and Ambrosio Toval},
  journal = {Journal of Universal Computer Science (JUCS)},
  title   = {{ModelSec: A Generative Architecture for Model-Driven Security}},
  year    = {2009},
  number  = {15},
  pages   = {2957--2980},
  volume  = {15},
}

@InProceedings{Machiry2017,
author = {Aravind Machiry and Chad Spensky and Jake Corina and Nick Stephens and Christopher Kruegel and Giovanni Vigna},
title = {{DR. CHECKER: A Soundy Analysis for Linux Kernel Drivers}},
booktitle = {USENIX Security Symposium},
year = {2017},
}

@InProceedings{Xu2020,
author = {Zhiwu Xu and Xiongya Hu and Yida Tao and Shengchao Qin},
title = {{Analyzing Cryptographic API Usages for Android Applications Using HMM and N-Gram}},
booktitle = {International Symposium on Theoretical Aspects of Software Engineering (TASE)},
year = {2020},
}

@Article{Snelting2014,
  author  = {Gregor Snelting and Dennis Giffhorn and Jürgen Graf and Christian Hammer and Martin Hecker and Martin Mohr and Daniel Wasserrab},
  journal = {Information Technology},
  title   = {{Checking Probabilistic Noninterference Using JOANA}},
  year    = {2014},
  volume  = {56},
}

@Misc{ErrorProne,
    title = {{ErrorProne Website}},
    URL = {https://errorprone.info/},
    author = {Google},
    year = {2024},
    note         = {last accessed February 10\textsuperscript{th}, 2025},
}

@Article{Pinto2015,
  author  = {Mónica Pinto and Nadia Gámez and Mercedes Amor and José Miguel Horcas and Inmaculada Ayala},
  journal = {Sensors},
  title   = {{Dynamic Reconfiguration of Security Policies in Wireless Sensor Networks}},
  year    = {2015},
  volume  = {15},
}

@Misc{Snyk,
  title = {{Snyk Website}},
author = {{Snyk Limited}},
year = {2025},
URL = {https://snyk.io/},
  note         = {last accessed February 10\textsuperscript{th}, 2025},
}

@Article{Gajrani2020,
  author  = {Jyoti Gajrani and Meenakshi Tripathi and Vijay Laxmi and Gaurav Somani and Akka Zemmari and Manoj Singh Gaur},
  journal = {Digital Threats: Research and Practice},
  title   = {{Vulvet: Vetting of Vulnerabilities in Android Apps to Thwart Exploitation}},
  year    = {2020},
  number  = {2},
  volume  = {1},
}

@InProceedings{Menzel2009,
author = {Michael Menzel and Christoph Meinel},
title = {{A Security Meta-model for Service-Oriented Architectures}},
booktitle = {International Conference on Services Computing},
year = {2009},
}

@inproceedings{Nazzal2024,
author = {Nazzal, Mahmoud and Khalil, Issa and Khreishah, Abdallah and Phan, NhatHai},
title = {{PromSec: Prompt Optimization for Secure Generation of Functional Source Code with Large Language Models (LLMs)}},
year = {2024},
doi = {10.1145/3658644.3690298},
booktitle = {Conference on Computer and Communications Security (CCS)},
pages = {2266--2280},
}

@inproceedings{Li2024,
author = {Li, Dong and Yan, Meng and Zhang, Yaosheng and Liu, Zhongxin and Liu, Chao and Zhang, Xiaohong and Chen, Ting and Lo, David},
title = {{CoSec: On-the-Fly Security Hardening of Code LLMs via Supervised Co-decoding}},
year = {2024},
doi = {10.1145/3650212.3680371},
booktitle = {International Symposium on Software Testing and Analysis (ISSTA)},
pages = {1428--1439},
}

@INPROCEEDINGS{Pearce2022,
  author={Pearce, Hammond and Ahmad, Baleegh and Tan, Benjamin and Dolan-Gavitt, Brendan and Karri, Ramesh},
  booktitle={Symposium on Security and Privacy (SP)}, 
  title={{Asleep at the Keyboard? Assessing the Security of GitHub Copilot’s Code Contributions}}, 
  year={2022},
  pages={754--768},
  doi={10.1109/SP46214.2022.9833571}}

@inproceedings{Krueger2020,
author = {Kr\"{u}ger, Stefan and Ali, Karim and Bodden, Eric},
title = {{CogniCrypt\textsubscript{GEN}: Generating Code for the Secure Usage of Crypto APIs}},
year = {2020},
doi = {10.1145/3368826.3377905},
booktitle = {International Symposium on Code Generation and Optimization (CGO)},
pages = {185--198},
}

@Article{McGraw2004,
  author  = {McGraw, G.},
  journal = {IEEE Security \& Privacy (SP)},
  title   = {{Software Security}},
  year    = {2004},
  number  = {2},
  pages   = {80--83},
  volume  = {2},
  doi     = {10.1109/MSECP.2004.1281254},
}

@ARTICLE{Sai2024,
  author={Sai, Siva and Yashvardhan, Utkarsh and Chamola, Vinay and Sikdar, Biplab},
  journal={IEEE Access}, 
  title={{Generative AI for Cyber Security: Analyzing the Potential of ChatGPT, DALL-E, and Other Models for Enhancing the Security Space}}, 
  year={2024},
  volume={12},
  pages={53497--53516},
  doi={10.1109/ACCESS.2024.3385107}}

@InProceedings{Toeberg2022,
  author    = {Töberg, Jan-Philipp and Schiffl, Jonas and Reiche, Frederik and Beckert, Bernhard and Heinrich, Robert and Reussner, Ralf},
  booktitle = {International Conference on Decentralized Applications and Infrastructures (DAPPS)},
  title     = {{Modeling and Enforcing Access Control Policies for Smart Contracts}},
  year      = {2022},
  pages     = {38-47},
  doi       = {10.1109/DAPPS55202.2022.00013},
}

@InProceedings{Arogundade2021,
    author = {Oluwasefunmi Tale Arogundade and Olutimi Onilede and Sanjay Misra and Olusola Abayomi-Alli and Modupe Odusami and Jonathan Oluranti},
    title = {{From Modeling to Code Generation: An Enhanced and Integrated Approach}},
    year = {2021},
    booktitle = { Innovations in Information and Communication Technologies (IICT)},
    pages = {421--427},
    doi = {10.1007/978-3-030-66218-9_50},
}

@inproceedings{Modesti2014,
    author = {Paolo Modesti},
    title = {{Efficient Java Code Generation of Security Protocols Specified in AnB/AnBx}},
    booktitle = {Security and Trust Management},
    year = {2014},
    pages = {204--208},
doi = {10.1007/978-3-319-11851-2_17},
}

@InProceedings{Katkalov2013,
  author    = {Katkalov, Kuzman and Stenzel, Kurt and Borek, Marian and Reif, Wolfgang},
  booktitle = {International Conference on Social Computing},
  title     = {{Model-Driven Development of Information Flow-Secure Systems with Iflow}},
  year      = {2013},
  pages     = {51--56},
  doi       = {10.1109/SocialCom.2013.14},
}

@Article{Peldszus2021,
  author    = {Sven Peldszus and Jens B{\"{u}}rger and Timo Kehrer and Jan J{\"{u}}rjens},
  journal   = {Data \& Knowledge Engineering (DKE)},
  title     = {{Ontology-Driven Evolution of Software Security}},
  year      = {2021},
  volume    = {134},
  doi       = {10.1016/j.datak.2021.101907},
}

@Article{Tuma2022,
  author  = {Katja Tuma and Sven Peldszus and Daniel Str{\"{u}}ber and Riccardo Scandariato and Jan J{\"{u}}rjens},
  journal = {Software \& Systems Modeling (SoSyM)},
  title   = {{Checking Security Compliance between Models and Code}},
  year    = {2022},
}

@InProceedings{Peldszus2019,
  author    = {Sven Peldszus and Katja Tuma and Daniel Str{\"{u}}ber and Jan J{\"{u}}rjens and Riccardo Scandariato},
  booktitle = {International Conference on Model Driven Engineering Languages and Systems (MODELS)},
  title     = {{Secure Data-Flow Compliance Checks between Models and Code Based on Automated Mappings}},
  year      = {2019},
  pages     = {23--33},
  doi       = {10.1109/MODELS.2019.00-18},
}

@Article{Herzog2007,
  author  = {Almut Herzog and Nahid Shahmehri and Claudiu Duma},
  journal = {International Journal of Information Security and Privacy (IJISP)},
  title   = {{An Ontology of Information Security}},
  year    = {2007},
  number  = {4},
  pages   = {1--23},
  volume  = {1},
  doi     = {10.4018/jisp.2007100101},
}

@Article{Busch2015,
  author    = {Marianne Busch and Martin Wirsing},
  journal   = {International Journal of Software and Informatics},
  title     = {{An Ontology for Secure Web Applications}},
  year      = {2015},
  number    = {2},
  pages     = {233--258},
  volume    = {9},
}

@InProceedings{Tuma2019a,
  author    = {Katja Tuma and Riccardo Scandariato and Musard Balliu},
  booktitle = {International Conference on Software Architecture (ICSA)},
  title     = {{Flaws in Flows: Unveiling Design Flaws via Information Flow Analysis}},
  year      = {2019},
  pages     = {191--200},
  doi       = {10.1109/ICSA.2019.00028},
}

@InProceedings{Heckman2018,
  author    = {Sarah Heckman and Kathryn T. Stolee and Christopher Parnin},
  booktitle = {International Conference on Software Engineering: Software Engineering Education and Training (ICSE-SEET)},
  title     = {{10+ Years of Teaching Software Engineering with iTrust: the Good, the Bad, and the Ugly}},
  year      = {2018},
  pages     = {1--4},
  doi       = {10.1145/3183377.3183393},
}

@Misc{Meneely,
  author = {Meneely, Andrew and Smith, Ben and Williams, Laurie},
  title  = {{iTrust Electronic Health Care System Case Study}},
  url    = {https://github.com/ncsu-csc326/iTrust2},
    note         = {last accessed February 10\textsuperscript{th}, 2025},
}

@Misc{Kummita2019,
  author    = {Kummita, Sriteja and Piskachev, Goran},
  title     = {Integration of the Static Analysis Results Interchange Format in Cognicrypt},
  year      = {2019},
  doi       = {10.48550/ARXIV.1907.02558},
  publisher = {arXiv},
}

@InProceedings{Kern2019,
  author    = {Kern, Matthias and Erata, Ferhat and Iser, Markus and Sinz, Carsten and Loiret, Frederic and Otten, Stefan and Sax, Eric},
  booktitle = {Computer Software and Applications Conference (COMPSAC)},
  title     = {{Integrating Static Code Analysis Toolchains}},
  year      = {2019},
  pages     = {523--528},
  doi       = {10.1109/COMPSAC.2019.00080},
}

@TechReport{SARIF,
  author = {David Keaton and ChairLuke Cartey},
  title  = {{Static Analysis Results Interchange Format (SARIF)}},
  year   = {2020},
  school = {OASIS},
}

@Article{Mohammed2017,
  author    = {Nabil M. Mohammed and Mahmood Niazi and Mohammad Alshayeb and Sajjad Mahmood},
  journal   = {Computer Standards \& Interfaces},
  title     = {Exploring Software Security Approaches in Software Development Lifecycle: {A} Systematic Mapping Study},
  year      = {2017},
  pages     = {107--115},
  volume    = {50},
  doi       = {10.1016/j.csi.2016.10.001},
}

@Article{Khan2021,
  author    = {Rafiq Ahmad Khan and Siffat Ullah Khan and Habib Ullah Khan and Muhammad Ilyas},
  journal   = {{IEEE} Access},
  title     = {Systematic Mapping Study on Security Approaches in Secure Software Engineering},
  year      = {2021},
  pages     = {19139--19160},
  volume    = {9},
  doi       = {10.1109/ACCESS.2021.3052311},
}

@InProceedings{Baker2012,
  author    = {Chase Baker and Michael Shin},
  booktitle = {International Conference on Software Security and Reliability (SERE)},
  title     = {{Mapping of Security Concerns in Design to Security Aspects in Code}},
  year      = {2012},
  pages     = {102--110},
  doi       = {10.1109/SERE-C.2012.20},
}

@Article{Berger2014,
  author    = {Thorsten Berger and Rolf{-}Helge Pfeiffer and Reinhard Tartler and Steffen Dienst and Krzysztof Czarnecki and Andrzej Wasowski and Steven She},
  journal   = {Information and Software Technology (IST)},
  title     = {{Variability Mechanisms in Software Ecosystems}},
  year      = {2014},
  number    = {11},
  pages     = {1520--1535},
  volume    = {56},
  doi       = {10.1016/j.infsof.2014.05.005},
}

@TechReport{ISO42010,
  title   = {{Systems and Software Engineering -- Architecture Description}},
  author = {ISO/IEC/IEEE},
  year    = {2011},
  doi     = {10.1109/IEEESTD.2011.6129467},
  type = {Standard},
  number = {42010:2011},
}

@Article{Nguyen2015,
  author    = {Phu Hong Nguyen and Max E. Kramer and Jacques Klein and Yves Le Traon},
  journal   = {Information and Software Technology (IST)},
  title     = {{An Extensive Systematic Review on the Model-Driven Development of Secure Systems}},
  year      = {2015},
  pages     = {62--81},
  volume    = {68},
  doi       = {10.1016/j.infsof.2015.08.006},
}

@phdthesis{Seifermann2022,
    author       = {Seifermann, Stephan},
    year         = {2022},
    title        = {Architectural Data Flow Analysis for Detecting Violations of Confidentiality Requirements},
    doi          = {10.5445/IR/1000148748},
    school       = {Karlsruher Institut für Technologie (KIT)},
    language     = {english}
}

@inproceedings{Alnaim19_VMescape,
author = {Alnaim, Abdulrahman and Alwakeel, Ahmed and Fernandez, Eduardo B.},
title = {A Misuse Pattern for Compromising VMs via Virtual Machine Escape in NFV},
year = {2019},
isbn = {9781450371643},
publisher = {Association for Computing Machinery},
address = {New York, NY, USA},
url = {https://doi.org/10.1145/3339252.3340530},
doi = {10.1145/3339252.3340530},
booktitle = {Proceedings of the 14th International Conference on Availability, Reliability and Security},
articleno = {77},
numpages = {6},
location = {Canterbury, CA, United Kingdom},
series = {ARES '19}
}

@TechReport{Swanson2006,
  author = {Swanson, M and Hash, J and Bowen, P},
  title  = {{Guide for Developing Security Plans for Federal Information Systems}},
  year   = {2006},
  number = {SP 800-18},
  type   = {Special Publication},
  doi    = {10.6028/NIST.SP.800-18r1},
  school = {NIST},
}

@InProceedings{Giffhorn2008,
  author    = {Giffhorn, Dennis and Hammer, Christian},
  booktitle = {International Working Conference on Source Code Analysis and Manipulation (SCAM)},
  title     = {{Precise Analysis of Java Programs Using JOANA}},
  year      = {2008},
  pages     = {267-268},
  doi       = {10.1109/SCAM.2008.17},
}

@InProceedings{Arzt2014a,
  author    = {Arzt, Steven and Rasthofer, Siegfried and Fritz, Christian and Bodden, Eric and Bartel, Alexandre and Klein, Jacques and Le Traon, Yves and Octeau, Damien and McDaniel, Patrick},
  booktitle = {Conference on Programming Language Design and Implementation (PLDI)},
  title     = {{FlowDroid: Precise Context, Flow, Field, Object-Sensitive and Lifecycle-Aware Taint Analysis for Android Apps}},
  year      = {2014},
  pages     = {259–269},
  doi       = {10.1145/2594291.2594299},
}

@InProceedings{Walden2014,
  author    = {Walden, James and Stuckman, Jeff and Scandariato, Riccardo},
  booktitle = {International Symposium on Software Reliability Engineering},
  title     = {{Predicting Vulnerable Components: Software Metrics Vs Text Mining}},
  year      = {2014},
  pages     = {23--33},
  doi       = {10.1109/ISSRE.2014.32},
}

@InProceedings{Zhioua2014,
  author    = {Zeineb Zhioua and Stuart Short and Yves Roudier},
  booktitle = {Computer Software and Applications Conference (COMPSAC)},
  title     = {{Static Code Analysis for Software Security Verification: Problems and Approaches}},
  year      = {2014},
  pages     = {102--109},
  doi       = {10.1109/COMPSACW.2014.22},
}

@InProceedings{Shostack2008,
  author    = {Adam Shostack},
  booktitle = {Workshop on Modeling Security (MODSEC)},
  title     = {{Experiences Threat Modeling at Microsoft}},
  year      = {2008},
}

@Book{Wasowski2023,
  author    = {Andrzej Wasowski and Thorsten Berger},
  publisher = {Springer},
  title     = {{Domain-Specific Languages} -- {Effective Modeling, Automation, and Reuse}},
  year      = {2023},
  isbn      = {978-3-031-23668-6},
  doi       = {10.1007/978-3-031-23669-3},
}

@Article{Peldszus2024,
  author    = {Sven Peldszus and Jens Bürger and Jan Jürjens},
  journal   = {Transactions on Software Engineering (TSE)},
  title     = {{UMLsecRT: Reactive Security Monitoring of Java Applications with Round-Trip Engineering}},
  year      = {2024},
  pages     = {16--47},
  volume    = {50},
  doi       = {10.1109/TSE.2023.3326366},
  issue     = {1},
}

@inproceedings{Rafnsson2020,
author = {Rafnsson, Willard and Giustolisi, Rosario and Kragerup, Mark and H\o{}yrup, Mathias},
title = {{Fixing Vulnerabilities Automatically with Linters}},
year = {2020},
doi = {10.1007/978-3-030-65745-1_13},
booktitle = {International Conference on Network and System Security (NSS)},
pages = {224--244},
}

@article{Peldszus2025,
    author = {Sven Peldszus and Katharina Großer and Marko Konersmann and Waja Brunotte and Maike Ahrens and Kurt Schneider and Jan Jürjens},
    title = {{Too Many Issues: Automatically Prioritizing Analyzer Findings by Tracing Security Importance}},
    journal = {Transactions on Software Engineering and Methodology (TOSEM)},
    year = {2025},
    doi = {10.1145/3744708},
}

@InProceedings{Lenarduzzi2018,
  author    = {Valentina Lenarduzzi and Alberto Sillitti and Davide Taibi},
  booktitle = {International Conference in Software Engineering for Defence Applications (SEDA)},
  title     = {{A Survey on Code Analysis Tools for Software Maintenance Prediction}},
  year      = {2018},
  pages     = {165--175},
  doi       = {10.1007/978-3-030-14687-0_15},
}

@Misc{SonarSource,
  author       = {{SonarSource S.A.}},
  url = {https://docs.sonarsource.com/sonarqube/latest/user-guide/rules/security-related-rules/},
  note         = {last accessed February 10\textsuperscript{th}, 2025},
  title        = {{Security-related Rules in SonarQube}},
}

@InProceedings{Ryan2023,
  author    = {Ita Ryan and Utz Roedig and Klaas{-}Jan Stol},
  booktitle = {Conference on Computer and Communications Security (CCS)},
  title     = {{Unhelpful Assumptions in Software Security Research}},
  year      = {2023},
  pages     = {3460--3474},
  doi       = {10.1145/3576915.3623122},
}

@Misc{cwe,
  author       = {{The MITRE Corporation}},
  url = {https://cwe.mitre.org/},
  note         = {last accessed February 10\textsuperscript{th}, 2025},
  title        = {{Common Weakness Enumeration (CWE)}},
}

@TechReport{SP800-30,
  author = {{Joint Task Force Transformation Initiative}},
  title  = {{Guide for ConductingRisk Assessments}},
  year   = {2012},
  number = {SP 800-30},
  type   = {Special Publication},
  school = {National Institute of Standards and Technology},
}

@Article{RoblesGonzalez2020,
  author    = {Antonio Robles{-}Gonz{\'{a}}lez and Javier Parra{-}Arnau and Jordi Forn{\'{e}}},
  journal   = {Computers \& Security},
  title     = {{A LINDDUN-Based Framework for Privacy Threat Analysis on Identification and Authentication Processes}},
  year      = {2020},
  volume    = {94},
  doi       = {10.1016/J.COSE.2020.101755},
}

@TechReport{SP800-160,
  author = {Ron Ross and Mark Winstead and Michael McEvilley},
  title  = {{Engineering Trustworthy Secure Systems}},
  year   = {2022},
  number = {SP 800-160},
  type   = {Special Publication},
  doi    = {10.6028/NIST.SP.800-160v1r1},
  school = {NIST},
}

@InCollection{Ebert2010,
  author    = {J{\"{u}}rgen Ebert and Daniel Bildhauer},
  booktitle = {{Graph Transformations} and {Model-}{Driven} {Engineering} -- {Essays} {Dedicated} to {Manfred} {Nagl} on the {Occasion of his 65\textsuperscript{th} Birthday}},
  publisher = {Springer},
  title     = {{Reverse Engineering Using Graph Queries}},
  year      = {2010},
  pages     = {335--362},
  doi       = {10.1007/978-3-642-17322-6\_{1}{5}},
}

@Book{xtext,
  author    = {Lorenzo Bettini},
  publisher = {Packt Publishing},
  title     = {{Implementing DSLs with Xtext and Xtend}},
  year      = {2016},
  edition   = {2},
}

@InProceedings{Barros2022,
  author    = {Djonathan Barros and Sven Peldszus and Wesley K. G. Assun{\c{c}}{\~{a}}o and Thorsten Berger},
  booktitle = {International Conference on Model Driven Engineering Languages and Systems (MODELS)},
  title     = {{Editing Support for Software Languages: Implementation Practices in Language Server Protocols}},
  year      = {2022},
  pages     = {232--243},
  doi       = {10.1145/3550355.3552452},
}

@InProceedings{Mazurek2022,
  author    = {Michelle L. Mazurek},
  booktitle = {Conference on Computer and Communications Security (CCS)},
  title     = {{We Are the Experts, and We Are the Problem: The Security Advice Fiasco}},
  year      = {2022},
  pages     = {7},
  doi       = {10.1145/3548606.3559394},
}

@InProceedings{Vassallo2018,
  author    = {Vassallo, Carmine and Palomba, Fabio and Bacchelli, Alberto and Gall, Harald C.},
  booktitle = {International Conference on Automated Software Engineering (ASE)},
  title     = {{Continuous Code Quality: Are We (Really) Doing That?}},
  year      = {2018},
  pages     = {790-795},
  doi       = {10.1145/3238147.3240729},
}

@Misc{OWASPTop10,
  author       = {{Open Web Application Security Project}},
  url = {https://owasp.org/Top10/A04_2021-Insecure_Design/},
  note         = {last visited February 10\textsuperscript{th}, 2025},
  title        = {{OWASP Top 10: A04 Insecure Design}},
  year         = {2021},
}

@Misc{OWASPFoundation2022,
  author = {{OWASP Foundation}},
  note   = {last visited February 10\textsuperscript{th}, 2025},
  title  = {{OWASP Source Code Analysis Tools List}},
  year   = {2022},
  url    = {https://owasp.org/www-community/Source_Code_Analysis_Tools},
}

@Misc{OWASPDependency,
  author = {{OWASP Foundation}},
  note   = {last visited February 10\textsuperscript{th}, 202},
  title  = {{OWASP Dependency-Check}},
  year   = {2022},
  url    = {https://owasp.org/www-project-dependency-check/},
}

@Article{Uzunov2012,
  author    = {Anton V. Uzunov and Eduardo B. Fern{\'{a}}ndez and Katrina Falkner},
  journal   = {Journal of Universal Computer Science},
  title     = {{Engineering Security into Distributed Systems: {A} Survey of Methodologies}},
  year      = {2012},
  number    = {20},
  pages     = {2920--3006},
  volume    = {18},
  doi       = {10.3217/JUCS-018-20-2920},
}

@Article{Mashkoor2023,
  author    = {Atif Mashkoor and Alexander Egyed and Robert Wille and Sebastian Stock},
  journal   = {Journal of Software: Evolution and Process},
  title     = {{Model-Driven Engineering of Safety and Security Software Systems: A Systematic Mapping Study and Future Research Directions}},
  year      = {2023},
  number    = {7},
  volume    = {35},
  doi       = {10.1002/SMR.2457},
}

@Article{DenBerghe2017,
  author    = {Alexander van Den Berghe and Riccardo Scandariato and Koen Yskout and Wouter Joosen},
  journal   = {Software \& Systems Modeling (SoSyM)},
  title     = {{Design Notations for Secure Software: A Systematic Literature Review}},
  year      = {2017},
  number    = {3},
  pages     = {809--831},
  volume    = {16},
  doi       = {10.1007/S10270-015-0486-9},
}

@InProceedings{Yurchenko2017,
  author    = {Yurchenko, Kateryna and Behr, Moritz and Klare, Heiko and Kramer, Max E and Reussner, Ralf H},
  booktitle = {MODELS (Satellite Events)},
  title     = {{Architecture-driven Reduction of Specification Overhead for Verifying Confidentiality in Component-based Software Systems}},
  year      = {2017},
  pages     = {321--323},
}

@Misc{Ralph2023,
  author = {Paul Ralph and others},
  url   = {https://acmsigsoft.github.io/EmpiricalStandards/docs/},
  title  = {{SIGSOFT Empirical Standards}},
  year   = {2023},
    note         = {last accessed February 10\textsuperscript{th}, 2025},
}

@Book{Steinberg2008,
  author    = {Dave Steinberg and Frank Budinsky and Marcelo Paternostro and Ed Merks},
  publisher = {Addison-Wesley Professional},
  title     = {{EMF: Eclipse Modeling Framework}},
  year      = {2008},
}

@TechReport{uml,
  author = {{Object Management Group}},
  title  = {{Unified Modeling Language (UML) Superstructure}},
  year   = {2011},
  number = {2.4.1},
}

@Article{Zhang2023,
  author   = {Zhang, Ying and Kabir, Md Mahir Asef and Xiao, Ya and Yao, Danfeng and Meng, Na},
  journal  = {Transactions on Software Engineering (TSE)},
  title    = {{Automatic Detection of Java Cryptographic API Misuses: Are We There Yet?}},
  year     = {2023},
  number   = {1},
  pages    = {288-303},
  volume   = {49},
  doi      = {10.1109/TSE.2022.3150302},
}

@InProceedings{Liu2023,
  author    = {Liu, Han and Chen, Sen and Feng, Ruitao and Liu, Chengwei and Li, Kaixuan and Xu, Zhengzi and Nie, Liming and Liu, Yang and Chen, Yixiang},
  booktitle = {International Symposium on Software Testing and Analysis (ISSTA)},
  title     = {{A Comprehensive Study on Quality Assurance Tools for Java}},
  year      = {2023},
  pages     = {285–297},
  doi       = {10.1145/3597926.3598056},
}

@InProceedings{Nachtigall2022,
  author    = {Nachtigall, Marcus and Schlichtig, Michael and Bodden, Eric},
  booktitle = {International Symposium on Software Testing and Analysis (ISSTA)},
  title     = {{A Large-Scale Study of Usability Criteria Addressed by Static Analysis Tools}},
  year      = {2022},
  pages     = {532–543},
  doi       = {10.1145/3533767.3534374},
}

@InProceedings{Moor2007,
  author    = {Moor, Oege de and Verbaere, Mathieu and Hajiyev, Elnar and Avgustinov, Pavel and Ekman, Torbjorn and Ongkingco, Neil and Sereni, Damien and Tibble, Julian},
  booktitle = {International Working Conference on Source Code Analysis and Manipulation (SCAM)},
  title     = {{Keynote Address: .QL for Source Code Analysis}},
  year      = {2007},
  pages     = {3-16},
  doi       = {10.1109/SCAM.2007.31},
}

@InProceedings{Greiner2017,
  author    = {Simon Greiner and Martin Mohr and Bernhard Beckert},
  booktitle = {International Conference on Software Engineering and Formal Methods (SEFM)},
  title     = {{Modular Verification of Information Flow Security in Component-Based Systems}},
  year      = {2017},
  pages     = {300--315},
  doi       = {10.1007/978-3-319-66197-1_19},
}

@phdthesis{Greiner2018_Diss,
    author       = {Greiner, Simon},
    year         = {2018},
    title        = {A Framework for Non-Interference in Component-Based Systems},
    doi          = {10.5445/IR/1000082042},
    publisher    = {{Karlsruher Institut für Technologie (KIT)}},
    pagetotal    = {229},
    school       = {Karlsruher Institut für Technologie (KIT)},
    language     = {english}
}

@Book{Ahrendt2016,
  author    = {Wolfgang Ahrendt and Bernhard Beckert and Richard Bubel and Reiner H\"{a}hnle and Peter H. Schmitt and Mattias Ulbrich},
  publisher = {Springer},
  title     = {{Deductive Software Verification -- The KeY Book} -- {From Theory to Practice}},
  year      = {2016},
  booktitle = {Deductive Software Verification - The KeY Book - From Theory to Practice},
  doi       = {10.1007/978-3-319-49812-6},
}

@InProceedings{Balliu2017,
  author    = {Balliu, Musard and Schoepe, Daniel and Sabelfeld, Andrei},
  booktitle = {European Symposium on Research in Computer Security (ESORICS)},
  title     = {{We Are Family: Relating Information-Flow Trackers}},
  year      = {2017},
  pages     = {124--145},
}

@InProceedings{Singh2017,
  author    = {Devarshi Singh and Varun Ramachandra Sekar and Kathryn T. Stolee and Brittany Johnson},
  booktitle = {Symposium on Visual Languages and Human-Centric Computing (VL/HCC)},
  title     = {{Evaluating How Static Analysis Tools Can Reduce Code Review Effort}},
  year      = {2017},
  pages     = {101--105},
  doi       = {10.1109/VLHCC.2017.8103456},
}

@misc{tiobe,
    title =  {{TIOBE Index for January 2025}},
    author = {Paul Jansen},
    year = {2025},
    url = {https://www.tiobe.com/tiobe-index/},
  note         = {last accessed February 10\textsuperscript{th}, 2025},
}

@inproceedings{Hermann2025,
    author = {Kevin Hermann and Sven Peldszus and Jan-Philipp Steghöfer and Thorsten Berger},
    title = {{An Exploratory Study on the Engineering of Security Features}},
    booktitle = {International Conference on Software Engineering (ICSE)},
    year = {2025},
    pages = {2470--2482},
    doi = {10.1109/ICSE55347.2025.00184},
}

@Misc{Replication,
  author       = {Sven Peldszus and Frederik Reiche and Kevin Hermann and Sophie Corallo and Thorsten Berger and Robert Heinrich},
  url = {https://www.dropbox.com/scl/fi/rwlph6i1yk4gbs8gieu9a/SecLan-SupplementaryMaterial.zip?rlkey=l8rdyscz3trvv8hdg84s6vivn&st=pki3tkhp&dl=0},
  note         = {We will upload the supplementary material to Zenodo upon acceptance.},
  title        = {Replication Package},
  year         = {2025},
}

@InProceedings{Massey2008,
  author    = {Aaron K. Massey and Paul N. Otto and Annie I. Ant{\'{o}}n},
  booktitle = {International Requirements Engineering Conference (RE)},
  title     = {{Aligning Requirements with HIPAA in the iTrust System}},
  year      = {2008},
  pages     = {335--336},
  doi       = {10.1109/RE.2008.53},
}

@Book{Shostack2014,
  author    = {Shostack, Adam},
  publisher = {Wiley},
  title     = {{Threat Modeling: Designing for Security}},
  year      = {2014},
}

@InProceedings{Ma2017,
  author    = {Siqi Ma and Ferdian Thung and David Lo and Cong Sun and Robert H. Deng},
  booktitle = {European Symposium on Research in Computer Security (ESORICS)},
  title     = {{VuRLE: Automatic Vulnerability Detection and Repair by Learning from Examples}},
  year      = {2017},
  pages     = {229--246},
  doi       = {10.1007/978-3-319-66399-9\_{1}{3}},
}

@InProceedings{Clavel2008,
  author    = {Manuel Clavel and Viviane Torres da Silva and Christiano Braga and Marina Egea},
  booktitle = {European Conference on Model Driven Architecture - Foundations and Applications (CMDA-FA)},
  title     = {{Model-Driven Security in Practice: An Industrial Experience}},
  year      = {2008},
  pages     = {326--337},
  doi       = {10.1007/978-3-540-69100-6\_{2}{2}},
}

@InProceedings{Juerjens2008,
  author    = {Jan J{\"{u}}rjens and J{\"{o}}rg Schreck and Peter Bartmann},
  booktitle = {International Conference on Software Engineering (ICSE)},
  title     = {{Model-based Security Analysis for Mobile Communications: An Industrial Application of UMLsec}},
  year      = {2008},
  pages     = {683--692},
  doi       = {10.1145/1368088.1368186},
}

@InProceedings{Edmundson2013,
  author    = {Anne Edmundson and Brian Holtkamp and Emanuel Rivera and Matthew Finifter and Adrian Mettler and David A. Wagner},
  booktitle = {International Symposium on Engineering Secure Software and Systems (ESSoS)},
  title     = {{An Empirical Study on the Effectiveness of Security Code Review}},
  year      = {2013},
  pages     = {197--212},
  doi       = {10.1007/978-3-642-36563-8\_{1}{4}},
}

@InProceedings{Baca2009,
  author    = {Dejan Baca and Kai Petersen and Bengt Carlsson and Lars Lundberg},
  booktitle = {International Conference on Availability, Reliability and Security (ARES)},
  title     = {{Static Code Analysis to Detect Software Security Vulnerabilities - Does Experience Matter?}},
  year      = {2009},
  pages     = {804--810},
  doi       = {10.1109/ARES.2009.163},
}

@Article{Hough2022,
  author  = {Katherine Hough and Jonathan Bell},
  journal = {Transactions Software Engineering and Methodology (TOSEM)},
  title   = {{A Practical Approach for Dynamic Taint Tracking with Control-flow Relationships}},
  year    = {2022},
  number  = {2},
  pages   = {26:1--26:43},
  volume  = {31},
  doi     = {10.1145/3485464},
}

@InProceedings{Wang2015,
  author    = {Wang, Haoyu and Guo, Yao and Tang, Zihao and Bai, Guangdong and Chen, Xiangqun},
  title     = {{Reevaluating Android Permission Gaps with Static and Dynamic Analysis}},
  booktitle = {Global Communications Conference (GLOBECOM)},
  year      = {2015},
  pages     = {1--6},
  doi       = {10.1109/GLOCOM.2015.7417621},
}

@Misc{Wheler2001,
  author  = {Wheler, David A.},
  title   = {Flawfinder},
  year    = {2001},
  url     = {https://dwheeler.com/flawfinder/},
  note         = {last accessed February 10\textsuperscript{th}, 2025},
}

@Misc{CodeQL,
  url = {https://codeql.github.com/},
  title        = {{CodeQL Website}},
    author = {GitHub},
  year         = {2024},
  note         = {last accessed February 10\textsuperscript{th}, 2025},
}

@Misc{pmd,
  author       = {{PMD Open Source Project}},
  url = {https://docs.pmd-code.org/latest/pmd_rules_apex_security.html},
  title        = {{PMD Security Rules}},
  year         = {2024},
  note         = {last accessed February 10\textsuperscript{th}, 2025},
}

@InProceedings{Geismann2021,
  author    = {Johannes Geismann and Bastian Haverkamp and Eric Bodden},
  booktitle = {International Workshop on Model-Driven Engineering for Software Architecture},
  title     = {{Ensuring Threat-Model Assumptions by Using Static Codeanalyses}},
  year      = {2021},
}

@InProceedings{Kulenovic2014,
  author    = {Kulenovic, Melina and Donko, Dzenana},
  booktitle = {2014 37th International Convention on Information and Communication Technology, Electronics and Microelectronics (MIPRO)},
  title     = {A survey of static code analysis methods for security vulnerabilities detection},
  year      = {2014},
  pages     = {1381-1386},
  doi       = {10.1109/MIPRO.2014.6859783},
}

@book{Allen2008,
  title={Software Security Engineering},
  author={Allen, Julia H. and Barnum, Sean and Ellison, Robert J. and McGraw, Gary and Mead, Nancy R.},
  isbn={978-8131726488},
  year={2008},
  publisher={Pearson Education}
}

@book{Bass2003,
author = {Bass, Len and Clements, Paul and Kazman, Rick},
year = {2003},
month = {01},
pages = {},
title = {Software Architecture In Practice},
isbn = {978-0321154958},
publisher = {Addison-Wesley Longman}
}

@Book{Schumacher2013,
  author    = {Schumacher, Markus and Fernandez-Buglioni, Eduardo and Hybertson, Duane and Buschmann, Frank and Sommerlad, Peter},
  publisher = {John Wiley \& Sons},
  title     = {{Security Patterns: Integrating Security and Systems Engineering}},
  year      = {2013},
}

@Article{Tsipenyuk2005,
  author  = {Tsipenyuk, Katrina and Chess, Brian and McGraw, Gary},
  journal = {IEEE Security \& Privacy},
  title   = {{Seven Pernicious Kingdoms: A Taxonomy of Software Security Errors}},
  year    = {2005},
  number  = {6},
  pages   = {81-84},
  volume  = {3},
  doi     = {10.1109/MSP.2005.159},
}

@InProceedings{Almorsy2013,
  author    = {Almorsy, Mohamed and Grundy, John and Ibrahim, Amani S.},
  booktitle = {International Conference on Software Engineering (ICSE)},
  title     = {Automated software architecture security risk analysis using formalized signatures},
  year      = {2013},
  pages     = {662--671},
  doi       = {10.1109/ICSE.2013.6606612},
}

@InProceedings{AlBreiki2014,
  author    = {AlBreiki, Hamda Hasan and Mahmoud, Qusay H.},
  booktitle = {International Conference on Innovations in Information Technology (IIT)},
  title     = {Evaluation of static analysis tools for software security},
  year      = {2014},
  pages     = {93--98},
  doi       = {10.1109/INNOVATIONS.2014.6987569},
}

@InProceedings{Lipp2022,
  author    = {Lipp, Stephan and Banescu, Sebastian and Pretschner, Alexander},
  booktitle = {International Symposium on Software Testing and Analysis (ISSTA)},
  title     = {An empirical study on the effectiveness of static C code analyzers for vulnerability detection},
  year      = {2022},
  pages     = {544--555},
  doi       = {10.1145/3533767.3534380},
}

@InProceedings{Imtiaz2021,
  author    = {Imtiaz, Nasif and Thorn, Seaver and Williams, Laurie},
  booktitle = {International Symposium on Empirical Software Engineering and Measurement (ESEM)},
  title     = {{A Comparative Study of Vulnerability Reporting by Software Composition Analysis Tools}},
  year      = {2021},
  doi       = {10.1145/3475716.3475769},
}

@Book{Chess2007,
  author    = {Chess, Brian and West, Jacob},
  publisher = {Pearson Education},
  title     = {{Secure Programming with Static Analysis}},
  year      = {2007},
}

@Article{Wang2020,
  author   = {Wang, Zuoguang and Sun, Limin and Zhu, Hongsong},
  journal  = {IEEE Access},
  title    = {{Defining Social Engineering in Cybersecurity}},
  year     = {2020},
  pages    = {85094--85115},
  volume   = {8},
  doi      = {10.1109/ACCESS.2020.2992807},
}

@PhdThesis{Hahner2024,
  author  = {Hahner, Sebastian},
  school  = {Karlsruhe Institute of Technology (KIT)},
  title   = {Architecture-Based and Uncertainty-Aware Confidentiality Analysis},
  year    = {2024},
  address = {Karlsruhe, Germany},
  type    = {Dissertation},
}

@InProceedings{Pawar2015,
  author    = {Mohan V. Pawar and J. Anuradha},
  booktitle = {International Conference on Computer, Communication and Convergence (ICCC)},
  title     = {{Network Security and Types of Attacks in Network}},
  year      = {2015},
  pages     = {503--506},
  doi       = {10.1016/j.procs.2015.04.126},
}

@Misc{findsecbugs,
    title = {{Find Security Bugs}},
    author = {Philippe Arteau},
    url   = {https://find-sec-bugs.github.io/},
    year = {2022},
    note = {last accessed February 10\textsuperscript{th}, 2025},
}

@InProceedings{Mousa2020,
  author    = {Mousa, Abdulazeez and Karabatak, Murat and Mustafa, Twana},
  booktitle = {International Symposium on Digital Forensics and Security (ISDFS)},
  title     = {{Database Security Threats and Challenges}},
  year      = {2020},
  pages     = {1-5},
  doi       = {10.1109/ISDFS49300.2020.9116436},

}

@InProceedings{Cowan1998,
  author    = {Crispan Cowan and Calton Pu and Dave Maier and Jonathan Walpole and Peat Bakke and Steve Beattie and Aaron Grier and Perry Wagle and Qian Zhang and Heather Hinton},
  booktitle = {USENIX Security Symposium},
  title     = {{StackGuard: Automatic Adaptive Detection and Prevention of Buffer-Overflow Attacks}},
  year      = {1998},
}

@InProceedings{Apvrille2013,
  author    = {Ludovic Apvrille and Yves Roudier},
  booktitle = {Asia-Pacific Council on Systems Engineering Conference (APCOSEC)},
  title     = {{SysML-Sec: A SysML Environment for the Design and Development of Secure Embedded Systems}},
  year      = {2013},
}

@InProceedings{Tang2017,
  author    = {Junwei Tang and Ruixuan Li and Hongmu Han and Heng Zhan and Xiwu Gu},
  booktitle = {International Conference on Trust, Security and Privacy in Computing and Communications (TrustCom)},
  title     = {{Detecting Permission Over-claim of Android Applications with Static and Semantic Analysis Approach}},
  year      = {2017},
  pages     = {706-713},
  doi       = {10.1109/Trustcom/BigDataSE/ICESS.2017.303},
}

@Article{Salnitri2015,
  author  = {Mattia Salnitri and Fabiano Dalpiaz and Paolo Giorgini},
  journal = {Software \& Systems Modeling (SoSyM)},
  title   = {{Designing Secure Business Processes with SecBPMN}},
  year    = {2015},
  number  = {3},
  pages   = {737--757},
  volume  = {16},
  doi     = {10.1007/S10270-015-0499-4},
}

@Article{Jin2021,
  author  = {Wenhui Jin and Sami Ullah and Dongmin Yoo and Heekuck Oh},
  journal = {IEEE Access},
  title   = {{NPDHunter: Efficient Null Pointer Dereference Vulnerability Detection in Binary}},
  year    = {2021},
  pages   = {90153--90169},
  volume  = {9},
  doi     = {10.1109/ACCESS.2021.3091209},
}

@TechReport{Kramer2017,
  author      = {Max E. Kramer and Martin Hecker and Simon Greiner and Kaibin Bao and Kateryna Yurchenko},
  institution = {Karlsruhe Institute of Technology (KIT)},
  title       = {{M}odel-{D}riven {S}pecification and {A}nalysis of {C}onfidentiality in {C}omponent-{B}ased {Systems}},
  year        = {2017},
  number      = {12},
  doi         = {10.5445/IR/1000076957},
}

@Book{Juerjens2005,
  author    = {Jan J{\"{u}}rjens},
  publisher = {Springer},
  title     = {{Secure Systems Development with UML}},
  year      = {2005},
  isbn      = {978-3-540-00701-2},
}

@InProceedings{Kumar2017,
  author    = {Rajesh Kumar and Marielle Stoelinga},
  booktitle = {International Symposium on High Assurance Systems Engineering (HASE)},
  title     = {{Quantitative Security and Safety Analysis with Attack-Fault Trees}},
  year      = {2017},
  pages     = {25--32},
  doi       = {10.1109/HASE.2017.12},
}

@InProceedings{Lodderstedt2002,
  author    = {Torsten Lodderstedt and David Basin and Jürgen Doser},
  booktitle = {Conference on the Unified Modeling Language},
  title     = {{SecureUML: A UML-Based Modeling Language for Model-Driven Security}},
  year      = {2002},
  pages     = {426--441},
}

@InProceedings{Zhang2019,
  author    = {Tong Zhang and Wenbo Shen and Dongyoon Lee and Changhee Jung and Ahmed M. Azab and Ruowen Wang},
  booktitle = {USENIX Security Symposium},
  title     = {{PeX: A Permission Check Analysis Framework for Linux Kernel}},
  year      = {2019},
  pages     = {1205--1220},
}

@Misc{Clang,
  title   = {{Clang Static Analyzer Security Checks}},
  author = {{The Clang Team}},
  url     = {https://clang.llvm.org/docs/ClangStaticAnalyzer.html},
  date = {2025},
  note         = {last accessed February 10\textsuperscript{th}, 2025},
}

@misc{cwe200,
    title = {{CWE-200: Exposure of Sensitive Information to an Unauthorized Actor}},
    author = {{Common Weakness Enumeration}},
    year = {2006},
    url = {https://cwe.mitre.org/data/definitions/200.html},
}

@INPROCEEDINGS{Hu2003,
  author={Yih-Chun Hu and Adrian Perrig and David B. Johnson},
  booktitle={Joint Conference of the IEEE Computer and Communications Societies (INFOCOM)}, 
  title={{Packet Leashes: A Defense Against Wormhole Attacks in Wireless Networks}}, 
  year={2003},
  pages={1976--1986},
  doi={10.1109/INFCOM.2003.1209219}
}

@techreport{Dougherty2009,
    author = {Chad Dougherty and Kirk Sayre and Robert C. Seacord and David Svoboda and Kazuya Togashi},
    title = {{Secure Design Patterns}},
    institution = {CERT Program},
    year = {2009},
    number ={CMU/SEI-2009-TR-010},
}

@article{shedden2011incorporating,
  title={{Incorporating a Knowledge Perspective into Security Risk Assessments}},
  author={Shedden, Piya and Scheepers, Rens and Smith, Wally and Ahmad, Atif},
  journal={Vine},
  volume={41},
  number={2},
  pages={152--166},
  year={2011},
  publisher={Emerald Group Publishing Limited}
}

@InProceedings{Walter2022,
  author    = {Maximilian Walter and Robert Heinrich and Ralf Reussner},
  booktitle = {International Conference on Software Architecture (ICSA)},
  title     = {{Architectural Attack Propagation Analysis for Identifying Confidentiality Issues}},
  year      = {2022},
  pages     = {1--12},
  doi       = {10.1109/ICSA53651.2022.00009},
}

@Misc{Bandit,
    title   = {{Bandit Website}},
    author = {PyCQA},
    url     = {https://github.com/PyCQA/bandit},
    year = {2025},
  note         = {last accessed February 10\textsuperscript{th}, 2025},
}

@InProceedings{Krueger2017,
  author    = {Stefan Kr{\"{u}}ger and Sarah Nadi and Michael Reif and Karim Ali and Mira Mezini and Eric Bodden and Florian G{\"{o}}pfert and Felix G{\"{u}}nther and Christian Weinert and Daniel Demmler and Ram Kamath},
  booktitle = {International Conference on Automated Software Engineering (ASE)},
  title     = {{CogniCrypt: Supporting Developers in Using Cryptography}},
  year      = {2017},
  pages     = {931--936},
  doi       = {10.1109/ASE.2017.8115707},
}

@Book{Peldszus2022a,
  author    = {Sven Peldszus},
  publisher = {Springer},
  title     = {{Security Compliance in Model-driven Development of Software Systems in Presence of Long-Term Evolution and Variants}},
  year      = {2022},
  isbn      = {978-3-658-37665-9},
  doi       = {10.1007/978-3-658-37665-9},
}

@article{hermann2025taxonomy,
    title={{A Taxonomy of Functional Security Features and How They Can Be Located}}, 
    author={Kevin Hermann and Simon Schneider and Catherine Tony and Asli Yardim and Sven Peldszus and Thorsten Berger and Riccardo Scandariato and M. Angela Sasse and Alena Naiakshina},
    year={2025},
    journal =  {Empirical Software Engineering (EMSE)}, 
    volume = 30, 
    number = 117,
    doi = {10.1007/s10664-025-10649-7},
}

@inproceedings {lawall18,
author = {Julia Lawall and Gilles Muller},
title = {Coccinelle: 10 Years of Automated Evolution in the Linux Kernel},
booktitle = {2018 USENIX Annual Technical Conference (USENIX ATC 18)},
year = {2018},
isbn = {978-1-939133-01-4},
address = {Boston, MA},
pages = {601--614},
url = {https://www.usenix.org/conference/atc18/presentation/lawall},
publisher = {USENIX Association},
month = jul
}

@InProceedings{calcagno15,
author="Calcagno, Cristiano
and Distefano, Dino
and Dubreil, Jeremy
and Gabi, Dominik
and Hooimeijer, Pieter
and Luca, Martino
and O'Hearn, Peter
and Papakonstantinou, Irene
and Purbrick, Jim
and Rodriguez, Dulma",
editor="Havelund, Klaus
and Holzmann, Gerard
and Joshi, Rajeev",
title="Moving Fast with Software Verification",
booktitle="NASA Formal Methods",
year="2015",
publisher="Springer International Publishing",
address="Cham",
pages="3--11",
isbn="978-3-319-17524-9"
}

@TechReport{CC,
  author = {{ISO/IEC JTC 1/SC 27}},
  institution = {{International Organization for Standardization (ISO)}},
  title  = {{Common Criteria for Information Technology Security Evaluation}},
  year   = {2009},
  type   = {International Standard},
  number = {ISO/IEC 15408},
}

@TechReport{ISO21434,
  author = {{ISO/SAE 21434:2021}},
  institution = {{International Organization for Standardization (ISO)}},
  title  = {{Road vehicles — Cybersecurity engineering}},
  year   = {2021},
  type   = {International Standard},
  number = {ISO/SAE 21434},
}

@misc{semgrep,
    author = {{Semgrep, Inc.}},
    title = {Semgrep Website},
    url = {https://semgrep.dev/},
	urldate = {2026-04-07},
}

@Article{Abramov2012,
  author  = {Jenny Abramov and Arnon Sturm and Peretz Shoval},
  journal = {Information and Software Technology (IST)},
  title   = {{Evaluation of the Pattern-based method for Secure Development (PbSD): A controlled experiment}},
  year    = {2012},
  number  = {9},
  pages   = {1029--1043},
  volume  = {54},
  doi     = {10.1016/J.INFSOF.2012.04.001},
}
